\documentclass[acmsmall, review=false, screen=true]{acmart}

\PassOptionsToPackage{cmex10}{amsmath}

\usepackage[utf8]{inputenc}
\usepackage[T1]{fontenc}
\usepackage{textcomp}
\usepackage{fixltx2e}
\usepackage{placeins}
\usepackage{microtype}
\usepackage{booktabs}
\usepackage{tabularx}
\definecolor{Gray}{gray}{0.9}
\newcolumntype{V}{>{\centering\arraybackslash}p{2.5cm}}
\makeatletter
\newcommand{\manuallabel}[2]{\def\@currentlabel{#2}\label{#1}}
\makeatother

\usepackage[cmex10]{amsmath}
\usepackage{amsfonts}
\newcommand{\overbar}[1]{\mkern 1.5mu\overline{\mkern-1.5mu#1\mkern-1.5mu}\mkern 1.5mu}

\usepackage{url}
\usepackage{multirow,array,arydshln}
\usepackage{acro}
\DeclareAcronym{PET}{
	short=PET,
	long=privacy-enhancing technology,
	long-plural-form=privacy-enhancing technologies
}
\DeclareAcronym{SNP}{
	short=SNP,
	long=single nucleotide polymorphism
}
\DeclareAcronym{MOOC}{
	short=MOOC,
	long=Massive Open Online Course
}

\usepackage{hyphenat}

\usepackage{todonotes}
\acmJournal{CSUR}
\acmVolume{51}
\acmNumber{3}
\acmArticle{57}
\acmYear{2018}
\acmMonth{6}
\copyrightyear{2018}

\setcopyright{acmlicensed}
\acmDOI{0000001.0000001}

\received{November 2015}
\received[revised]{January 2017}
\received[revised]{October 2017}
\received[accepted]{November 2017}

\begin{document}

\title{Technical Privacy Metrics: a Systematic Survey}
% I like 'technical' as an indicator that we don't look at 'soft' user-attitude/concern type metrics
% But I don't think it has been used before. Is there a better term?

% A toolbox of privacy metrics
% Privcy Metrics in a Nutshell
\author{Isabel Wagner}
\orcid{0000-0003-0242-6278}
\affiliation{
  \institution{De Montfort University}
  \department{Cyber Security Centre}
  \city{Leicester}
  \postcode{LE1 9BH}
  \country{UK}
}
\email{isabel.wagner@dmu.ac.uk}
\author{David Eckhoff}
\orcid{}
\affiliation{
  \institution{TUMCREATE Ltd.}
  \department{Area-Interlinking Design Analysis}
  \postcode{138602}
  \country{Singapore}
}
\email{david.eckhoff@tum-create.edu.sg}

\begin{abstract}\nohyphens{%
The goal of privacy metrics is to measure the degree of privacy enjoyed by users in a system and the amount of protection offered by privacy-enhancing technologies.
In this way, privacy metrics contribute to improving user privacy in the digital world.
The diversity and complexity of privacy metrics in the literature makes an informed choice of metrics challenging.
As a result, instead of using existing metrics, new metrics are proposed frequently, and privacy studies are often incomparable.
In this survey we alleviate these problems by structuring the landscape of privacy metrics.
To this end, we explain and discuss a selection of over eighty privacy metrics and introduce categorizations based on the aspect of privacy they measure, their required inputs, and the type of data that needs protection.
In addition, we present a method on how to choose privacy metrics based on nine questions that help identify the right privacy metrics for a given scenario, and highlight topics where additional work on privacy metrics is needed.
Our survey spans multiple privacy domains and can be understood as a general framework for privacy measurement.
}\end{abstract}

% \category{D.2.8}{Software Engineering}{Metrics}
% \category{C.4}{Performance of Systems}{Measurement techniques, Performance attributes}
% % D.2.8 SOFTWARE ENGINEERING : Metrics
% % D.4.8 OPERATING SYSTEMS : Performance - Measurements
% \category{K.4.1}{Computers and Society}{Public Policy Issues---\emph{privacy}}
% 
% \terms{Measurement, Security}

\begin{tikzpicture}[overlay, remember picture]
\path (current page.north east) ++(-3,-0.6) node[below left, color=red, text width=12cm] (A){\footnotesize © ACM 2018. This is the authors' version of the work. It is posted here for your personal use. Not for redistribution.
The definitive Version of Record was published in ACM Computing Surveys (CSUR), http://dx.doi.org/10.1145/3168389.};
\node (C)[below=0.2cm of A, color=red, text width=12cm]{\footnotesize Please cite this work as: Isabel Wagner and David Eckhoff. 2018. Technical Privacy Metrics: A Systematic Survey. ACM Computing Surveys, Vol 51, No 3, Article 57 (June 2018)};
\end{tikzpicture}



% \acmformat{Isabel Wagner and David Eckhoff, 2017. Technical Privacy Metrics: a Systematic Survey.}

% \begin{bottomstuff}
% % This work is supported by the National Science Foundation, under
% % grant CNS-0435060, grant CCR-0325197 and grant EN-CS-0329609.
% % 
% Author's current addresses: I. Wagner (isabel.wagner@dmu.ac.uk, iw@ieee.org), De Montfort University, The Gateway, Leicester, LE1 9BH, United Kingdom;
% D. Eckhoff (david.eckhoff@tum-create.edu.sg), TUMCREATE, 1 Create Way, Singapore 138602.
% \end{bottomstuff}

\maketitle

\section{Introduction}
\label{sec:intro}

%European Convention on Human Right
%~\cite[Art.~8]{council_of_europe_european_2010}.
%has the right to respect for his private and family
%life, his home and his correspondence

Privacy is a fundamental human right codified in the United Nations Universal Declaration of Human Rights, which states that ``no one shall be subjected to arbitrary interference with his privacy, family, home or correspondence''~\cite[Art.~12]{UN1948universal}. 
However, it is difficult to define what exactly privacy is.
As early as 1967, \citeANP{westin1967privacy}~\cite{westin1967privacy} defined privacy as ``the ability of an individual to control the terms under which personal information is acquired and used.''
Personal information, according to the EU General Data Protection Regulation (and the OECD privacy framework~\cite{oecd2013privacy}), is ``any information relating to an [...] identifiable natural person''~\cite{european2016general}.

% A more practical definition of privacy is contextual integrity~\cite{nissenbaum2004privacy}, which states that all information is associated with a specific context such as a hospital visit, so that social norms for each context determine how information may be used.
\citeANP{nissenbaum2004privacy}~\cite{nissenbaum2004privacy} makes these definitions more practical and defines privacy in terms of contextual integrity, where information is associated with a specific context (e.g., a hospital visit), and social norms for this context dictate how information may be used or shared.
A privacy violation is then the use of personal information other than the norm allows.
Although contextual integrity clearly defines when a privacy violation has occurred, it provides no protection mechanism other than policy and regulations.

\Acfp{PET} protect privacy based on technology rather than policy, and can thus offer much stronger protection.
To judge the efficacy of \acp{PET}, privacy metrics are needed that can measure the level of privacy in a system, or the privacy provided by a given \ac{PET}.
A technical privacy metric takes properties of a system as an input (e.g., the amount of sensitive information leaked or the number of users who are indistinguishable with respect to some characteristic) and yields a numerical (or sometimes canonical) value, which allows to quantify the privacy level in a system and subsequently the comparison of different \acp{PET}.
Equally, the parameters of some privacy methods can be regarded as privacy metrics, e.g. the $k$ in $k$-anonymity (see Section \ref{metric:k-anonymity}).
Privacy metrics can be used in different contexts (or domains), and they can differ with regard to the kind of adversary they consider, the data sources they assume to be available to the adversary, and the aspects of privacy they measure.
% , e.g., the overall level of privacy protection in a system or privacy on a per-user basis.
% Many metrics assume some kind of adversary who is trying to learn sensitive information, others evaluate the characteristics (e.g., the diversity) of disclosed information.

%Many metrics have been proposed in the literature, using different approaches and measuring different aspects of privacy.
%
% However, the availability of a multitude of metrics brings two problems. 
% True, this is a problem; but our paper does not address it.
% First, it is difficult to compare \acp{PET} from different publications because the authors are very likely to have used different metrics.
Despite the large number of metrics in the literature, a structured and comprehensive overview of privacy metrics does not yet exist.
This makes informed decisions about which metrics to select for the evaluation of \acp{PET} difficult.
This in turn can lead to the choice of ineffective \acp{PET}, which is worrisome considering the pervasiveness of systems that can violate privacy~\cite{eckhoff2017privacy}.
% how will this paper help?
In this paper, we structure the landscape of privacy metrics, focusing on technical metrics that measure the degree of privacy in a system or the effectiveness of \acp{PET}.
% Metrics that measure the privacy attitudes and behaviors of users are out of scope for this paper; instead, we point to \citeN{preibusch_guide_2013} as an excellent starting point.
In detail, our contributions are as follows:

\begin{itemize}
	\item We review conditions for the quality of privacy metrics (Section \ref{sec:requirements}). These are essential as a basis for an informed decision about privacy metrics.
	\item We describe a selection of privacy domains including communication systems and databases to provide context and examples throughout the survey (Section \ref{sec:domains}).
	\item We identify four common characteristics that can classify privacy metrics (Section \ref{sec:characteristics}):
	\begin{itemize}
		\item \textit{Adversary models} describe the capabilities the adversary is assumed to have.
		\item \textit{Data sources} describe how the adversary might obtain the information a \ac{PET} aims to protect: from public data, observable data, re-purposed data, or other sources.
		\item \textit{Inputs} describe what information is used to compute a metric: the adversary's estimate, resources available to the adversary, the true outcome, prior knowledge, and parameters.
		\item \textit{Output measures} describe the properties that are measured by privacy metrics. Our taxonomy introduces eight categories: a) uncertainty, b) information gain or loss, c) data similarity, d) indistinguishability, e) adversary's success probability, f) error, g) time, and h) accuracy/precision.
	\end{itemize}
	\item We describe and classify over eighty privacy metrics in Section \ref{sec:metrics}. We focus our selection on popular metrics (in terms of citations) and metrics we found conceptually promising. Where possible, we unify and simplify metric notation and, when appropriate, we discuss advantages and disadvantages of metrics as well as application scenarios.
	\item We give recommendations on how to choose privacy  metrics in Section \ref{sec:recommendations}. We structure our recommendations along a series of questions, answers to which will highlight particularly suitable metrics and narrow down the number of candidates.
	\item We identify areas for future work in Section \ref{sec:futurework}. In particular, we believe that more work is needed on metrics for interdependent privacy, combinations of metrics, and evaluations of the quality of metrics.
\end{itemize}

In summary, we systematize the literature on privacy measurement.
Our survey can thus serve as a reference guide for privacy metrics and as a framework that enables privacy researchers to make informed decisions on which metrics to choose in a particular setting. This will contribute to the advancement of \acp{PET} and privacy protection in general.

\section{Conditions for Privacy Metrics}
\label{sec:requirements}

There is no general consensus which conditions privacy metrics have to fulfill.
% In this section we review which characteristics contribute to the quality of a privacy metric.
% mathematical requirements: a metric on a set is a distance function $d : X \times X \rightarrow R$. Properties: non-negativity, identity of indiscernibles, symmetry, triangle inequality
In the mathematical sense, a metric is a measure for the distance between two elements of a set and needs to fulfill four conditions to qualify as a metric (non-negativity, identity of indiscernibles, symmetry, and triangle inequality).
%, i.e., a function that maps each combination of two set members onto the set of real numbers: $d : X \times X \rightarrow \mathbb{R}$. 
%Metrics have to fulfill four conditions, namely non-negativity $d(x,y) \geq 0$, identity of indiscernibles $d(x,y) = 0$ iff $x=y$, symmetry $d(x,y)=d(y,x)$, and the triangle inequality $d(x,z) \leq d(x,y)+d(y,z)$. 
%If either the second, third, or fourth of these conditions are dropped, the functions are instead referred to as pseudometrics, quasimetrics, and semimetrics, respectively.
However, many of the metrics discussed in this survey are not metrics in the mathematical sense, as they do not fulfill all four conditions.
Nevertheless, to remain consistent with the literature (e.g., \cite{bertino_survey_2008,bezzi2010information,claus_structuring_2006,chatzikokolakis2015constructing,andersson_fundamentals_2008,kelly_survey_2008,murdoch_metrics_2008}), we will consider as privacy metrics all measures that in some way describe the level of privacy.

%However, to keep consistent with existing literature, we will use the term 'privacy metric' to refer to all measures that in some way describe the level of privacy provided by a system, e.g., through a direct measurement of privacy or through privacy parameters.
%It could be argued that the term \textit{metric} is therefore inadequate, however, for the sake of readability and conformity to existing literature we will refer to all privacy measures as metrics.
% note: the triangle inequality implies monotonicity, which I stated as desirable in the GenoPri paper.

Many authors have proposed requirements and recommendations for privacy metrics.
% in order to increase their meaningfulness, e.g., how well they represent the actual level of privacy in a system or how comprehensive their output is. %that are more specific to the measurement of privacy.
% 
For example, \citeANP{alexander_engineering_2003}~\cite{alexander_engineering_2003} require that privacy metrics are understandable by mathematically inclined laypeople, are orthogonal to cost and utility metrics, and give bounds on how effectively the adversary can succeed in identifying individuals.
\citeANP{andersson_fundamentals_2008}~\cite{andersson_fundamentals_2008} require that privacy metrics are based on probabilities (e.g., the probability of an adversary identifying a given individual) and have well defined and intuitive endpoints.
They argue that a metric should measure privacy based on the number of individuals an adversary cannot distinguish and how evenly spread the adversary's guesses are.
% indistinguishable they are.
%In addition, anonymity should be rated higher the more uniform the probability distribution, and the more users there are in the anonymity set.
%Finally, they require that the metric's value domain should be ordered, not too coarse, and contain well-defined elements.

In contrast to that, \citeANP{syverson_why_2013}~\cite{syverson_why_2013} requires that privacy metrics reflect how difficult it is for an adversary to succeed, that they do not depend on variables that cannot be determined or predicted, and that they reflect the resources needed for successful attacks on privacy instead of relying on cardinalities or probabilities.
\citeANP{bertino_survey_2008}~\cite{bertino_survey_2008} require that privacy metrics indicate the privacy level, the portion of sensitive data that is not hidden, and the data quality after application of the \ac{PET}.
% 
% borcea2006towards:
% “technical design principles” for privacy: “Design must start from maximum privacy. Explicit privacy rules govern system usage. Privacy rules must be enforced, not just stated. Privacy enforcement must be trustworthy. Users need easy and intuitive abstractions of privacy. Privacy needs an integrated approach. Privacy must be integrated with applications.”
% 
\citeANP{shokri_quantifying_2011}~\cite{shokri_quantifying_2011} require that privacy metrics consider three aspects of the adversary's success: accuracy, uncertainty, and correctness. 
%The adversary can determine both his accuracy and uncertainty, but since he does not know the ground truth, he cannot determine the correctness of his estimate.
% 

In an earlier publication, we required that privacy metrics should be monotone with increasing adversary strength~\cite{wagner2017evaluating}. While the discussed conditions in this section cannot be seen as strict requirements for a measure to qualify as a privacy metric, they can serve as a guideline to increase the strength, usability, and meaningfulness of newly proposed metrics.

% 

%This condition, implied by the triangle inequality, is important because otherwise stronger adversaries could be assigned higher privacy, which could lead to erroneous designs of \acp{PET}.

% common ideas: adversary; utility; understandability;
% privacy-specific requirements: accurate, computable, comparable, understandable (this is isa's wish list)

\section{Privacy Domains}
\label{sec:domains}

Privacy domains are areas where \acfp{PET} can be applied.
With the increasing use of information technology, \acp{PET} are being researched in a growing number of domains.
We describe six domains to provide context and examples for the remainder of the paper.

\subsection{Communication Systems} 
The main privacy challenge in communication systems is anonymous communication, which aims to hide which (or even that) two users communicated, not just the contents of their communication.
Maintaining the confidentiality of communication contents is an orthogonal problem that can be solved via public-key encryption~\cite{chaum_dining_1988}.
Adversaries typically try to identify either the sender of a message, its receiver, or sender-receiver relationships.
Metrics for communication systems have been previously reviewed by \citeANP{kelly_survey_2008}~\cite{kelly_survey_2008}.
% are one of the oldest application areas for \acp{PET}, 

% databases
\subsection{Databases} %are another well-researched privacy domain.
There are two typical scenarios in the database domain: in the interactive setting, users issue queries to a database; in the non-interactive setting, a sanitized database is released to the public.
In both scenarios, adversaries attempt to identify individuals in the database and reveal sensitive attributes, for example, health information contained in a patient record.
Databases can include microdata (i.e., information about individuals) or aggregate data that masks information about individuals, for example by presenting only the averages of multiple values.
Surveys that review metrics for this domain include \citeANP{fung_privacy-preserving_2010}~\cite{fung_privacy-preserving_2010}, \citeANP{shabtai2012survey}~\cite{shabtai2012survey}, \citeANP{xu2014survey}~\cite{xu2014survey} (privacy preserving data publishing), \citeANP{bertino_survey_2008}~\cite{bertino_survey_2008} (data mining), and \citeANP{kelly_survey_2008}~\cite{kelly_survey_2008} (databases).

% location-based services
\subsection{Location-based Services} %is another privacy domain. 
Location-based services provide context-aware services to mobile users, such as information about nearby points of interest. 
Adversaries with access to location information can infer sensitive attributes like home and work locations, and create  movement profiles that can be sold or used for marketing purposes.
Metrics for location privacy are discussed by \citeANP{shokri_unified_2010}~\cite{shokri_unified_2010} and \citeANP{krumm_survey_2009}~\cite{krumm_survey_2009}.
In previous work, we reviewed metrics for vehicular networks~\cite{wagner2014privacy}.

% smart metering
\subsection{Smart Metering}
Smart meters record fine-grained electricity consumption data in a user's home and send this data to the energy provider.
The energy provider can use this data for billing and network optimization, but can also act as an adversary who infers behavioral profiles above and beyond the stated purpose.
Metrics and mechanisms for smart metering are reviewed by \citeANP{zeadally_towards_2013}~\cite{zeadally_towards_2013}.

% social networks
\subsection{Social Networks} 
Social networks allow users to share updates about their daily lives. 
Adversaries in this domain try to identify users in anonymized social graphs, or infer sensitive attributes from private profiles.
\citeANP{yang_stalking_2012}~\cite{yang_stalking_2012} survey privacy risks in social networks.

% genomic privacy
\subsection{Genome Privacy}
Advances in whole genome sequencing have raised new questions regarding the privacy of a person's genome.
The genome uniquely identifies an individual, and at the same time reveals highly sensitive information, like susceptibility to diseases. 
An adversary with access to genomic data could engage in genetic discrimination (e.g., denial of insurance) or blackmail (e.g., planting fake evidence at crime scenes).
In previous work, we reviewed privacy metrics for genomics~\cite{wagner2015genomic}.

% \include{table2}

% \section{Privacy Domains}
% \label{sec:background}
\section{Characteristics of Privacy Metrics}
\label{sec:characteristics}

Despite their diversity, privacy metrics share common characteristics.
Here, we describe four characteristics that can classify privacy metrics and can thus serve as an initial guideline for choosing privacy metrics for specific scenarios (we give detailed recommendations in Section \ref{sec:recommendations}).

\subsection{Adversary Goals}
\label{sec:hiding}

The goal of privacy metrics is to quantify the level of privacy in a system or the privacy provided by a \ac{PET}, often under consideration of a specific adversary.
The adversary aims to compromise users' privacy and to learn sensitive information.
This sensitive information can be user identities (e.g. by deanonymizing data sets), user properties (e.g. location or energy consumption), or both \cite{heurix2015taxonomy}.
%their goal may be to deanonymize data sets to reveal user identities or to expose user properties.
%, e.g., by deanonymizing data sets or by profiling users.
%To defend against these different adversaries,  \acp{PET} are needed to hide user identities, to hide user properties (e.g., location or energy consumption), or both \cite{heurix2015taxonomy}.
It is therefore important to select metrics that are able to measure the relevant aspect.
For example, a metric in location-based services can indicate whether the adversary can identify a user, given a location (identity hiding), or whether the adversary can identify the location, given a user (property hiding).
We indicate which metrics are suitable to measure identity or property hiding in Tables~\ref{tab:privacymetrics1} and \ref{tab:privacymetrics2} (pages \pageref{tab:privacymetrics1} and \pageref{tab:privacymetrics2}, column \emph{Identity/Property}).
The distinction between identity and property hiding can be blurry because it depends on the adversary and the employed \ac{PET}, and because metrics that were originally proposed for one setting are often applied in other settings as well.
Therefore, a missing entry in Tables~\ref{tab:privacymetrics1} and \ref{tab:privacymetrics2} does not necessarily mean that a metric cannot be applied, only that, to the best of our knowledge, no research has done so.
%\todo[inline]{sometimes blurry because metrics are applied in new settings (applied to user instead of property), 
%and sometimes question of perspective}

\subsection{Adversary Capabilities}
\label{sec:adversaries}
%this section can be shortened if we need space. Adversary models are not considered in the remainder of this paper nor are they part of our overview tables
Naturally, a stronger adversary, such as one with more resources or prior knowledge, might be able to attack privacy more successfully.
The value of a privacy metric therefore depends on the adversary model, and evaluating a \ac{PET} with a weak adversary model can lead to an overestimation of privacy.
Essentially, \acp{PET} that provide protection against a stronger adversary model can give stronger privacy guarantees.
As a result, metrics can only be used to compare two different \acp{PET} if they use the same adversary model.

Metrics that do not account for any type of adversary %and, for example, measure certain properties of data.
implicitly assume an adversary with limited capabilities.
For example, metrics that measure privacy purely based on certain properties of data assume that every attack on the system will only rely on these properties.
Attacks that exploit other properties of the data may be able to disclose sensitive information nevertheless.
%While this makes them easy to compare, they may give a false sense of privacy protection as they inherently assume a certain attack on the system using only these chosen properties.
%Unaccounted characteristics of the data could then be used for other attacks to disclose private information.

The literature reflects the importance of adversary models by considering adversaries with diverse characteristics.
To allow for a better interpretation of the outcome of privacy metrics, studies should always include a detailed description of the used adversary model.
%Thus, we believe that the adversary model needs to be included in any discussion about privacy, regardless of the privacy metric used.
%This view is reflected in the literature, where adversaries with diverse characteristics are considered.
% introduce adversaries (taxonomy from diaz paper)
To this end, we extend the taxonomy of adversary types described by \citeANP{diaz_towards_2003}~\cite{diaz_towards_2003} (and later refined in \citeANP{diaz2006anonymity}~\cite{diaz2006anonymity}), and classify adversaries as follows:

\subsubsection{Local--Global}
Local adversaries can only act on a restricted part of the system, for example a geographical location or a subset of nodes.
Global adversaries have access to the entire system.

\subsubsection{Active--Passive}
Active adversaries can interfere with the system by adding, removing or modifying information or communication.
Passive adversaries can only read and observe.

\subsubsection{Internal--External}
Internal adversaries are part of the system, for example servers providing location-based services, energy providers in smart metering, or third parties controlling nodes in the system.
External adversaries are not part of the system, but are able to attack it, e.g., via shared communication links or publicly available data.

\subsubsection{Static--Adaptive}
Static adversaries choose which strategy and resources to use prior to an attack and stick to their choice irrespective of how the attack progresses.
Adaptive adversaries can adapt their strategy while the attack is ongoing, e.g., by learning system parameters through observation.

\subsubsection{Prior Knowledge}
Some adversaries may have additional knowledge about the system, such as general domain-specific knowledge -- knowledge about the world -- or scenario-specific knowledge, for example in the form of a prior probability distribution or specific information about users in the system, such as their home and work addresses.
Prior information can considerably strengthen the adversary, and thus it is important that privacy metrics can account for it.

%\todo[inline,color=blue!20]{Reviewer 2: page 5, line 8, section 4.1.5: Prior knowledge is crucial and may include more than "knowledge about the world". If the adversary is, say, a neighbour, relative, or co-worker, then the prior knowledge may include very specific information. Thus it is important to understand and model prior knowledge (a.k.a. side information).}

\subsubsection{Resources}
\label{sec:adversary-resources}
Adversaries can also be classified according to the resources available to them.
For computational resources, \textit{efficient} adversaries are restricted to probabilistic polynomial time (PPT) algorithms, while \textit{unbounded} adversaries are not restricted to any computational model.
Other types of resources include the bandwidth or number of malicious nodes available to the adversary \cite{murdoch_quantifying_2014}.
%Although PPT adversaries are most common, many other models can be found in the literature, for example non-deterministic state machines (Dolev-Yao model) \cite{herzog_computational_2005}, or probabilistic interactive Turing machines \cite{dwork_calibrating_2006}.

\subsection{Data Sources}
\label{sec:sources}

%\todo[inline, color=blue!20]{Data sources are defined in section 4.2, but are not regularly used
%throughout the text. In some cases (especially in location privacy) the
%application domain of the measures is mentioned, but it would be helpful
%if the domains in which the measure was used would be mentioned.
%Otherwise, the existence of section 4.2 is not really justified.}

Data sources describe which data needs to be protected, and how the adversary is assumed to gain access to the data.
We indicate the primary data sources for each metric in Tables~\ref{tab:privacymetrics1} and \ref{tab:privacymetrics2} (pages \pageref{tab:privacymetrics1} and \pageref{tab:privacymetrics2}, column \emph{Primary data source}).
%Another characteristic that may influence the choice of privacy metric is which sources of data need to be protected.
%This determines which \acp{PET} and which metrics can be applied.

%In many cases, using only information about application areas to select privacy metrics is too narrow a scope because it disregards metrics from other scenarios that share common characteristics.
%A classification for privacy domains which is wider than a specific scenario would widen the scope and help with metric selection.
%\citeN{finn2013seven} present such a classification by defining \textit{seven types of privacy} (privacy of person; behavior; communication; data; thoughts; location; and association). 
%This classification, although insightful, does not provide much guidance because the seven categories overlap significantly.
%Instead, we found it helpful to classify privacy areas into domains depending on the data source, or kind of information that \acp{PET} are trying to protect.

\subsubsection{Published Data} Published data refers to information that has been willingly and persistently made available to the public.
This includes statistical databases as well as information individuals choose to disclose, e.g., on social networks.
In both cases, adversaries attempt to identify anonymized individuals or reveal sensitive attributes.

\subsubsection{Observable Data} Observable data is transient information that requires the adversary to be present in order to gain access to it.
This category includes information that can be obtained by a passive adversary who can access data without compromising the underlying system.
In communication systems, for example, adversaries overhear communications to identify message senders and receivers.

\subsubsection{Re-purposed Data} Re-purposed data is used for a different purpose than the purpose for which it was initially acquired.
Examples are service providers who obtain user information to offer location-based services, smart metering, or social networks, but then use this information for purposes other than providing the service.
Having access to non-public user information (regardless of the users' privacy setting) allows for tailored advertising and other forms of marketing or monetization.

% secret, shielded, confidential, protected, access/inaccessible
\subsubsection{All Other Data} All other data refers to information that was not made public, was not observable and that the adversary was not intended to have access to.
% For \acp{PET} this data source is mostly considered to be out of scope, to be protected by other means
This data is typically not anonymized or protected, and can be obtained using methods such as wiretapping, hacking into a system, blackmailing, or buying off the black market.
Implications for users can be severe, including financial losses and publication of medical records or confidential communication.
% Privacy protections often neglect this category, because data is not expected to be disclosed.
% The main problem is that in this context privacy protection is often neglected as this data is not expected to be disclosed, thus increasing the potential negative impact on the affected users.
%Although the same metrics and methods as for published or observable data could be applied to measure and improve privacy protections for this data, 
\acp{PET} are often not deployed by the original owner as they can make it less convenient to work with the data.

\subsection{Inputs for Computation of Metrics}
\label{sec:inputs}

Privacy metrics rely on different kinds of input data to compute privacy values.
The availability of input data or appropriate assumptions determine whether a metric can be used in a specific scenario.
We indicate which of the input categories each metric relies on in Tables~\ref{tab:privacymetrics1} and \ref{tab:privacymetrics2} (column group \textit{Inputs}).

\subsubsection{Adversary's Estimate}

The adversary's estimate is the result of the adversary's effort to breach privacy.
It often takes the form of a posterior probability distribution.
For example, in a communication system the estimate can describe how likely each user is to have sent a message.
In smart metering, the estimate can describe how much energy a user is likely to have consumed during a specific time period.

% \subsubsection{Adversary's Observations or Perturbed Data}
% 
% This category contains data that the adversary uses to compute his estimate.
% Observations are often the result of eavesdropping in a network, but can also be physical observations of locations.
% Perturbed data is data that has been sanitized and made publicly available.

\subsubsection{Adversary's Resources}

The resources available to the adversary can be given, for example, in terms of computational power, time, bandwidth, or physical nodes (see Section \ref{sec:adversary-resources}).

\subsubsection{True Outcome}

The true outcome, or ground truth, is often used to judge how good the adversary's estimate is.
However, this information is not available to the adversary, so they cannot compute metrics that use the true outcome.
For example, in location-based services the true outcome corresponds to a user's true location, and in social networks it corresponds to the true connections in a social graph.
The ground truth is usually assumed to describe sensitive data.

\subsubsection{Prior Knowledge}

Prior knowledge describes concrete, scenario-specific knowledge that the adversary has.
It usually takes the form of a prior probability distribution.
In genome privacy, for example, prior knowledge can include information about a user's population group, which influences how likely a user is to have specific genetic variations.

\subsubsection{Parameters}

Parameters configure privacy metrics.
They describe threshold values, the sensitivity of attributes, which attributes are sensitive, or desired privacy levels.

\subsection{Output Measures}
\label{sec:taxonomy}

The output of a privacy metric refers to the kind of property that a privacy metric measures.
We introduce a taxonomy with eight output properties, each of which represents a different aspect of privacy.
This is an important categorization because it shows that a single metric cannot capture the entire concept of privacy.
A more complete estimate of privacy can only be obtained by using metrics from different output categories. 

\begin{figure}
\centering
\includegraphics[width=\textwidth]{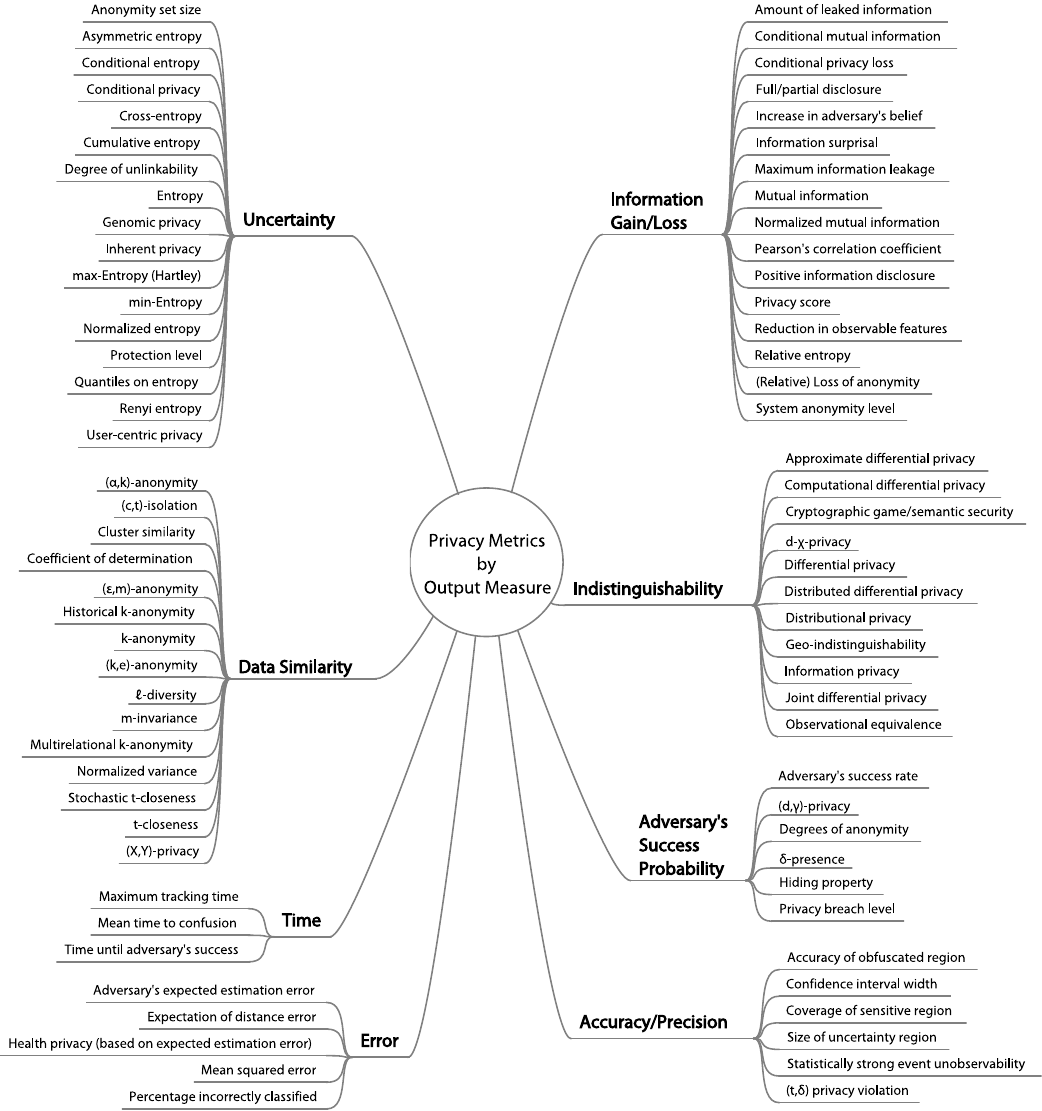}
\caption{Taxonomy of privacy metrics, classified by output}
\label{fig_outcomes}
\vspace{-.5cm}
\end{figure}

Figure~\ref{fig_outcomes} gives an overview of the output measures and the metrics associated with each.

While there exist many possible categorizations for metrics, e.g., based on domain or data source, we believe that a classification based on the output is the most intuitive.
We note that, as for any classification, the boundaries between categories can be blurred and some metrics could also be assigned to other categories.
For example, \citeANP{bezzi2010information}~\cite{bezzi2010information} describe metrics from the data similarity category in terms of metrics from the uncertainty and information gain/loss categories, and \citeANP{soria-comas_differential_2013}~\cite{soria-comas_differential_2013} showed that data similarity metrics can be related with metrics from the indistinguishability category.
%\todo[inline,color=green!20]{Of course,  there are some gray areas, (e.g., k-anonymity could have
%gone in the Indistinguishability category), but this is a general
%problem whenever we try to classify complex objects/phenomena.}
In this survey, we assigned metrics to the output which they seem to measure the most directly.

%more than one aspect of privacy may be relevant in a given scenario, and thus highlights that scenarios will typically benefit from using multiple metrics from different output categories.
%We created a taxonomy consisting of nine categories that allow for a classification of all privacy metrics we found in the literature.
%This classification was surprisingly straight-forward, giving strong indication that our choice of categories is meaningful and complete.

%Detailed descriptions of each category are provided in the next section.

%\cite{domingoferrer2015t-closeness} (we already cited their conference paper, )

% \todo{We found this to be the least ambiguous classification in that }
%\todo[inline,color=red!20]{we could add another graphic visualizing which metrics have been used in which domains (limiting ourselves to references and domains we use anyway). could be a venn diagram -- there are R packages to generate them}
% http://stackoverflow.com/questions/8713994/venn-diagram-proportional-and-color-shading-with-semi-transparency

% Some metrics combine more than one output property. In these cases, we explain the metrics in the category that appears first in the text \todo{last may be more useful}, and mention it in Table \ref{tab:privacymetrics1}. This fact underlines the usefulness of the taxonomy, because it demonstrates clearly that the analysis of more than one output property can be beneficial in many scenarios.

% \begin{figure}
% \centering
% \includegraphics[width=\textwidth]{taxonomy-mindmap}
% \caption{Taxonomy of privacy metrics}
% \label{fig_taxonomy}
% \end{figure}

\subsubsection{Uncertainty}
%Metrics in this category quantify the adversary's uncertainty. %, in many cases by using information theory~\citeN{shannon_mathematical_1948} to compute some quantity related to entropy.
Uncertainty metrics assume that high uncertainty in the adversary's estimate correlates with high privacy, because the adversary cannot base his guesses on information known with certainty.
However, even guesses based on uncertain information can be correct, and thus individual users may suffer privacy losses even in scenarios with a highly uncertain adversary.

\subsubsection{Information Gain or Loss}
Metrics that measure information gain or loss quantify the amount of information gained by the adversary, or the amount of privacy lost by users due to the disclosure of information.

\subsubsection{Data Similarity}
Data similarity metrics measure similarity either within a dataset, for example by forming equivalence classes, or between two sets of data, for example between a private dataset and its public, sanitized counterpart.
These metrics abstract away from an adversary and focus on the properties of the data.
For example, similarity can refer to the frequencies of data values, numerical similarity, or the (lack of) variation in published data.

% \todo[inline,color=blue!20]{Reviewer 2: page 6, line 32: What is "the diversity between two sets of data"?}

% Low similarity, i.e. large equivalence classes or  and high diversity between the two datasets correlates with high privacy.
%Most metrics in this category can be computed without considering the adversary; privacy is evaluated solely based on the data itself.

\subsubsection{Indistinguishability}
Indistinguishability is a classic notion in the security community.
Metrics based on indistinguishability analyze whether the adversary is able to distinguish between two outcomes of a privacy mechanism.
Privacy is high if the adversary cannot distinguish between any pair of outcomes.
Metrics in this category are usually binary; they indicate whether two outcomes are indistinguishable or not, but do not quantify the privacy levels in-between.
% A challenge-response game is set up in which the adversary is given publicly observable information and then has to guess the value of a secretly and randomly chosen bit.
% The adversary has an advantage if he is able to distinguish between the two alternatives with a non-negligible probability; otherwise, the two outcomes are indistinguishable.
% Indistinguishability is assumed to correlate with high privacy.

\subsubsection{Adversary's Success Probability}
Metrics using the adversary's success probability to quantify privacy indicate how likely it is for the adversary to succeed in any one attempt, or how often they would succeed in a large number of attempts.
Low success probabilities correlate with high privacy.
While this assumption holds for an averaged population of users, an individual user may still suffer a loss of privacy even when the adversary's success probability is low.

\subsubsection{Error}
Error-based metrics measure how correct the adversary's estimate is, for example using the distance between the true outcome and the estimate.
High correctness and small errors correlate with low privacy.

\subsubsection{Time}
Time-based metrics either measure the time until the adversary's success, or the time until the adversary's confusion. %can be divided into two groups.
%These two approaches are based on two opposing assumptions.
In the first case, metrics assume that the adversary will succeed eventually, and so a longer time correlates with higher privacy.
In the second case, metrics assume that the privacy mechanism will eventually confuse the adversary, and so a shorter time correlates with higher privacy.
% The first group measures the time it takes for the adversary to succeed.
% This assumes that the adversary will succeed eventually, a pessimistic assumption.
% The second group measures the time until the privacy mechanism confuses the adversary.
% This assumes that the adversary will not have lasting success, a more optimistic assumption than before.
% Time-based metrics assume that a longer time before the adversary's success means higher privacy.
% These metrics can be particularly useful in scenarios where information becomes meaningless after some time has passed \todo{find an example, if one exists}.

\subsubsection{Accuracy or Precision}
These metrics quantify how precise the adversary's estimate is without considering the estimate's correctness.
% As an example, consider a confidence interval.
% Correctness is high if the center of the interval matches the estimate closely.
% These metrics, in contrast, are concerned with the width, not the position, of the confidence interval, a smaller width indicating higher precision.
More precise estimates correlate with lower privacy.

\section{Privacy Metrics}
\label{sec:metrics}

We now describe over eighty privacy metrics from the literature, grouped by the outputs they measure\footnote{For the first read, we suggest to only focus on the first 2-3 metrics in each category. This will provide an understanding of the most important metrics in each category as well as the differences between categories.}.
Where possible, we point out their advantages or disadvantages, point out similarities or differences between related metrics, and give examples for application scenarios.
% , and sorted roughly by their popularity, as indicated by the number of citations.
We also simplify and unify metric notation (see Table \ref{tab:notation}), however, we did not alter notation that occurs in a metric's name (e.g., $t$-closeness or $(X,Y)$-Privacy).
% To recognize these metrics easier and for the sake of readability, we decided to not change the symbols for these parameters.}

At the end of the section, Tables~\ref{tab:privacymetrics1} and \ref{tab:privacymetrics2} summarize how each metric can be classified according to the characteristics introduced in Section \ref{sec:characteristics}.
The tables also provide information about value ranges, and an indication whether higher or lower values represent better privacy.
We will refer to Tables~\ref{tab:privacymetrics1} and \ref{tab:privacymetrics2} again in Section \ref{sec:recommendations}, when we give recommendations on how to select metrics.
% Table \ref{tab:notation} gives an overview of the notation, which we have unified wherever possible.
%Since most papers do not specify the value ranges of their metrics, the value ranges in our table do not represent tight bounds, but rather coarse estimates.

\begin{table}[!ht]
\caption{Unified notation for all privacy metrics in this paper}
\label{tab:notation}
\rowcolors{1}{white}{gray!20}
\resizebox{.8\textwidth}{!}{%
\begin{tabular}[t]{lp{12cm}}
\toprule
$B$ & Base metric \\
$d()$ & Distance function \\
$D$ & Database or database table\\
$E$ & Equivalence class\\
$H(\cdot)$ & Entropy \\
$I(\cdot;\cdot)$ & Mutual Information \\
$\mathcal{K}$ & Privacy mechanism \\
$L$ & Locations \\ 
$M$ & Messages, requests\\
$p(x)$ & Equivalent to $p(X=x)$\\
$q$ & Quasi-identifiers\\
$R$ & Regions \\
$S$ & Sensitive values or sets of query responses (differential privacy)\\
$T$ & Time \\
$\vec{T}$ & Time series \\
$U$ & Set of users $u \in U$\\
$V$ & Genetic variations (or SNPs) \\
$X$ & Discrete random variable that represents the adversary's estimated probabilities for each member of the anonymity set \\
$X^*$ & True distribution of (hidden) data \\
$Y$ & Data observed by the adversary (which may be obfuscated) \\
$Z$ & Prior information \\
$\beta()$ & Loss function \\
$\tau$ & Thresholds \\
$\omega$ & Weights \\
% \bottomrule
\end{tabular}%
}
\end{table}

\begin{subsection}{Uncertainty}

Uncertainty metrics assume that an adversary who is uncertain of his estimate cannot breach privacy as effectively as one who is certain.
Many uncertainty metrics build on entropy, an information-theoretic notion to measure uncertainty~\cite{shannon_mathematical_1948}.
Most metrics in this category originate from the communication domain, where, for example, they can be used to assess an adversary's uncertainty of associating different users and messages.
In location-based services, they have been applied to measure the uncertainty of an adversary in associating a user with a location or to distinguish between different users.

%\begin{table}[htp]
 %\tbl{Uncertainty Metrics by Domain\label{tab:uncertainty-metricsbydomain}}{
%%  \resizebox{.7\textwidth}{!}{%
  %\resizebox*{\textwidth}{!}{%
  	%\rowcolors{2}{gray!20}{white}
 %\begin{tabular}{>{\raggedleft\slshape}p{3.3cm}VVVVVV}
 %\upshape\bfseries Metric & \bfseries Communication & \bfseries Databases & \bfseries Location & \bfseries Smart Metering & \bfseries Social Networks & \bfseries Genome Privacy \\
%\midrule
     %Anonymity Set Size & Chaum 1988, Kesdogan 1998 &      & Duckham 2005, Golle 2009   &   &      & \\
     %Asymmetric Entropy &  &      & & & & Ayday 2013    \\
     %Conditional entropy & Diaz 2007 &  & & &       &     \\
     %Conditional privacy & & Agrawal 2001     &  & &  &\\
     %Cross-entropy &  & Merugu 2003     &  & & & \\
     %Cumulative entropy & &      & Freudiger 2007 & & &  \\
     %Degree of unlinkability & Steinbrecher 2003 &  & & & & \\
     %Entropy & Serjantov 2003 & & & & & \\
     %Entropy + Bayes &  &      & Ma 2010 & & & \\
     %Genomic Privacy &  &      & &   &       &       Ayday 2013 \\
     %Inherent privacy & Andersson 2008 & Agrawal 2001     &  &  & & \\
     %Max-entropy (Hartley) & Clauss 2006 & & & & &\\
     %Min-entropy & Clauss 2006 & & & & & \\
     %Normalized conditional entropy &  & & & & & \\
     %Normalized entropy & Diaz 2003 &  & & & & \\
     %Protection Level &  &      & Xu 2009 & & & \\
     %Quantiles on entropy & Clauss 2006 & & & & & \\
     %R\'{e}nyi entropy & Clauss 2006 & & & & & \\
     %User-centric privacy & &      & Freudiger 2009 &  & & \\
 %\end{tabular}}}
%\end{table}     

\begin{table}[htp]
 \caption{Metrics and references in the uncertainty category and the domains they originated in}
 \vspace{-.2cm}
 \label{tab:uncertainty-metricsbydomain}
%  \resizebox{.7\textwidth}{!}{%
%   \resizebox*{\textwidth}{!}{%
  \rowcolors{2}{gray!20}{white}
	\resizebox{.85\textwidth}{!}{%
 \begin{tabular}{p{1.2cm}p{6cm}p{2.8cm}p{2cm}}
 \upshape\bfseries Section & \bfseries Metric & \bfseries Original Domain & \bfseries Reference \\
\midrule
 \ref{metric:anonymity-set-size} & Anonymity set size & Communication & \cite{chaum_dining_1988} \\
 \ref{metric:entropy} & Entropy & Communication & \cite{serjantov_towards_2003} \\
 \ref{metric:renyi-entropy} & R\'{e}nyi entropy & Communication & \cite{claus_structuring_2006} \\
 \ref{metric:max-entropy} & Max-entropy (Hartley) & Communication & \cite{claus_structuring_2006} \\
 \ref{metric:min-entropy} & Min-entropy & Communication & \cite{claus_structuring_2006} \\
 \ref{metric:normalized-entropy} & Normalized entropy & Communication & \cite{diaz_towards_2003} \\
 \ref{metric:degree-of-unlinkability} & Degree of unlinkability & Communication & \cite{steinbrecher_modelling_2003} \\
 \ref{metric:quantiles-on-entropy} & Quantiles on entropy & Communication & \cite{claus_structuring_2006} \\
 \ref{metric:conditional_entropy} & Conditional entropy & Communication & \cite{diaz_does_2007} \\
 \ref{metric:conditional-privacy} & Conditional privacy & Databases & \cite{agrawal_design_2001} \\
 \ref{metric:inherent_privacy} & Inherent privacy & Databases & \cite{agrawal_design_2001} \\
 \ref{metric:cross-entropy} & Cross-entropy & Databases & \cite{merugu_privacy-preserving_2003} \\
 \ref{metric:cumulative-entropy} & Cumulative entropy & Location & \cite{freudiger_mix-zones_2007} \\
 \ref{metric:protection-level} & Protection level & Location & \cite{xu_feeling-based_2009} \\
 \ref{metric:asymmetric-entropy} & Asymmetric entropy & Genome privacy & \cite{ayday_protecting_2013} \\
 \ref{metric:genomic-privacy} & Genomic privacy & Genome privacy & \cite{ayday_personal_2013} \\
 \ref{metric:user-centric-privacy} & User-centric privacy & Location & \cite{freudiger_non-cooperative_2009} \\
 \end{tabular}%
}
\end{table}

\subsubsection{Anonymity Set Size}
\label{metric:anonymity-set-size}
The anonymity set for an individual $u$, denoted $AS_u$ is the set of users that the adversary cannot distinguish
from $u$~\cite{chaum_dining_1988,kesdogan_stop-_1998}.
It can be seen as the size of the crowd into which the target $u$ can blend.
\[priv_{\text{ASS}} \equiv |AS_u|\]

Instead of users, anonymity sets can also be applied to locations \cite{duckham_formal_2005}, location pairs (e.g., home/work) \cite{golle_anonymity_2009}, or radio frequency identification (RFID) devices \cite{heydt-benjamin_privacy_2006}.
As a result of its simplicity, the anonymity set size is widely used in the literature.

% papers that argue against anonymity set size: diaz2002towards, serjantov2002towards
The main criticism of the anonymity set size is that it only depends on the number of users in the system.
This means that it does not take into account prior knowledge, information the adversary has gathered by observing the system, or how likely each member of the anonymity set is to be the target~\cite{serjantov_towards_2003,diaz_towards_2003}.
% papers that argue for anonymity set size: steinbrecher2003modelling
However, it can be argued that the size of the anonymity set is useful in combination with other metrics such as normalized entropy (Section~\ref{metric:normalized-entropy})~\cite{steinbrecher_modelling_2003}.

% $k$-Approximate Beyond Suspicion extends the anonymity set size by filtering the anonymity set to only include members that are equally likely ($\pm \epsilon$) to be the target so the anonymity set size is less influenced by unlikely members~\cite{chen_measuring_2012}.
% \todo{i still dont get that, what if p(target) = 0.9 and four more elements have 0.025 each?! -- simple: in that case the ASS is 1, instead of 5 in the original}

\subsubsection{Entropy}
\label{metric:entropy}

Shannon entropy is the basis for many other metrics. 
In general, entropy measures the uncertainty associated with predicting the value of a random variable.
As a privacy metric, it can be interpreted as the effective size of the anonymity set, or as the number of bits of additional information the adversary needs to identify a user~\cite{serjantov_towards_2003}.
% The discrete random variable $X$ indicates the adversary's estimated probabilities for each member of the anonymity set.

% whether through Bayesian inference, random guessing, or prior knowledge.

For example, the adversary may be interested in identifying which member of the anonymity set took a specific action, e.g., who sent a particular message, or who visited a particular location.
The adversary would then estimate a probability $p(x)$ for each member $x$ of the anonymity set $AS_u$ which indicates the likelihood that $x$ is the targeted user $u$ (ensuring that $\sum_{x \in AS_u} p(x) = 1$).
To use the entropy metric, it does not matter how the adversary estimates $p(x)$.
Attacks could, for example, be based on Bayesian inference, random guessing, prior knowledge, or a combination of methods.
% For example, when tracking an individual $t$, each member of the anonymity set $x \in AS_t$ is assigned a probability $p$ to be the target $t$, with $\sum_{x \in AS_t} p(x) = 1$.
% How these probabilities are derived depends on the deployed adversary model.
% The adversary could be confused when the paths of $t$ and another user $u$ intersect.
% The level of confusion can then be represented by adding $u$ (that is, an element representing the future locations of what the adversary thinks $u$ is) to the anonymity set and assigning a probability of this user to actually be the target $t$.
% If the adversary is confused to a point where it can no longer distinguish the two users (and assuming the anonymity set has only 2 member at this point), the probabilities for both members would be $0.5$.

More generally, each value $\{x_1, ..., x_n\}$ of the discrete random variable $X$ represents a member of the anonymity set and $p(x_i)$ is the (estimated) probability of this member to be the target.
Then, the entropy of $X$ can be expressed as:

\[priv_{\text{ENT}} \equiv H(X) = - \sum_{x\in X}p(x) \log_2 p(x)\]

% papers that argue against entropy: clauss2006structuring, shmatikov2006measuring, toth2004measuring, syverson_why_2013, murdoch2008metrics, hamel2011misentropists, hoh2005protecting, shokri2011quantifying2

Entropy has also been used in cases where privacy is measured at more than one point in time, for example in location privacy, where the adversary tracks users during a period of time.
In this case, entropy is computed at every point in time, and the underlying probabilities are updated after each timestep using Bayesian belief tables~\cite{ma_measuring_2010}.
After the first timestep, this accounts for the prior knowledge the adversary has acquired during previous timesteps.

Many papers argue against the use of entropy as a privacy metric. %, or against the use of information theoretic metrics in general. 
Entropy is strongly influenced by outlier values, i.e., users in the anonymity set that are very unlikely to be the target~\cite{claus_structuring_2006}.
% Entropy is strongly influenced by outlier values, i.e., users in the anonymity set that are very unlikely to be the target~\cite{claus_structuring_2006}.
Even if an adversary is able to identify a target with high probability, the remaining low probability members of the anonymity set can still lead to high values of entropy and thus indicate high privacy~\cite{toth_measuring_2004}.
It is easy to construct different probability distributions that yield the same entropy value, for example a uniform distribution over 20 users, and an almost uniform distribution over 101 users where one user has a probability of $\frac{1}{2}$~\cite{toth_measuring_2004,murdoch_quantifying_2014}.
This makes it difficult to compare different systems.

In the case of location privacy, entropy measures how well an adversary can disclose the position of a user.
However, if two positions are very close to each other, locations may be revealed despite high entropy~\cite{hoh_protecting_2005}.

Although entropy has an intuitive interpretation as the number of additional bits of information the adversary needs, it can be argued that the absolute value of entropy does not convey much meaning~\cite{hamel_misentropists:_2011}.
Entropy gives an indication of the adversary's uncertainty, but does not state how correct or accurate the adversary's estimates are~\cite{shokri_quantifying_2011}.
For example, the adversary could be certain but wrong (low correctness) if the estimate indicates that the wrong member of the anonymity set is the target.
The adversary could also be certain but with low accuracy if the confidence intervals for the estimated probabilities are very large. 
Low certainty is usually correlated with low correctness, but otherwise, correctness and certainty are not correlated~\cite{shokri_quantifying_2011}.
Entropy also does not indicate how many resources (e.g. in terms of computation or bandwidth, see Section \ref{sec:adversary-resources}) the adversary has to expend to succeed~\cite{syverson_why_2013,murdoch_metrics_2008}.

%\todo[inline,color=blue!20]{Reviewer 2: page 9, line 18: "Entropy gives an indication of the adversary's
%uncertainty, but does not state how correct or accurate the adversary's
%estimates are" This seems crucial and should be expanded upon.}

\subsubsection{R\'{e}nyi Entropy}
\label{metric:renyi-entropy}

R\'{e}nyi entropy is a generalization of Shannon entropy that also quantifies the uncertainty in a random variable.
It uses an additional parameter $\alpha$, and Shannon entropy is the special case with $\alpha \rightarrow 1$.
\[priv_{\text{RE}} \equiv H_{\alpha}(X) = \frac{1}{1-\alpha} \log_2 \sum_{x \in X} p(x)^{\alpha} \]

%\todo[inline]{Why are some metrics defined as $priv_{bla} \equiv H(0) = DEF$ and others are $priv_{bla} = DEF$}

Hartley entropy $H_0$ or max-entropy is the special case with $\alpha=0$.
It depends only on the number of users and is therefore a best-case scenario because it represents the ideal privacy situation for a user.
Min-entropy $H_{\infty}$ is the special case with $\alpha=\infty$ which is a worst-case scenario because it only depends on the user for whom the adversary has the highest probability~\cite{claus_structuring_2006}.
\label{metric:max-entropy}
\label{metric:min-entropy}
\begin{align*}
 priv_{\text{MXE}} &\equiv H_0(X) = \log_2 |X| = \log_2 priv_{\text{ASS}} \\
 priv_{\text{MNE}} &\equiv H_{\infty}(X) = -\log_2 \max_{x \in X} p(x) 
\end{align*}

\subsubsection{Normalized Entropy (Degree of Anonymity)}
\label{metric:normalized-entropy}

Because the value range of entropy depends on the number of elements in the anonymity set, the absolute value cannot always be used to compare entropy values.
This is why entropy is frequently normalized using Hartley entropy (i.e., the maximum value entropy takes when all elements in the anonymity set are equally likely).
Normalized entropy can be interpreted as the amount of information the system is leaking~\cite{diaz_towards_2003}.
\[priv_{\text{NE}} \equiv \frac{H(X)}{H_{0}(X)} \]

\subsubsection{(Degree of) Unlinkability}
\label{metric:degree-of-unlinkability}

Unlinkability measures the adversary's uncertainty about which items are related, for example which users are related by anonymous communication.
In this case, the adversary does not assign probabilities to members of the anonymity set, but to the relationships between them.
The set of partitions $\Pi$ of users $U$ contains all possible relationships.
Unlinkability is then computed as the entropy over the set of partitions $\Pi$ \cite{steinbrecher_modelling_2003}.
% Each partition $\pi$ defines an equivalence class of related users, and the adversary aims to find the true partition $\tau$~\cite{steinbrecher_modelling_2003}.
% of users $U$ partitioned by an equivalence relation $\pi$.
\[priv_{\text{DUE}} \equiv  H(\Pi) = - \sum_{\pi \in \Pi} p(\pi) \log_2 p(\pi) \]

% \todo[inline]{Also it says $\Pi \subset U$. Why are we switching to greek letters now?
% The story of this metric has to change. the partitions are actually X and the cool thing here is to not use the anonymity set but partitions of the Users. Just shows another application of the entropy }

The degree of unlinkability takes into account the prior knowledge of an adversary by computing the ratio of unlinkability for an adversary with ($H(\Pi_Z)$) and without ($H(\Pi)$) prior knowledge~\cite{Franz2007}.
\[priv_{\text{DUP}} \equiv \frac{H(\Pi_Z)}{H(\Pi)} \]
Using a ratio to compute the degree of unlinkability makes sure that the values represent the \textit{degree} of unlinkability, i.e., the metric is in the range $[0,1]$, and indicates the portion of unlinkability that remains even if the adversary has prior knowledge.
Other options to account for prior information are taking the difference (see increase in adversary's belief, Section \ref{metric:increase_adv_belief}) or the conditional entropy (see Section \ref{metric:conditional_entropy}).

% \todo[inline,color=blue!20]{Reviewer 2: page 10, line 7: Why a ratio instead of, say, a difference? Linking two people is a privacy violation, but may have a small impact on the ratio.}

% \todo[inline,color=red!20]{what's the interpretation for this metric? what do its values mean? is it the increase in the adversary's probabilities from prior to posterior?}

\subsubsection{Quantiles on Entropy}
\label{metric:quantiles-on-entropy}

Quantiles on entropy compute the entropy of a chosen percentile of the random variable $X$.
To account for the fact that entropy is strongly influenced by outlier values and to avoid overestimating the level of privacy, this metric ignores all members $x \in X$ whose assigned probability $p(x)$ is smaller than the threshold $\tau$~\cite{claus_structuring_2006}.
% For example, when tracking a target $t$, this would remove all users from the anonymity set who have a probability smaller than $c$ to be the target $t$.

\[priv_{\text{QE}} \equiv H(\hat X)\text{, where } \hat X = \{x : x \in X, p(x) \geq \tau\} \]

\subsubsection{Conditional Entropy}
\label{metric:conditional_entropy}

The conditional entropy, or equivocation, of a random variable $X^*$, given a random variable $Y$, measures how much information is needed to describe $X^*$ if the value of $Y$ is known.
The random variable $X^*$ represents the true distribution, for example a sender's true sending profile (in communications) or the true distribution of a data attribute (in databases).
$Y$ can then be taken to describe the adversary's observations, for example information about messages in a communications network \cite{diaz_does_2007}, or a perturbed data release \cite{agrawal_design_2001}. 
% a different layer in the network stack , about biometric characteristics \cite{lai_privacy_2011}, or about residual information in a perturbed data release .
However, care must be taken to distinguish conditional entropy from the entropy of a conditional probability distribution~\cite{diaz_does_2007}.
\[priv_{\text{COE}} \equiv H(X^*|Y) = - \sum_{y \in Y} \sum_{x^* \in X^*} p(y, x^*) \log_2 p(x^*|y) \]

% \todo[inline]{The definition of the conditional entropy does not seem to be how the estimate is affected by Z, but how much prior knowledge is needed to describe the true X? check this}
% \todo[inline]{we need a clearer example for Y. Also, this sounds very similar to the asymmetric entropy you described above, needs clarification about the difference between the two}
% \todo[inline,color=red!20]{diaz: x=truth (e.g. senders' sending profile), y=observations; 
% lai: x=biometric measurement during enrollment stage, y=biometric measurement during release stage;
% agrawal: x=distribution of database attribute; y=perturbed release;
% SO: x is truth again (as in infogain metrics below), y seems to be observation in all cases}
%\subsubsection{Normalized Conditional Entropy}
\label{metric:normalized-conditional-entropy}
Normalized conditional entropy uses the entropy of $X^*$ (because entropy is the maximum of conditional entropy) to normalize conditional entropy~\cite{lai_privacy_2011}.
%Conditional entropy can be normalized to enable comparisons between entropy values.
%Because the maximum of the conditional entropy of $X$ given $Y$ is the entropy of $X$, entropy can be used to normalize~\cite{lai_privacy_2011}.
\[priv_{\text{NCE}} \equiv \frac{H(X^*|Y)}{H(X^*)} \]

\subsubsection{Inherent Privacy}
\label{metric:inherent_privacy}
Inherent privacy (also called scaled anonymity set size) is derived from entropy and describes the privacy inherent in the random variable $X$ as the number of possible outcomes given the expected amount of binary questions the adversary needs to answer~\cite{agrawal_design_2001,andersson_fundamentals_2008}.
\[priv_{\text{IP}} \equiv 2^{H(X)} \]

% \subsubsection{Conditional Privacy}
\label{metric:conditional-privacy}
In a similar way, conditional privacy is based on conditional entropy and measures the privacy inherent in a random variable $X$, given random variable $Y$~\cite{agrawal_design_2001}.
\[priv_{\text{CP}} \equiv 2^{H(X|Y)} \]

\subsubsection{Cross-entropy / Likelihood}
\label{metric:cross-entropy}

In data clustering, cross-entropy measures the uncertainty in predicting the original dataset from the clustered model~\cite{merugu_privacy-preserving_2003}.
Generally, cross-entropy measures the amount of information needed to identify an object in the data set if the original data are coded in terms of the model's distribution $X$, rather than their true distribution $X^*$.
Cross-entropy is derived from entropy, which indicates the uncertainty in a probability distribution (Section \ref{metric:entropy}), and the relative entropy $D_\text{KL}$, which indicates the distance between two probability distributions (Section \ref{metric:relative-entropy}).
% \[priv_{\text{CE}} \equiv 2^{(-\frac{1}{|Z|} \sum_{z \in Z} \log_2 p_\lambda(z))} \]
\[priv_\text{CE} \equiv H(X^*) + D_\text{KL}(X^*||X) \]

% \todo[inline]{This is a forward reference, and as such not helpful. I have no idea what $D_{KL}$ is. Can we at least say what $D_{KL}$ is measuring?}

\subsubsection{Cumulative Entropy}
\label{metric:cumulative-entropy}

In location privacy, cumulative entropy measures how much entropy can be gathered on a route through a series of independent mix zones.
A mix zone $R$ is a region where several nodes are close to each other at the same time, such that the adversary cannot distinguish the nodes as they leave the mix zone in different directions.
Cumulative entropy adds up the entropy gathered in each mix zone $r$ on a node's path~\cite{freudiger_mix-zones_2007}. $X_r$ indicates the adversary's estimate at the time when the user is in mix zone $r$.
\[priv_{\text{CUE}} \equiv \sum_{r \in R} H(X_r) \]

\subsubsection{Protection Level}
\label{metric:protection-level}

The protection level is a metric from location privacy which is based on the popularity of regions $r \in R$.
The popularity of a region $r$ with respect to a set of users, $\text{Pop}(r,U)$, is defined as the inherent privacy (Section \ref{metric:inherent_privacy}) computed over the frequencies $f^r_U$ of location samples from all users in this region.
A user $u$ in the system can specify a public reference region $r^{\text{ref}}_u$ to define how private they want to be.
The protection level is then the ratio of the average popularity of all regions $R_u$ along the user's trajectory (with respect to the set of users $\hat{U}$ common to all these regions) and the popularity of the reference region~\cite{xu_feeling-based_2009}.
A protection level of at least 1 indicates adequate protection.
%\[priv_{\text{PL}} \equiv \frac{\sum_{r \in R} Pop(r,U)}{|U| Pop(r_{\text{ref}},U)} \text{, where popularity } Pop(r,U) = 2^{H(f_r)} \]

\[priv_{\text{PL}} \equiv \frac{\sum_{r \in R_u} \text{Pop}(r,\hat{U})}{|R_u| \text{Pop}(r^{\text{ref}}_u,U)} 
\text{, where } \text{Pop}(r,U) = 2^{H(f^r_U)} \]

% \todo[inline]{Whats with the $\sum_i$? is there something missing?}
% \todo[inline,color=red!20]{I think that's all regions on the user's path. another problem with the description: we say 'average popularity' -- but we only use the sum of popularities without averaging (they actually refer to the average wrt the users common to all regions
% $\sum_{R_i \in R_u}$
% }
% \todo[inline,color=red!20]{what's the interpretation for this metric? what do its values mean?}

\subsubsection{Asymmetric Entropy}
\label{metric:asymmetric-entropy}

When the adversary has access to prior information about the distribution of the random variable $X$, the point $\alpha$ where uncertainty is highest can differ from equiprobability.
For example, in genomics, information about the population-wide average probabilities of genetic variations are readily available and determine where the adversary's uncertainty is highest.
In this case, asymmetric entropy can be used instead of entropy to account for this prior information~\cite{ayday_protecting_2013,marcellin_asymmetric_2006}.
Asymmetric entropy uses $p(x)$ as the adversary's probability of inferring the target correctly, and does not take into account individual probabilities for the other members of the anonymity set.

%\[priv_{\text{AE}} \equiv \sum_i \frac{p(x_i)(1-p(x_i))}{(-2w_i+1)p(x_i)+w_i^2} \]
\[priv_{\text{AE}} \equiv \frac{p(x)(1-p(x))}{(-2\alpha+1)p(x)+\alpha^2} \]
%\todo[inline]{i dont understand this. how can entropy have a different maximum for each element? i'd also move this down to conditional entropy, and the notation has to change. its completely different from the rest}
%\todo[inline,color=red!20]{for SNPs, we have population-wide minor allele frequencies as baseline guesses. So for each SNP, uncertainty depends on whether it has 0, 1, or 2 minor alleles. asymmetric entropy then combines the probability for inferring the snp correctly with a value w (where uncertainty is highest) that depends on the minor allele frequency and the ground truth for this SNP}
%\todo[inline,color=red!20]{the idea for the sum is that we have many random variables (one for each SNP) and want to aggregate them, but each random variable has a different maximum entropy.}
%\todo[inline,color=red!20]{vanilla entropy: maximum uncertainty for equiprobability}
%\todo[inline,color=red!20]{for some reason, we decided to not give asymmetric entropy its own section. remember why? I'd put it back into its own section again and move further down (to a less exposed place), because in genomics the metric is actually not very good.
%And of course, this metric combines two things: an asymmetric entropy, and the cumulation over SNPs (which is just cumulative entropy with AE instead of E as base metric).
%}

In genomic privacy, asymmetric entropy can be applied to each genetic variation separately (with separate parameters $\alpha_i$) and then summed up to give {\sffamily\itshape{cumulative asymmetric entropy}} (similar to cumulative entropy in Section \ref{metric:cumulative-entropy}).

\subsubsection{Genomic Privacy}
\label{metric:genomic-privacy}

Genomic privacy assumes that the adversary has estimated probabilities for all genetic variations $V$ (so-called \aclp{SNP}, or \acp{SNP}) that occur in a person's genome. 
Most \acp{SNP} have two variants, one of which is less common than the other in human populations.
The metric uses the probabilities for the cases where a \ac{SNP} $v$ is present with the less common variant and weights these probabilities with a rating $\omega_v$ of each \ac{SNP}'s severity, which indicates, for example, how much a \ac{SNP} contributes to a disease~\cite{ayday_personal_2013}.
The value of genomic privacy does not have an intuitive interpretation and depends strongly on the number of \acp{SNP} studied and the magnitude of the severities.
%The severity can be determined by the SNP's contribution to diseases and tables of disease severities provided by insurance companies
\[priv_{\text{GP}} \equiv - \sum_{v \in V} \log_2 (p(v \text{ has less common variant})) \cdot \omega_v \]

%\todo[inline,color=red!20]{what's the interpretation for this metric? what do its values mean?}

% \subsubsection{Conditional Computational Entropy}
% 
% This metric combines conditional entropy with the computational model of an efficient adversary. 
% The metric states that sensitive information $S$ has $r$ bits of computational entropy, given leaked $L \subset S$ and cryptographically hidden information $C \subset S$, if there is a distribution $F$ that is computationally close to the distribution of the hidden information $C$, so that the information theoretic entropy of $S$, given $F$ and $L$, is at least $r$~\cite{bernhard_measuring_2012}.
% \[priv_{\text{CCE}} \equiv r \text{, where } \exists F=(F_k)_{k \in \mathbb{N}} \text{ so that } (S, L, C) \approx^C (S, L, F) \text{ and } (\forall k \in \mathbb{N}) \mathbb{F}(S_k|L_k,F_k) \geq r \]

% \subsubsection{Entropy of Trip Probabilities + Bayes}
% This metric combines information theory with probability theory.
% It is based on the entropy of trip probabilities at snapshots (or single points in time) and updates the probabilities between them using Bayesian belief tables~\cite{ma_measuring_2010}.
% \begin{align*}
%  priv_{\text{ETP}} \equiv -\sum_k p_k \log(p_k) \text{, where $p_k$ are updated as } p_k^{H_i^+} = \frac{p_k^{Sn_i} p_k^B}{\sum_k p_k^{Sn_i} p_k^B} \\ \text{with }p_k^B = p_k^{H_j^+} \text{ (trip in belief table) or } p_k^B = \frac{1}{n_i} \text{ (otherwise)}
% \end{align*}

\subsubsection{User-centric Privacy}
\label{metric:user-centric-privacy}

User-centric privacy assumes that the privacy of a user decreases linearly over time with speed $\omega$.
This decay can be expressed through the privacy loss function $\beta(\Delta t)$, with
$\Delta t$ being the time elapsed since $t'$, the time of the last successful activation of a privacy protection mechanism~\cite{freudiger_non-cooperative_2009}.
This metric makes use of a base privacy metric $B$, with $B_{t'}$ giving the level of privacy enjoyed by the user at time $t'$.
To avoid a negative level of privacy, the metric is capped at zero.
%Note that the formulation here assumes that higher values of the base metric indicate higher privacy.
Note that for base metrics where lower values indicate higher privacy, the privacy loss function $\beta(\Delta t)$ has to be added to the base metric instead of subtracting it.

\begin{align*}
priv_{\text{UCP}} &  \equiv \max(0,B_{t'} - \beta(\Delta t))\\
\beta(\Delta t) &= \omega \cdot \Delta t,~\Delta t \geq 0
\end{align*}

User-centric privacy assumes a linear decay of privacy, which may not hold for all base metrics.
In addition, the metric assumes that successive activations of a privacy mechanism are independent from each other.

%\todo[inline,color=red!20]{subtracting $\beta$ only works for higher-better metrics, lower-better metrics would have to add $\beta$}
%\todo[inline,color=red!20]{mention our criticisms: linear decrease by a 'magic number' may not be sensible; re-application of privacy mechanism not guaranteed to work and have to be independent}
%\todo[inline,color=blue!20]{Reviewer 2: page 11, line 30: I have no idea what is going on here.}
\end{subsection}

%%%%%%%%%%%%%%%%%%%%%%%%%%%%%%%%%%%%%%%%%%%%%%%%%%
% Information Gain or Loss
%%%%%%%%%%%%%%%%%%%%%%%%%%%%%%%%%%%%%%%%%%%%%%%%%%
\begin{subsection}{Information Gain or Loss}

Metrics in this category measure the amount of information an adversary can gain, assuming that privacy is higher the less information an adversary can obtain.
Similar to uncertainty metrics, many information gain metrics are based on information theory.
However, information gain metrics explicitly consider the amount of prior information.

While frequently used in the context of communication systems or databases, metrics in this category have found wide application across all domains, including genome privacy, smart metering, and social networks.

%\begin{table}[htp]
 %\tbl{Information Gain/Loss Metrics by Domain\label{tab:infgain-metricsbydomain}}{
%%  \resizebox{.7\textwidth}{!}{%
  %\resizebox*{\textwidth}{!}{%
  	%\rowcolors{2}{gray!20}{white}
 %\begin{tabular}{>{\raggedleft\slshape}p{3.3cm}VVVVVV}
 %\upshape\bfseries Metric & \bfseries Communication & \bfseries Databases & \bfseries Location & \bfseries Smart Metering & \bfseries Social Networks & \bfseries Genome Privacy \\
%\midrule
  %Amount of leaked information &  & & & & Backstrom 2007 & Wang 2009 \\
     %Conditional Mutual Information & Coble 2008  & & & & & \\
     %Conditional privacy loss &  & Agrawal 2001 & & & & \\
     %Increase in adversary's belief & & Narayanan 2008  & & & & \\
     %Information Disclosure & & Fawaz 2016 & & & & \\
     %Information Surprisal & & & & & Chen 2013  & \\
     %Maximum information leakage & & du Pin Calmon 2012  & & & & \\
     %Mutual information &  & & & & & Lin 2002 \\
     %Normalized mutual information  &  & & & & & Humbert 2013  \\
     %Pearson's correlation coefficient & & & & Kim 2011 & & \\
     %Privacy Score & & & & & Liu 2010 & \\
     %Reduction in observable features & & & & McLaughlin 2011  & & \\
     %Relative entropy & Deng 2007    & & & & & \\
     %(Relative) Loss of anonymity & Chatzikokolakis 2007 & & & & & \\
     %System anonymity level & Gierlichs 2008  & & & & & \\
 %\end{tabular}}}
%\end{table}

\begin{table}[htp]
 \caption{Metrics and references in the information gain/loss category and the domains they originated in}
 \label{tab:infgain-metricsbydomain}
%  \resizebox{.7\textwidth}{!}{%
%   \resizebox*{\textwidth}{!}{%
  	\rowcolors{2}{gray!20}{white}
			\resizebox{.85\textwidth}{!}{%
 \begin{tabular}{p{1.2cm}p{6cm}p{2.8cm}p{2cm}}
 \upshape\bfseries Section & \bfseries Metric & \bfseries Original Domain & \bfseries Reference \\
\midrule
 \ref{metric:amount-of-leaked-information} & Amount of leaked information & Social networks & \cite{backstrom_wherefore_2007} \\
 \ref{metric:relative-entropy} & Relative entropy & Communication & \cite{deng_measuring_2007} \\
 \ref{metric:mutual_information} & Mutual information & Genome privacy & \cite{lin_using_2002} \\
 \ref{metric:normalized-mutual-information} & Normalized mutual information  & Genome privacy & \cite{humbert_addressing_2013} \\
 \ref{metric:conditional-privacy-loss} & Conditional privacy loss & Databases & \cite{agrawal_design_2001} \\
 \ref{metric:conditional_mutual_information} & Conditional mutual information & Communication & \cite{coble_formalized_2008} \\
 \ref{metric:loss-of-anonymity} & (Relative) Loss of anonymity & Communication & \cite{chatzikokolakis_anonymity_2007} \\
 \ref{metric:maximum-information-leakage} & Maximum information leakage & Databases & \cite{du_pin_calmon_privacy_2012} \\
 \ref{metric:system-anonymity-level} & System anonymity level & Communication & \cite{gierlichs_revisiting_2008} \\
 \ref{metric:information-surprisal} & Information surprisal & Social networks & \cite{chen_how_2013} \\
 \ref{metric:privacy-score} & Privacy score & Social networks & \cite{liu_framework_2010} \\
 \ref{metric:positive-information-disclosure} & Positive information disclosure & Databases & \cite{miklau2004formal} \\
 \ref{metric:increase_adv_belief} & Increase in adversary's belief & Databases & \cite{narayanan_robust_2008} \\
 \ref{metric:reduction-in-observable-features} & Reduction in observable features & Smart metering & \cite{mclaughlin_protecting_2011} \\
 \ref{metric:pearson-correlation-coefficient} & Pearson's correlation coefficient & Smart metering & \cite{kim_cooperative_2011} \\
 \ref{metric:full-partial-disclosure} & Full/Partial disclosure & Databases & \cite{kenthapadi2005simulatable} \\
 \end{tabular}%
}
\end{table}

\subsubsection{Amount of Leaked Information}
\label{metric:amount-of-leaked-information}

This metric counts the information items $S$ disclosed by a system, e.g., the number of compromised users~\cite{backstrom_wherefore_2007} or the number of leaked DNA base pairs~\cite{wang_privacy-preserving_2009}.
However, this metric does not indicate the severity of a leak because it does not account for the sensitivity of the leaked information.
\[priv_{\text{ALI}} \equiv |S|\]

\subsubsection{Relative Entropy}
\label{metric:relative-entropy}

Relative entropy (also called Kullback-Leibler divergence $D_\text{KL}$) measures the distance between two probability distributions.
The two distributions must fulfill absolute continuity, i.e. if $q(x)=0$, then $p(x^*)=0$ as well.
As a privacy metric, the two distributions usually represent the true distribution $X^*$ and the adversary's estimate $X$, and relative entropy gives the amount (bits) of probabilistic information revealed to the adversary~\cite{deng_measuring_2007}.
For example, in a location privacy scenario, the adversary may aim to find out which points of interest a user has visited. Relative entropy then indicates how far the adversary's estimate is from the truth.
\[priv_{\text{RLE}} \equiv D_\text{KL}(X^*||X) = \sum_{x, x^*}p(x^*) \log_2 \frac{p(x^*)}{q(x)} \]

Instead of the adversary's estimate $X$, some applications of relative entropy use the adversary's observations $Y$, for example of obfuscated data in smart metering \cite{kalogridis_privacy_2010}.
In this case, relative entropy indicates how far the distribution of obfuscated data is from the true distribution.
% D_{KL}(p(X)||q(X))
%\todo[inline,color=red!20]{how do we make clear that p(x*) draws from the X* distribution instead of the X distribution? because that was the purpose of using p() and q() in the original definition}
%\todo[inline,color=blue!20]{Reviewer 2: page 11, line 52: How should $priv_\text{RLE}$ be interpreted? It ranges from 0
%to infinity, with higher being better. But I think more context is
%needed to interpret the value.}
%\todo[inline,color=red!20]{what's the interpretation for this metric? what do its values mean? give examples for location (freudiger 2012) and smart metering?}

\subsubsection{Mutual Information}
\label{metric:mutual_information}

Mutual information quantifies how much information is shared between two random variables.
It can be computed as the difference between entropy (Section \ref{metric:entropy}) and conditional entropy (Section \ref{metric:conditional_entropy}).
In most cases, mutual information is computed between the true distribution of data $X^*$ and the adversary's (obfuscated) observations $Y$, and it measures the amount of information leaked from a privacy mechanism~\cite{lin_using_2002}.
\[priv_{\text{MI}} \equiv I(X^*;Y) = H(X^*) - H(X^*|Y) = \sum_{x^* \in X^*} \sum_{y \in Y} p(x^*,y) \log_2 \frac{p(x^*,y)}{p(x^*)p(y)}\]

%\subsubsection{Normalized Mutual Information}
\label{metric:normalized-mutual-information}

To allow comparisons between scenarios, mutual information between $X^*$ and $Y$ can be normalized using the entropy of $X^*$.
%o make the value range of mutual information comparable between scenarios, 
This can be interpreted as the degree of dependence between hidden data $X^*$ and observed data $Y$~\cite{humbert_addressing_2013}.
\[priv_{\text{NMI}} \equiv 1 - \frac{I(X^*;Y)}{H(X^*)} \]
Alternatively, mutual information can be normalized using the number of entries in $X^*$, for example the number of rows in a database.
In this case, normalized mutual information measures the number of bits leaked on average from any entry \cite{sankar2013utility-privacy}.
 
% \todo[inline]{an introductory sentence why it needs to be normalized would be great.}
% \todo[inline,color=red!20]{I(X;Y) -- lin: original data, binned data; humbert: truth, prior knowledge; agrawal: truth, perturbed data; chatzikokolakis: anonymous events, observed events; calmon: private features, privacy preserving output. I(X;Y|Z) -- coble: truth, observations, prior knowledge. Looks like X is always the true data, Y is obfuscated data/observed data or prior knowledge, Z is prior knowledge (only one paper where it is used)}

\subsubsection{Conditional Privacy Loss}
\label{metric:conditional-privacy-loss}

Another way of normalizing mutual information is the conditional privacy loss, which measures the fraction of privacy of $X^*$ which is lost by revealing $Y$~\cite{agrawal_design_2001}.
\[priv_{\text{CPL}} \equiv 1-2^{-I(X^*;Y)} \]

\subsubsection{Conditional Mutual Information}
\label{metric:conditional_mutual_information}

Mutual information can also be applied when the adversary has access to prior knowledge.
Conditional mutual information measures the amount of information about $X^*$ that can be learned by observing $Y$, given prior knowledge $Z$.
It measures the correlation between $X^*$ and $Y$ given $Z$~\cite{coble_formalized_2008}.
\[priv_{\text{CMI}} \equiv I(X^*; Y|Z) = H(X^*|Z)-H(X^*|Y,Z) \]

\subsubsection{(Relative) Loss of Anonymity}
\label{metric:loss-of-anonymity}

Loss of anonymity describes the amount of information that can be learned about a set of anonymous events $X^*$, given a set of observed events $Y$, for the least private distribution of $X^*$~\cite{chatzikokolakis_anonymity_2007}.
In an anonymity protocol for example, $X^*$ indicates a user's sending profile and $p(y|x^*)$ describes the probability that the output $y$ is produced by the anonymity protocol, given a specific user input $x^*$.
To characterize the worst-case behavior of the anonymity protocol, the metric computes the maximum mutual information (Section \ref{metric:mutual_information}), i.e., the maximum amount of information that can leak from the anonymity protocol, over all possible distributions of user sending profiles.
%The metric is based on mutual information (Section \ref{metric:mutual_information}), and finds the least private distribution by taking a maximum over all possible distributions of $X^*$.
\[priv_{\text{LA}} \equiv \max_{p(x^*)} I(X^*;Y) \]

%\todo[inline,color=blue!20]{Reviewer 2: page 12, line 33: What does the notation $max_p(x) I(X; Y)$ mean?}
%\todo[inline,color=red!20]{need an example, explain what the values mean}
%\todo[inline,color=red!20]{decompose the joint probability used in MI: $p(x,y)=p(x)p(y|x)$. $p(x)$ depends on the user, in their case indicating how often the user sends messages. $p(y|x)$ is a characteristic of the anonymity protocol, i.e. given a specific user input $x$, $y$ is the observable output produced. then using $\max p(x)$ means looking at the worst possible user input distribution. so they take the maximum of mutual information over all possible input distributions $p(x)$. this is the maximum amount of information revealed by the anonymity protocol, regardless of user input.}

Relative loss of anonymity extends loss of anonymity by taking into account that the adversary has access to certain revealed information $Z$. 
Instead of mutual information, this metric is based on conditional mutual information (Section \ref{metric:conditional_mutual_information}) and indicates the maximum amount of information that can leak from a privacy mechanism over all distributions of anonymous events $X^*$, given observations $Y$ and prior knowledge $Z$.
\[priv_{\text{RLA}} \equiv \max_{p(x^*)} I(X^*; Y|Z) \]

%\todo[inline]{Verstehe ich nicht leider. was ist ein Beispiel fuer Z? Ist das Y wieder eine beobachtung hier? p von I von X und Y abhaengig von Z. uff}
%\todo[inline,color=red!20]{I think 'revealed information' == some kind of prior knowledge (in this case, information that has been revealed, can be revealed through the privacy mechanism 'by design', their examples is the cardinality of x). }

\subsubsection{Maximum Information Leakage}
\label{metric:maximum-information-leakage}

Maximum information leakage modifies mutual information to consider only a single realization of the random variable $Y$. 
It quantifies the maximum amount of information about private events or data $X^*$ that can be gained by an adversary observing a single output $y$, where the maximum is taken over all possible outputs~\cite{du_pin_calmon_privacy_2012}.
In communications, for example, maximum information leakage can refer to the amount of information the adversary gains by observing a single message, taking the maximum information gain over all possible messages that the adversary could observe.
%\[priv_{\text{MIL}} \equiv \max_{y \in Y} H(X^*) - H(X^*|Y=y) \]
\[priv_{\text{MIL}} \equiv \max_{y \in Y} I(X^*;Y=y) \]

%\todo[inline]{a small example would really help here. Also, couldn't this be $\max_{y \in Y} I(X;Y)$?}

\subsubsection{System Anonymity Level}
\label{metric:system-anonymity-level}

In anonymous communication, the system's anonymity level describes the amount of additional information needed to reveal all sender-receiver relationships.
Sender-receiver relationships are described in the adjacency matrix $A$.
%, and the system's anonymity level corresponds to finding the equivalence class containing the correct perfect matching between senders and receivers.
%The metric is based on computing entropy using the frequency of perfect matchings $C_p$ in the permanent of the adjacency matrix $per(A)$, normalized with the number of users $|U|$~\cite{gierlichs_revisiting_2008}.
Among all possible combinations between senders and receivers, the adversary aims to find the correct combination that corresponds to the messages sent in a communication round.
If each sender/receiver can only send/receive one message, then the number of combinations that the adversary has to choose from is the permanent of the adjacency matrix $per(A)$, and the adversary's estimated probability for each combination would be $p(x) = \frac{1}{per(A)}$.
Multiplicities on the sender or receiver side (i.e. one sender sending multiple messages, or a receiver receiving multiple messages) partition the possible combinations into equivalence classes $E$.
The cardinality $|E|$ of each equivalence class indicates how many combinations it contains.
The adversary's estimate thus improves depending on the cardinalities: $p(x) = \frac{|E|}{per(A)}$.
The system anonymity level then computes the entropy based on this adversary estimate and normalizes with the number of users $|U|$~\cite{gierlichs_revisiting_2008}.
% number of equivalent perfect matchings in an equivalence class $C_p$
% number of equivalence classes $\Theta$
% set of all users $U$
% permanent of the adjacency matrix gives count of possible perfect matchings
% \[
%  priv_{\text{SAL}} \equiv
%  \begin{cases}
%   0, &\text{ if } |U|=1 \\
%   \frac{1}{\log(|U|!)} (-\sum_{p=1}^{\Theta} \frac{C_p}{per(A)} \cdot \log(  \frac{C_p}{per(A)} )), &\text{ if }|U|>1
%  \end{cases}
% \]
\[
 priv_{\text{SAL}} \equiv
 \begin{cases}
  0, &\text{ if } |U|=1 \\
  \frac{1}{\log(|U|!)} H(\frac{|E|}{per(A)}), &\text{ if }|U|>1
 \end{cases}
\]

%\todo[inline]{How can the entropy computed here? It needs a random variable as input. Why does $C$ have a subscript?}
%\todo[inline,color=red!20]{instead of p(x), this metric uses a frequency -- the frequency of $C_p$ in $per(A)$}
%\todo[inline,color=red!20]{perfect matching: one-to-one relation between senders and receivers, indicates that edge between input and output in communication graph exists, i.e. a perfect matching is one possible assignment of senders to receivers, and the correct perfect matching is the correct association between senders and receivers in one communication round. 
%if a sender can send more than one message in a round, there may be several perfect matchings that indicate the same combination of sender+receiver. multiplicities in the sender/receiver thus partition the perfect matchings into equivalence classes $C_p$.
%
%permanent of $t \times t$ adjacency matrix for a fully connected graph is $t!$. 
%subscript $p$ denotes the equivalence classes, i.e. instead of $x \in X$, entropy in this case sums over $p \in \text{equivalence classes}$
%}

\subsubsection{Information Surprisal}
\label{metric:information-surprisal}

Information surprisal is a measure of self-information.
It quantifies how much information is contained in a specific outcome $x$ of a random variable $X$.
In social networks, $X$ represents user profiles that contain a set of attributes, and $p(x)$ is the frequency of a specific user's combination of attribute values within the set of all social network users.
Information surprisal measures how surprised the adversary would be upon learning the user's attributes~\cite{chen_how_2013}.
% % how much information the adversary gains if he gets to know the value of one specific attribute in a user profile~\cite{chen_how_2013}.
% \[priv_{\text{IS}} \equiv - \log_2 P(X=x) = - \log_2 (\frac{|x|}{|X|}) \]
\[priv_{\text{IS}} \equiv - \log_2 p(x)\]

% \todo[inline]{Can this still be the same $X$ as for the entropy? does this metric give a value for each of the $x \in X$ then or a single value}
% \todo[inline,color=red!20]{X is an individual user's profile with a set of attributes A; p(x) is the frequency of the user's attribute combination within the set of all social network users. metric measures the adversary's surprise when learning the attributes of a specific user profile, or the uniqueness of that user profile.}

\subsubsection{Privacy Score}
\label{metric:privacy-score}

The privacy score in a social network indicates a user $u$'s potential privacy risk.
It increases with the sensitivity $\omega_{x^*}$ of information items $x^* \in X^*$ and their visibility $\text{Vis}(x^*,u)$, e.g., the number of users knowing about each item~\cite{liu_framework_2010}.
Any information on a user's profile can be an information item, for example the user's gender or mother's maiden name.
To make the privacy score comparable between users, the sensitivity $\omega_{x^*}$ is independent of the user (for example, computed from the privacy settings of a large number of users).

\[priv_{\text{PS}} \equiv \sum_{x^* \in X^*} \omega_{x^*} \cdot \text{Vis}(x^*,u) \]

%\todo[inline]{Paper nochmal lesen ;) kick user out, as this is a per user metric. Is this aggregated over all users in the paper? is $\beta$ actually a per-user setting?}
% \todo[inline]{$beta$ is the loss function and also a horrible letter for sensitivity. }
% \todo[inline,color=red!20]{$\phi$ was one of the few letters not yet used.}
%What are the information items? example needed}
% \todo[inline,color=red!20]{examples for information items: gender (least sensitivity, according to their paper), mother's maiden name (highest sensitivity)}
%\todo[inline,color=red!20]{what's the interpretation for this metric? what do its values mean? -- the values seem to be meaningless and cannot be compared if different methods are used for computing sensitivity/visibility for the same dataset. the interpretation is that high numbers mean high privacy risk (i.e. users are more comfortable putting stuff on social networks).}
%\todo[inline,color=red!20]{the sensitivity of an item $i$ is independent of the user. it is computed from the privacy settings for the item $i$ of a large number of users. the visibility of an item depends on the user's position in the social network, so it might make sense to leave it in. leaving it in would also highlight the fact that sensitivity is independent of users.}

\subsubsection{Positive Information Disclosure}
\label{metric:positive-information-disclosure}

Shannon's criterion for perfect secrecy \cite{shannon1949communication} demands that the adversary's prior probability for the secret $x^*$ equals the posterior probability that takes into account a new observation $y$, i.e. $p(x^*) = p(x^*|y)$, expressing that the adversary gains no additional information.
(For encryption, it has been shown that the one-time pad is the only cipher that satisfies perfect secrecy).
Building on Shannon's perfect secrecy, the positive information disclosure metric \cite{miklau2004formal} quantifies how much the adversary's posterior probability improves.
The metric indicates the highest improvement across all secrets $x^*$.

\[priv_{\text{PID}} \equiv \sup_{x^* \in X^*}{\frac{p(x^*|y) - p(x^*)}{p(x^*)}} \]

In location privacy, for example, the secret is the path that a user travels on, and new observations are geographic locations disclosed to the adversary \cite{fawaz2016privacy}.

\subsubsection{Increase in Adversary's Belief}
\label{metric:increase_adv_belief}

The increase in adversary's belief measures the difference between the adversary's prior and posterior probabilities  (e.g., of identifying an individual in a set of users).
Privacy is breached if this difference is greater than the privacy parameter $\tau$~\cite{narayanan_-anonymizing_2009}.
\[priv_{\text{IAB}} \equiv \tau \text{, where } p(x|y) - p(x) > \tau \]

\subsubsection{Reduction in Observable Features}
\label{metric:reduction-in-observable-features}

In smart metering, load hiding algorithms try to hide load transitions from the energy provider, because these can disclose at what time which appliance was used.
The reduction in observable features measures how many transitions are hidden successfully by a privacy protection mechanism~\cite{mclaughlin_protecting_2011}.
Load transitions form a time-series $\vec{T}$, and the feature mass ${F(\vec{T})}$ condenses this time series to a single value, for example the number of transitions in $\vec{T}$ with a certain property, such as a minimum power level.
The metric then relates the feature masses with ($\vec{T}_Y$) and without ($\vec{T}_{X^*}$) privacy protection.
%However, a reduction in observable features does not necessarily mean that sensitive information cannot be inferred.
%\todo[inline]{is the last sentence something you concluded, or is this in the paper? either way needs an example}
%\todo[inline,color=red!20]{for example, 90\% of features may be removed, but the remaining 10\% give away sensitive information. but: this sounds like a given for most metrics, maybe we don't need to mention it (or not here)?}
\[priv_{\text{ROF}} \equiv \frac{F(\vec{T}_{Y})}{F(\vec{T}_{X^*})} \]
% - there may be some features left
% - other features may not have been hidden, but useful to the adversary

%\todo[inline]{this is like the frantz one, isn't it?} 
%\todo[inline,color=red!20]{except that franz used entropy, this one uses ``feature mass''}

\subsubsection{Pearson's Correlation Coefficient}
\label{metric:pearson-correlation-coefficient}

In statistics, Pearson's correlation coefficient measures the degree of linear dependence between two random variables.
It is computed as the covariance between $X^*$ and $Y$, normalized with the standard deviations $\sigma_{X^*}$ and $\sigma_Y$.
In smart metering, this can be used to measure the correlation between original and obfuscated load data~\cite{kim_cooperative_2011}.
\[priv_{\text{PCC}} \equiv \frac{cov(X^*,Y)}{\sigma_{X^*}\cdot \sigma_Y} \]

%\todo[inline]{this is a good example for a description. only thing i'd add is to "`[...] and obfuscated load data to analyse the effectivity of a protection mechanism"`}
%\todo[inline,color=red!20]{I wouldn't, because that's what all metrics aim for. I think it would just confuse readers at this point}

\subsubsection{Full/Partial Disclosure}
\label{metric:full-partial-disclosure}

In query auditing, full disclosure indicates whether a set of database queries uniquely determines a sensitive value \cite{nabar2008survey}.
For example, if a database only permits aggregate queries to protect sensitive values, then a series of sum queries may allow to infer sensitive values.
However, the full disclosure metric has important limitations.
For example, if the adversary can infer that a sensitive value falls in a small interval, then full disclosure would not be violated because the sensitive value was not uniquely determined, but privacy may be violated nevertheless~\cite{kenthapadi2005simulatable}.

Partial disclosure addresses these limitations and is also applicable to online query auditing, i.e., the problem whether a new query should be answered or not, given a set of past database queries and answers.
The partial disclosure metric bounds the change in the adversary's confidence of inferring sensitive values. %, and answers a database query if this change in confidence is below a threshold $\tau$.
Specifically, a series of queries $q$ and query responses $y$ is called $\tau$-\textit{Safe} with regard to a particular numeric sensitive value $s_i$ and an interval $Int$ if this change in confidence is below a threshold $\tau$.
\[ priv_{\text{PD}} \equiv \text{Safe}_{\tau,i,Int} = 
\begin{cases}
1, & \text{if } \frac{1}{1+\tau} \leq \frac{p(s_i \in Int | q_1,...,q_t,y_1,...,y_t)}{p(s_i \in Int)} \leq (1+\tau) \\
0, & \text{otherwise}\end{cases}
\]
To apply this metric, the \textit{Safe} predicate has to hold for all sensitive items and all intervals.
This \textit{AllSafe} predicate can then be used to define the adversary's success, and an auditing mechanism is called private if the probability for the adversary's success is below a threshold $\tau'$ \cite{kenthapadi2005simulatable}.
This definition assumes that both adversary and auditor hold the same information about the distribution of sensitive values in the database.

% \todo[inline,color=blue!20]{Reviewer 2: The survey does not mention query auditing, which is also a form of
% privacy metric and is a well-developed area of research.}

% -- Simulatable auditing captures the notion that denying to answer certain queries also leaks information. (but probably not relevant for the metric)
% \todo[inline,color=red!20]{classic compromise vs probabilistic compromise}

\end{subsection}

%%%%%%%%%%%%%%%%%%%%%%%%%%%%%%%%%%%%%%%%%%%%%%%%%%
% Similarity or Diversity
%%%%%%%%%%%%%%%%%%%%%%%%%%%%%%%%%%%%%%%%%%%%%%%%%%
\begin{subsection}{Data Similarity}
\label{sec:metrics:similarity}

Data similarity metrics measure properties of observable or published data.
They are usually independent of the adversary and derive the privacy level solely from the features of disclosed data.
Almost all of these metrics originate from the database domain, where they are commonly applied in the context of data sanitization and data publishing.

%\begin{table}[htp]
 %\tbl{Data Similarity Metrics by Domain\label{tab:similarity-metricsbydomain}}{
%%  \resizebox{.7\textwidth}{!}{%
  %\resizebox*{\textwidth}{!}{%
  	%\rowcolors{2}{gray!20}{white}
 %\begin{tabular}{>{\raggedleft\slshape}p{3.3cm}VVVVVV}
 %\upshape\bfseries Metric & \bfseries Communication & \bfseries Databases & \bfseries Location & \bfseries Smart Metering & \bfseries Social Networks & \bfseries Genome Privacy \\
%\midrule
    %($\alpha$,k)-anonymity & & Wong 2006, Wang 2007 & & & &  \\
     %(c,t)-isolation & & Chawla 2005 & & & & \\
     %Cluster similarity & & &  & Kalogridis 2010 & & \\
     %Coefficient of determination $R^2$ & & &  & Kalogridis 2010 & & \\
     %($\epsilon$,m)-anonymity & & Li 2008 & & & &  \\
%%      Haplotype-SNP-test & & & & & & Zhou 2011 \\
     %Historical $k$-Anonymity & & & Bettini 2005 & & & \\
     %$k$-anonymity & & Samarati 2001 & & & & \\
     %(k,e)-anonymity & & Zhang 2007 & & & & \\
     %$\ell$-diversity & & Machanavajjhala 2007 & & & &  \\
     %$m$-invariance & & Xiao 2007 & & & & \\
     %Multirelational & & Nergiz 2009 & & & & \\
     %Normalized variance & & Oliveira 2003& & & & \\
     %Stochastic $t$-closeness & & Domingo-Ferrer 2015 & & & & \\
     %$t$-closeness & & Li 2007 & & & & \\
     %(X,Y)-privacy & & Wang 2006 & & & & \\
 %\end{tabular}}}
%\end{table}

\begin{table}[htp]
 \caption{Metrics and references in the data similarity category and the domains they originated in}
 \label{tab:similarity-metricsbydomain}
%  \resizebox{.7\textwidth}{!}{%
%   \resizebox*{\textwidth}{!}{%
  	\rowcolors{2}{gray!20}{white}
			\resizebox{.85\textwidth}{!}{%
 \begin{tabular}{p{1.2cm}p{6cm}p{2.8cm}p{2cm}}
 \upshape\bfseries Section & \bfseries Metric & \bfseries Original Domain & \bfseries Reference \\
\midrule
 \ref{metric:k-anonymity} & $k$-anonymity & Databases & \cite{samarati_protecting_1998} \\
 \ref{metric:alpha-k-anonymity} & ($\alpha$,k)-anonymity & Databases & \cite{wong__2006} \\
 \ref{metric:l-diversity} & $\ell$-diversity & Databases & \cite{machanavajjhala_l-diversity:_2007} \\
 \ref{metric:m-invariance} & $m$-invariance & Databases & \cite{xiao_m-invariance:_2007} \\
 \ref{metric:t-closeness} & $t$-closeness & Databases & \cite{li_t-closeness:_2007} \\
 \ref{metric:stochastic-t-closeness} & Stochastic $t$-closeness & Databases & \cite{domingoferrer2015t-closeness} \\
 \ref{metric:ct-isolation} & (c,t)-isolation & Databases & \cite{chawla_toward_2005} \\
 \ref{metric:ke-anonymity} & (k,e)-anonymity & Databases & \cite{zhang_aggregate_2007} \\
 \ref{metric:em-anonymity} & ($\epsilon$,m)-anonymity & Databases & \cite{li_preservation_2008} \\
 \ref{metric:multirelational-k-anonymity} & Multirelational $k$-anonymity & Databases & \cite{nergiz_multirelational_2009} \\
 \ref{metric:xy-privacy} & (X,Y)-privacy & Databases & \cite{wang_anonymizing_2006} \\
 \ref{metric:historical-k-anonymity} & Historical $k$-anonymity & Location & \cite{bettini_protecting_2005} \\
 \ref{metric:cluster-similarity} & Cluster similarity & Smart metering & \cite{kalogridis_privacy_2010} \\
 \ref{metric:coefficient-of-determination} & Coefficient of determination $R^2$ & Smart metering & \cite{kalogridis_privacy_2010} \\
 \ref{metric:normalized-variance} & Normalized variance & Databases & \cite{oliveira_privacy_2003} \\
 \end{tabular}%
}
\end{table}

\subsubsection{$k$-Anonymity}
\label{metric:k-anonymity}

$k$-Anonymity is conceptually similar to the size of the anonymity set (Section~\ref{metric:anonymity-set-size}), but does not consider the adversary.
% \todo[inline]{Technically, $|AS_u|$ does also not consider the adversary. The probabilities do}
% \todo[inline,color=red!20]{well, AS = users indistinguishable to the adversary, whereas k-anon = rows indistinguishable wrt quasi-identifiers}
It was originally proposed to prepare statistical databases for publication.
A medical database, for example, would contain both identifying information (e.g., the names of individuals) and sensitive information (e.g., their medical conditions).
$k$-Anonymity assumes that identifying columns are removed from a database before publication, and then demands that the database table~$D$ can be grouped into equivalence classes with at least~$k$ rows that are indistinguishable with respect to their quasi-identifiers~$q$~\cite{samarati2001protecting,sweeney_k-anonymity:_2002}.
Quasi-identifiers by themselves do not identify users, but can do so when correlated with other data.
For example, the combination of the three quasi-identifiers zip code, date of birth, and gender identifies 87\% of the American population \cite{sweeney_k-anonymity:_2002}.
% The sets of indistinguishable quasi-identifiers partition a table into equivalence classes.
Each equivalence class $E$ contains all rows that have the same values for each quasi-identifier $q$, for example all individuals with the same zip code, date of birth, and gender.
To increase the size of equivalence classes to a minimum of $k$ rows, several algorithms exist to transform a given database to make it $k$-anonymous, for example using suppression or generalization~\cite{samarati_protecting_1998} or random sampling~\cite{li2012sampling} (the latter is interesting because it also satisfies approximate differential privacy, see Section \ref{metric:approximate-differential-privacy}).

%\todo[inline,color=blue!20]{Rev1: Xu, Y., Ma, T., Tang, M., \& Tian, W. (2014). A survey of privacy
%preserving data publishing using generalization and suppression. Applied
%Mathematics \& Information Sciences, 8(3), 1103.}
%\[priv_{\text{KA}} \equiv k \text{, where } \forall (c = \text{ combination of values of }q): |D[c]| \geq k\]
\[priv_{\text{KA}} \equiv k \text{, where } \forall E: |E| \geq k\]

%\todo[inline]{the notation $|D[c]|$ isn't really clear}
%
% papers that argue against k-anonymity: shokri2010unraveling, machanavajjhala2007ldiversity, aggarwal2005kanonymity, xiao2007minvariance, xiao2006personalized
%Originally, the purpose of $k$-anonymity was identity hiding, i.e., that the release of an anonymized data set does not allow to draw conclusions about the record owner of a row in the data set.\todo{we already say that above}
%When an adversary knows a certain user is included in an equivalence class, it must also be ensured that the user's sensitive value is sufficiently hidden.\todo[inline]{this is what all the extensions take care of...}
%Furthermore, in settings where the sensitive values are also quasi-identifiers (e.g. in location privacy or for high-dimensional data such as movie ratings \cite{narayanan_robust_2008}), care must be taken to also protect these properties accordingly.\todo[inline]{i'm not sure whether this is actually correct. just seems like generalization to hide user identities. The more I think about it, vanilla $k$-anon should be only identity hiding, and the extensions should be also property-hiding --- (Isa) we already argue below that k-anon is unsuitable for high-dimensional data}

However, studies have shown $k$-anonymity to be insufficient, especially for high-dimensional data~\cite{aggarwal_k-anonymity_2005} and against correlation with other data sets~\cite{machanavajjhala_l-diversity:_2007}, because it fails to protect against attribute disclosure~\cite{xiao_personalized_2006}, i.e. it does not provide property hiding.
In addition, $k$-anonymous data releases do not offer protection across multiple releases of the same data set~\cite{xiao_m-invariance:_2007}, or when sensitive data, such as location data, are semantically close~\cite{shokri_unraveling_2010}. 
Despite this criticism, $k$-anonymity is still widely used today, and is routinely applied to new privacy domains.

\subsubsection{$(\alpha,k)$-Anonymity / Privacy Templates}
\label{metric:alpha-k-anonymity}

To prevent attribute disclosure and thus allow for property hiding, $(\alpha,k)$-anonymity extends $k$-anonymity with the additional requirement that in any equivalence class $E$ (rows that have the same quasi-identifier values), the frequency of a sensitive value $s$ has to be less than $\alpha$~\cite{wong__2006,wang_handicapping_2007}.
As a result, no single sensitive attribute can be dominant in an equivalence class.
\[priv_{\text{AK}} \equiv (\alpha,k) \text{, where } \forall E: |E| \geq k \wedge \frac{|(E,s)|}{|E|} \leq \alpha \]

% \subsubsection{Privacy Templates}
% 
% Privacy templates are specified by the data owner to limit the confidence of an adversary to a threshold value $h$.
% This means that the percentage of records containing a sensitive attribute $s$ among those with the same quasi-identifiers $q$ must not exceed $h$~\cite{wang_handicapping_2007}.
% 
% \[priv_{\text{PT}} \equiv h \text{, where } \forall q: \max \frac{|q \wedge s|}{|q|} \leq h \]
However, it has been shown that attribute linkage can occur even when the frequency of $s$ is less than $\alpha$~\cite{fung_privacy-preserving_2010}.

\subsubsection{$\ell$-Diversity}
\label{metric:l-diversity}

The $\ell$-diversity principle modifies $k$-anonymity to bound the diversity of published sensitive information. 
It states that every equivalence class $E$ must contain at least $\ell$ \textit{well-represented} sensitive values.
This general principle can be instantiated in different ways. 
In the simplest form, the $\ell$-diversity principle requires $\ell$ \textit{distinct} values in each equivalence class.
However, this simple instantiation does not prevent probabilistic inference attacks \cite{li_t-closeness:_2007}.

Stronger instantiations are based on the idea that in each equivalence class, the $\ell$ most frequent values of the sensitive attribute $s$ must have roughly the same frequencies~\cite{machanavajjhala_l-diversity:_2007}.
In an instantiation based on entropy (Section \ref{metric:entropy}), for example, similar frequencies are indicated by a high entropy $H(S_E)$ of the sensitive attribute frequencies.
\[priv_{\text{LE}} \equiv \ell \text{, where } \forall E : H(S_E) \geq \log (\ell)\]
In an instantiation based on recursion, the most frequent value $s_1$ must occur less often than all other values $s_i$ combined, within a multiplicative factor $\omega$.
\[priv_{\text{LR}} \equiv \ell \text{, where } \forall E : s_1 < \omega(s_{\ell}+s_{\ell+1}+...+s_n)\]

%\todo[inline]{what exactly is a quasi-identifier tuple? can we start with a very easy example where the frequency is just the count? the entropy is unnecessarily confusing here. after that we can show other instantiations}
%\todo[inline,color=red!20]{tuple not really necessary here, it's enough to say "for every equivalence class"}

% papers that argue against l-diversity: zhang2007information, li2007tcloseness, xiao2007minvariance, zhang_aggregate_2007
Although $\ell$-diversity is an improvement to $k$-anonymity, it has been shown to offer insufficient protection against some attacks.
In particular, it does not protect privacy when multiple releases of statistical data are available~\cite{xiao_m-invariance:_2007}, when the distribution of sensitive values is skewed~\cite{li_t-closeness:_2007}, or when sensitive attributes are semantically similar~\cite{li_t-closeness:_2007}, for example numerical values that are close to each other \cite{zhang_aggregate_2007}. 
In addition, the adversary may be able to reconstruct sensitive attributes if he knows the algorithm used for data sanitization~\cite{zhang_information_2007}. 
%Like $k$-anonymity, $\ell$-diversity does not sufficiently protect sensitive attributes that are numerical rather than categorical~\cite{zhang_aggregate_2007}.

%\todo[inline]{isn't it the other way round? example needed. Is this weakness different from the other 4 we mention before?}

\subsubsection{$m$-Invariance}
\label{metric:m-invariance}

$m$-Invariance modifies $k$-anonymity to allow for multiple releases of the same data set that may contain added, modified, or deleted rows. 
Given two $k$-anonymous data releases, an adversary can correlate the insertions and deletions between two releases to infer sensitive values.
To avoid this attack, $m$-Invariance states that every equivalence class $E$ must have at least $m$ rows, and the values for sensitive attributes $s$ must all be different~\cite{xiao_m-invariance:_2007}. 
In addition, the set of distinct sensitive values in each equivalence class must be the same in every release.
\begin{align*}
 priv_{\text{MI}} \equiv m \text{, where } &\forall E: |E| \geq m \ \wedge \forall s_i,s_j \in E: s_i \neq s_j \ \wedge \\
				    &\forall E: \text{ distinct } s \text{ must be the same in all releases}
\end{align*}

%\todo[inline]{we should explain how this is designed to avoid overlap attacks, i.e., that a missing or added row in another release can give away information}

\subsubsection{$t$-Closeness}
\label{metric:t-closeness}

To prevent attribute disclosure by an adversary with knowledge about the global distribution of sensitive attributes, $t$-closeness modifies $k$-anonymity to bound the distribution of sensitive values.
It states that the distribution $S_E$ of sensitive values in any equivalence class $E$ must be close to their distribution $S$ in the overall table.
In particular, the distance between distributions $d(S,S_E)$, measured using the Earth Mover Distance metric, must be smaller than a threshold $t$~\cite{li_t-closeness:_2007}.
\[priv_{\text{TC}} \equiv t \text{, where } \forall E: d(S,S_E) \leq t \]

%\todo[inline]{
%Violation of notation: overall table was never S, but i guess we mean that $S$ is the overall distribution. Maybe we should rephrase it to get rid of the ambiguity. 
%We should also mention against which of the weakness of $k$-anon this protects. 
%Also, distribution of values in an equ. class could be something like $S_E$. Do you think getting rid of a general $d()$ for distance would help? We could do something like $|S-S_i|_{EMD}$
%}

%\todo[inline,color=red!20]{prevents attribute disclosure via the skewness and similarity attacks in li\_t-closeness:\_2007}

\subsubsection{Stochastic $t$-Closeness}
\label{metric:stochastic-t-closeness}

Stochastic $t$-closeness was introduced to bridge the gap between $k$-anonymity based metrics and differential privacy (Section \ref{metric:differential-privacy}) \cite{domingoferrer2015t-closeness}.
$t$-Closeness in its original form leaves the sensitive values in a data table intact, whereas stochastic $t$-closeness allows stochastic modification of the sensitive values.
In particular, it can be shown that if the distribution of the sensitive values satisfies $\epsilon$-differential privacy (see Section \ref{metric:differential-privacy}), then the data table satisfies stochastic $t$-closeness, where the value of $t$ depends on the data table and $\epsilon$.

\subsubsection{$(c,t)$-Isolation}
\label{metric:ct-isolation}

This metric extends $k$-anonymity to consider an adversary.
The metric measures how well an adversary can isolate points in a database $D$~\cite{chawla_toward_2005}.
The difference between the adversary's estimate $x$ and the target point $x^*$ is given by $\delta_x$.
A target point $x^*$ is $(c,t)$-isolated, i.e., the adversary succeeds, if a ball~$\mathcal{B}$ with radius $c\delta_x$ around the adversary's estimate includes fewer than $t$ other points.
$c$ can be seen as isolation parameter, determining the size of the ball, whereas $t$ is a privacy threshold.
\[priv_{\text{CT}} \equiv (c,t) \text{, where } |\mathcal{B}(x,c\delta_x) \cap D| < t\text{ and }\delta_x = \|x-x^*\|\]

%\todo[inline]{has to be more clear that isolation is bad thing, right? it's not fully clear from which perspective this metric is approaching the problem. The equation uses $<$, so it means there are less than $t$ other points.}
%\todo[inline,color=red!20]{S was distribution of sensitive values before, should not suddenly become a sphere}

\subsubsection{$(k,e)$-Anonymity}
\label{metric:ke-anonymity}

To modify $k$-anonymity to apply to numerical instead of categorical attributes, $(k,e)$-anonymity additionally requires that the range of sensitive attributes in any equivalence class $E$ must be greater than $e$~\cite{zhang_aggregate_2007}.
\[priv_{\text{KE}} \equiv (k,e) \text{, where } \forall E : |E| \geq k \wedge range(E)>e \]
% papers that argue against it: fung2010privacypreserving and li2008preservation)
However, $(k,e)$-anonymity does not take into account how values within the range $e$ are distributed, which can lead to attribute disclosure via a proximity attack~\cite{li_preservation_2008}.
For example, if 90\% of sensitive values are within a short interval at one end of the range $e$, and the remaining 10\% are at the other end of $e$, then the adversary can infer with 90\% confidence that a user's sensitive value is in the short interval~\cite{fung_privacy-preserving_2010}.

%\todo[inline]{not a huge fan of $range$, but i guess $\max(E) - \min(E) > e$ isn't much better}

\subsubsection{$(\epsilon,m)$-Anonymity}
\label{metric:em-anonymity}

Another extension of $k$-anonymity to numerical attributes is $(\epsilon,m)$-anonymity.
It addresses the proximity attack against $(k,e)$-anonymity by bounding the probability of inferring the value of a sensitive attribute to at most $1/m$.
To achieve this bound, $(\epsilon,m)$-anonymity limits the number of members $e$ in each equivalance class $E$ with numerically $\epsilon$-similar sensitive values $s$ \cite{li_preservation_2008}.

% defines an $\epsilon$-neighborhood around each row's sensitive value $s(e)$, i.e. an interval of $\pm \epsilon$ around $s(e)$, and for each row $r$ limits the number of rows in its equivalence class $E_r$ whose sensitive values $s_{E_r}$ fall into $r$'s $\epsilon$-neighborhood.
%To achieve this bound, $(\epsilon,m)$-anonymity demands that in each equivalence class $E$ and for each sensitive value $x^*$ in $E$, at most $1/m$ of the rows can have a sensitive value $s$ that is similar to $x^*$, i.e. within an interval of $\pm \epsilon$ around $s$~\cite{li_preservation_2008}.
%\[priv_\text{EM} \equiv (\epsilon,m) \text{, where } \forall E \in D, \forall x \in S: \frac{| x \in [s-\epsilon, s+\epsilon] |}{|E|} \leq \frac{1}{m} \]

\begin{align*}
priv_\text{EM} \equiv  \forall E: \forall e \in E: \frac{|\hat{E}|}{|E|}  \leq \frac{1}{m}\text{,} &\text{ where } \hat{E} \text{ are the members of } E \text{ whose}\\
&\text{~sensitive values } s \text{ fall in } [s(e)-\epsilon, s(e)+\epsilon] 
\end{align*}

% \hat{E} = \{\hat{e}: \hat{e} \in E, \hat{e} \in I(\hat{e}) \} \]

% \[ priv_\text{EM} \equiv \forall E: \frac{|S|}{|E|} \leq \frac{1}{m} \text{, where } S \text{ is the number of tuples in } E \text{ whose sensitive values fall in } \[s-\epsilon, s+\epsilon\] \]

% \[ priv_\text{EM} \equiv \forall r \in D: \frac{|S_{E_r}|}{|E_{r}|} \leq \frac{1}{m} \text{, where } S_{E_r} = \{s_{E_r} : s_{E_r} \in S, s_r-\epsilon \leq s_{E_r} \leq s_r+\epsilon\}\]
% $| x \in [s-\epsilon, s+\epsilon] | $
% $[s-\epsilon, s+\epsilon]$ to be at most $1/m$.

% \todo[inline,color=blue!20]{Reviewer 2: page 15, line 34: I do not understand the notation $|x \in [s-\epsilon,s+\epsilon]|$}
%\todo[inline]{Not a great description. $1/m$ of the rows? rows in $E$? Last sentence is also a bit confusing. $S$ was not defined here.
%
%For the new notation, could we do something like that, or is this incorrect?
%
%$\forall E, \forall x \in S: |\hat S| \leq \frac{1}{m}$
%
%$\hat S \subseteq S = \{s| x-\epsilon \leq s \leq x+\epsilon\}$}

\subsubsection{Multirelational $k$-Anonymity}
\label{metric:multirelational-k-anonymity}

Multirelational $k$-anonymity modifies $k$-anonymity to apply to the record owner level instead of the record level, thus extending it to tables in a relational database~\cite{nergiz_multirelational_2009}.
To do this, multirelational $k$-anonymity joins the database table identifying the record owners $D_{pers}$ with all tables containing database records $D_i$, and then applies $k$-anonymity to the result of the join $J$.
For every record owner in $D_{pers}$, the resulting join needs to have at least $k-1$ other record owners with the same quasi-identifier values, and so the equivalence classes $E_{pers}$ contain all record owners with the same quasi-identifier values (instead of all records with the same quasi-identifier values, as in $k$-anonymity).
\begin{align*}
priv_{\text{MK}} &\equiv k \text{, where } J=D_{pers} \Join D_1 \Join \cdots \Join D_n \text{ and } \forall E_{pers} \in J: |E_{pers}| \geq k 
\end{align*}

% \todo[inline]{actually quite a nice explanation and notation in first line. second line is a bit chaotic and could be improved similar to what i mentioned for $k$-anon}
% \todo[inline,color=red!20]{this is the crucial idea: For every record owner in $D_{pers}$, the resulting join needs to have at least $k-1$ other record owners with the same quasi-identifier values.
% }

\subsubsection{$(X,Y)$-Privacy}
\label{metric:xy-privacy}

% the following is (X,Y)-Anonymity:
% $(X,Y)$-privacy groups database columns into identifying columns $X$ and sensitive columns $Y$, and requires that each value on $X$ is linked to at least $k$ distinct values on $Y$~\cite{wang_anonymizing_2006}.
% This is a generalization of $k$-anonymity, if $X$ corresponds to the quasi-identifiers, and $Y$ to a sensitive property.
% $(X,Y)$-privacy can be used to strengthen $k$-anonymity, for example when a database table has multiple records (e.g., multiple test results) for each individual.

$(X,Y)$-privacy modifies $k$-anonymity to bound the confidence with which sensitive values can be inferred~\cite{wang_anonymizing_2006}.
% to cope with sequential data releases by limiting the amount of linkage between two data sets~\cite{wang_anonymizing_2006}.
$X$ and $Y$ denote groups of database columns with quasi-identifiers and sensitive properties, respectively, and $|D[x]|$ denotes the number of records in database $D$ containing the value $x$.
% $(X,Y)$-privacy limits the confidence with which sensitive values in $Y$ can be inferred by requiring that for any values $x \in X$ and $y \in Y$, the percentage of records containing $x$ and $y$, among those containing $x$, be less than $k$.
$(X,Y)$-privacy then requires that for any values $x \in X$ and $y \in Y$, the percentage of records containing both $x$ and $y$, among those containing $x$, be less than $k$.
% $|D[x]|$ number of records in database $D$ containing $x$
% $|D[y,x]|$ number of records containing both $x$ and $y$
% \[priv_{\text{XY}} \equiv k \text{, where } \max_{d_2 \in D_2} \max_{d_1 \in D_1} \frac{|d_2,d_1|}{|d_1|} \leq k\]
\[priv_{\text{XY}} \equiv k \text{, where } \max_{y \in Y} \left\{ \max_{x \in X} \left\{ \frac{|D[y,x]|}{|D[x]|} \right\} \right\} \leq k \text{, and } 0 < k  \leq 1 \]
Applied to sequential data releases, $(X,Y)$-privacy uses columns that are common between two releases as $X$ and can thus ensure that sequential releases are $(X,Y)$-private.
% can be used to specify generalization rules that make sequential releases $(X,Y)$-private.

% \todo[inline]{Eigentlich ne recht einfache Metrik. Kann man die Beschreibung etwas geradliniger machen? Direkt auf multiple releases eingehen und dann beschreiben wie diese releases beschaffen sein muessen. das $\max$ vestehe ich nicht. ist das nicht einfach wieder ein $\forall x, \forall y$?}
% \todo[inline,color=red!20]{the original paper defined it using max, meaning that the condition has to be satisfied for the largest percentage across all x and y (not for all percentages). For all percentages would mostly say the same thing I think (but if you have to compute it, the max allows you to just store one value instead of one for each combination)}

\subsubsection{Historical $k$-Anonymity}
\label{metric:historical-k-anonymity}

% This metric extends $k$-anonymity to location privacy.
In location-based services, users include their location in every request they send to the service, which can allow the server to track users.
Thus, historical $k$-anonymity defines \textit{(time, location)} pairs as quasi-identifiers and requires that the adversary cannot link a request to an individual user, but only to $k$ or more users~\cite{bettini_protecting_2005}.
To formalize this requirement, a user's personal history of locations $L$ is a sequence of \textit{(time, location)} pairs, and requests $M$ are (potentially obfuscated) times and locations from which user requests were sent.
$L$ is time-location consistent with a request $m$ if there is an entry in $L$ whose time and location are within the time interval and location area given in $m$.
Historical $k$-anonymity is satisfied if a user's set of requests $M_u$ is location-time consistent with the location history of $k-1$ other users $U$.
\[ priv_\text{HKA} \equiv k \text{, where } \forall u,u' \in U: | L_{u'} \text{ is location-time consistent with } M_u | \geq k \]

\subsubsection{Cluster Similarity}
\label{metric:cluster-similarity}

In smart metering, the time series of differences in load measurements, so-called transitions, can be obfuscated by a load hiding algorithm.
Cluster similarity is based on the idea that an adversary may use clustering to retrieve information about patterns in energy consumption.
To compute cluster similarity, a clustering algorithm is applied to both the original time series of load transitions $\vec{T}_{X^*}$ and the obfuscated time series $\vec{T}_Y$, resulting in two sets of $n$ clusters $C_{X^*}$ and $C_Y$, respectively.
The element-wise subtraction of $C_{X^*}$ from $C_Y$ reveals all transitions that were not placed in the correct cluster.
After normalizing with the number of original load transitions, cluster similarity then indicates the percentage of correctly clustered transitions to show how effectively the original values have been hidden~\cite{kalogridis_privacy_2010}.
% Cluster similarity measures how close original time series $D_T$ and obfuscated time series $D_P$ are~\cite{kalogridis_privacy_2010}.
% To compute cluster similarity, original and obfuscated time series are clustered into clusters $C_T$ and $C_P$, respectively. The ratio of incorrectly classified transitions measures how effectively the original values have been hidden.
\[priv_{\text{CS}} \equiv 1 - \frac{| \forall i: C_{Yi} - C_{{X^*}i} |}{|\vec{T}_{X^*}|} \]

% \todo[inline]{reads good, but explain the normalization?}
% \todo[inline,color=red!20]{we have a number of transitions in the numerator (the incorrectly clustered ones), so to get a percentage we have to divide by another number of transitions (all transitions)
%$D_T$ -> $D_{X*^}$
%$D_P$ -> $D_{Y}$
%$\hat{D_P}$ -> $D_X$
% }

\subsubsection{Coefficient of Determination $R^2$}
\label{metric:coefficient-of-determination}

The coefficient of determination $R^2$ measures how much variability in data is accounted for by a model for the data.
In smart metering, for example, the data is the obfuscated time series of differences in load measurements $\vec{T}_{Y}$ (with $\overbar{\vec{T}_Y}$ indicating the mean value), and the model is a linear regression fitted to these obfuscated load transitions, resulting in predicted values $\vec{T}_X$~\cite{kalogridis_privacy_2010}.
The coefficient of determination compares the error sum of squares $SS_E$ and the regression sum of squares $SS_R$.
\[priv_{\text{R2}} \equiv 1-\frac{SS_E}{SS_R+SS_E} \text{, where } SS_E=\sum_t (\vec{T}_{Y}-\vec{T}_X)^2 \text{ and } SS_R=\sum_t (\vec{T}_X - \overbar{\vec{T}_{Y}}) \]

%\[priv_{\text{R2}} \equiv \frac{1}{N} \sum \frac{(x_i-\bar{x})*(y_i-\bar{y})}{\sigma_x*\sigma_y} ^2 \]

\subsubsection{Normalized Variance}
\label{metric:normalized-variance}

In privacy-preserving data publishing that uses data perturbation, normalized variance is derived from the statistical variance $\sigma^2$ and measures the dispersion between the original data $X^*$ and perturbed data $Y$~\cite{oliveira_privacy_2003}.
However, this metric does not account for the nature of the data and assumes that high variance means better privacy.
\[priv_{\text{VAR}} \equiv \frac{\sigma^2(X^*-Y)}{\sigma^2(X^*)} \]

% \todo[inline]{I like this, but can we discuss some weaknesses? this seems awfully general as it doesn't account for the nature of the data at all. It assumes that high variance means better privacy. Really convenient that my wife doesn't know exactly which room i visited in the red light hotel. 
%Also, $Y$ was never perturbed data, it was mostly prior knowledge or additional observations.
% }

% \subsubsection{Gini Coefficient}
% \todo{David: the more I think about this, the more I believe we could kick it out. It has never been advocated as a *privacy* metric, only as a metric for equality (of path selection in Tor). What do you think?}
% The Gini coefficient is a statistical measure of equality; it has been used to measure the equality of path selection in Tor~\cite{dingledine_tor:_2004}, where more equality (smaller Gini coefficient) corresponds to higher anonymity~\cite{snader_tune-up_2008}.
% \todo{formel vielleicht in 2 zeilen ohne das 'where'? braucht mehr erklaerung im text bzgl variablen}
% \[priv_{\text{GC}} \equiv 1 - \frac{\sum_{i=1}^n p(y_i)(S_{i-1}+S_i)}{S_n} \text{, where } S_i = \sum_{j=1}^i p(y_j)y_j \text{ and } S_0=0 \]
\end{subsection}

%%%%%%%%%%%%%%%%%%%%%%%%%%%%%%%%%%%%%%%%%%%%%%%%%%
% Indistinguishability
%%%%%%%%%%%%%%%%%%%%%%%%%%%%%%%%%%%%%%%%%%%%%%%%%%
\begin{subsection}{Indistinguishability}

Indistinguishability metrics indicate whether the adversary can distinguish between two items of interest (such as recipients of a message, or sensitive attributes in a database).
Many of these metrics are associated with privacy mechanisms that provide formal privacy guarantees.
While many come from the database domain, they have also found application in communication systems, location-based systems, and smart metering.

%\newcolumntype{V}{>{\centering\arraybackslash}p{2.5cm}}
%\begin{table}[htp]
 %\tbl{Indistinguishability Metrics by Domain\label{tab:indistinguishability-metricsbydomain}}{
%%  \resizebox{.7\textwidth}{!}{%
  %\resizebox*{\textwidth}{!}{%
  	%\rowcolors{2}{gray!20}{white}
 %\begin{tabular}{>{\raggedleft\slshape}p{3.3cm}VVVVVV}
 %\upshape\bfseries Metric & \bfseries Communication & \bfseries Databases & \bfseries Location & \bfseries Smart Metering & \bfseries Social Networks & \bfseries Genome Privacy \\
%\midrule
    %Approximate differential privacy & & Dwork 2006 & & & & \\
     %Computational differential privacy &  & Mironov 2009 & & & & \\
     %Cryptographic game & Juels 2009 (RFID) & & & & & \\
     %Differential privacy & & Dwork 2006 & & & & \\
     %Distributed differential privacy &  & & & Shi 2011  & & \\
     %Distributional privacy & & & & Jelasity 2014 & & \\
     %Geo-indistinguishability & & & Andres 2013 & & & \\
     %Joint differential privacy & & Hsu 2014, Kearns 2014 & & & & \\
     %Information privacy & & du Pin Calmon 2012& & & & \\
     %Observational equivalence & Hughes 2004 & & & & & \\
%%      Unconditional / computational privacy & Hermans 2011 (RFID) & & & & & \\
 %\end{tabular}}}
%\end{table}

\begin{table}[htp]
 \caption{Metrics and references in the indistinguishability category and the domains they originated in}
 \label{tab:indistinguishability-metricsbydomain}
%  \resizebox{.7\textwidth}{!}{%
%   \resizebox*{\textwidth}{!}{%
  	\rowcolors{2}{gray!20}{white}
			\resizebox{.85\textwidth}{!}{%
 \begin{tabular}{p{1.2cm}p{6cm}p{2.8cm}p{2cm}}
 \upshape\bfseries Section & \bfseries Metric & \bfseries Original Domain & \bfseries Reference \\
\midrule
 \ref{metric:cryptographic-game} & Cryptographic game & Communication & \cite{juels_defining_2009} \\
 \ref{metric:differential-privacy} & Differential privacy & Databases & \cite{dwork_differential_2006} \\
 \ref{metric:approximate-differential-privacy} & Approximate differential privacy & Databases & \cite{dwork_our_2006} \\
 \ref{metric:distributed-differential-privacy} & Distributed differential privacy & Smart metering & \cite{shi_privacy-preserving_2011} \\
 \ref{metric:distributional-privacy} & Distributional privacy & Smart metering & \cite{jelasity_distributional_2014} \\
 \ref{metric:geo-indistinguishability} & Geo-indistinguishability & Location & \cite{andres_geo-indistinguishability:_2013} \\
 \ref{metric:d-chi-privacy} & d-$\chi$-privacy & Databases & \cite{chatzikokolakis2013broadening} \\
 \ref{metric:joint-differential-privacy} & Joint differential privacy & Databases & \cite{kearns2014mechanism} \\
 \ref{metric:computational-differential-privacy} & Computational differential privacy & Databases & \cite{mironov_computational_2009} \\
 \ref{metric:information-privacy} & Information privacy & Databases & \cite{du_pin_calmon_privacy_2012} \\
 \ref{metric:observational-equivalence} & Observational equivalence & Communication & \cite{hughes_information_2004} \\
 \end{tabular}%
}
\end{table}

% \subsubsection{Privacy Approximation Ratio}
% 
% The privacy approximation ratio is a metric based on communication complexity.
% It expresses the amount of indistinguishability lost in the distributed computation of a function~\cite{ada_hardness_2014}.
% \[priv_{\text{PAR}} \equiv max_{(x,y)} \frac{|R_{x,y}|}{|P_{x,y}|} \]

\subsubsection{Cryptographic Games/Semantic Security}
\label{metric:cryptographic-game}

The classic definition of semantic security can be used to prove privacy properties of cryptographic protocols.
To this end, a challenge-response game, or cryptographic game, is set up in which the adversary selects the inputs for a protocol and is given the output and two alternative outcomes $y_1$ and $y_2$ after the protocol has been executed.
The adversary then has to make an estimate, $x$, indicating whether $y_1$ or $y_2$ is the correct outcome $x^*$.
The adversary has an advantage if they can do this with a probability that is non-negligibly greater than $\frac{1}{2}$, that is, if their probability is better than a random guess~\cite{juels_defining_2009}.

% \todo[inline,color=blue!20]{Reviewer 2: page 17, line 8: This seems like it is just the definition of semantic
% security.}

If the adversary's advantage is smaller than a negligible function $\epsilon(k)$ ($k$ is a security parameter), then the protocol provides {\sffamily\itshape{computational privacy}}, and {\sffamily\itshape{unconditional privacy}} if the advantage is zero~\cite{hermans_new_2011}.

%The security or privacy property under investigation holds if the adversary's advantage is smaller than a negligible function $\epsilon(k)$ ($k$ is a security parameter).
% A negible probability is defined as being smaller than the negligible function defined by the inverse of a polynomial function $p(k)$ ($k$ is a security parameter).
% The adversary has an advantage $Adv$ if the probability that his guess $s_{Adv}$ corresponds to the correct alternative $s$ is non-negligibly greater than $\frac{1}{2}$~\cite{hughes_information_2004}. 
%The outcome of this game is binary: the property either holds or not. 
%Beyond that, no quantification of privacy is provided.
\begin{align*}
priv_{\text{CG}} \equiv \begin{cases}
1 \text{ if } p(x=x^*) \leq \frac{1}{2} + \epsilon(k) \\
0 \text{ otherwise}
\end{cases}\end{align*}

\subsubsection{Differential Privacy}
\label{metric:differential-privacy}

In statistical databases, differential privacy guarantees that any disclosure is equally likely (within a small multiplicative factor $\epsilon$) regardless of whether or not an item is in the database~\cite{dwork_differential_2006}.
For example, the result of a database query should be roughly the same regardless of whether the database contains an individual's record or not.
This guarantee is usually achieved by adding a small amount of random noise to the results of database queries.
Formally, differential privacy is defined using two data sets $D_1$ and $D_2$ that differ in at most a single row, i.e., the Hamming distance between the two data sets is at most $1$.
A privacy mechanism, realized as a randomized function $\mathcal{K}$, operating on these data sets is $\epsilon$-differentially private if for all sets of query responses $S$, the output random variables (query responses) for the two data sets differ by at most $exp(\epsilon)$.
\[priv_{\text{DP}} \equiv \forall S \subseteq Range(\mathcal{K}): p(\mathcal{K}(D_1) \in S) \leq \exp(\epsilon) \cdot p(\mathcal{K}(D_2) \in S) \]

%\todo[inline]{We should make this easier to understand. Start with something like: When a provider releases information about a certain dataset (and not the dataset itself), e.g., in the form of query responses..
%Also mention that $\mathcal{K}$ is not just a function but a privacy mechanism that the provider runs before answering queries. The metric then evaluates if this mechanisms offers protection in the form of differential privacy when..
%It should also be more clear that $S \subseteq Range(K)$
%I'd even go away from the original definition and put it $\forall S \subseteq Range(\mathcal{K}): P(\mathcal{K}(D_1) \in S) \leq e^\epsilon \cdot P(\mathcal{K}(D_2) \in S)$. Problem: We'd have to change the entire section, and I believe that most of the other papers on diff privacy build upon the notation of Dwork. For me it wasn't really clear that S is the variable in this because of the 'missing' $\forall S$.
%Also, in the first line, the small multiplicative factor is $e^\epsilon$ for $\epsilon > 0$. This guarantees that the factor is $> 1$
%Also, it should say differ \textbf{at most} in one row. I think they can be the same and diff privacy would obviously also hold}
%\todo[inline,color=red!20]{$\epsilon$ does NOT have to be positive.}

In the interactive setting, differential privacy provides privacy guarantees if the allowed number of queries is limited~\cite{mcsherry_privacy_2009} (each subsequent query reduces the strength of the privacy guarantee by adding its privacy parameter $\epsilon$).
In the non-interactive setting~\cite{dwork_complexity_2009}, differential privacy provides guarantees only for a certain class of queries~\cite{soria-comas_differential_2013}.
In the local setting, differential privacy can protect properties in addition to identities, e.g. settings in a client software \cite{erlingsson2014rappor} or arbitrary strings \cite{fanti2016building}.
However, the choice of the parameter $\epsilon$ is difficult: values reported in the literature vary from 0.01~\cite{hsu_differential_2014} to 100~\cite{yu_scalable_2014}.
A no-free-lunch theorem shows that differential privacy's guarantees degrade in the case of correlated data, for example when nodes are added to a social network graph~\cite{kifer_no_2011}.

\subsubsection{Approximate Differential Privacy}
\label{metric:approximate-differential-privacy}

Approximate differential privacy relaxes differential privacy by allowing an additional small additive constant $\delta$~\cite{dwork_our_2006}.
Approximate differential privacy weakens the privacy guarantee, but allows data releases/query responses with higher utility, e.g. by allowing a wider range of query types~\cite{blum_learning_2013}, or by reducing the sample complexity of private learning \cite{beimel2013private}.
The parameter $\delta$ should be chosen to be smaller than the inverse of any polynomial in the size of the database $\|D\|$ \cite{dwork_algorithmic_2014}.
In particular, $\delta \approx \frac{1}{\|D\|}$ would allow to publish complete records of a small number of individuals, while still meeting the differential privacy requirement.
\citeANP{abadi_deep_2016}~\cite{abadi_deep_2016}, for example, use $\delta \in [10^{-5},1]$.
\[priv_{\text{ADP}} \equiv \forall S \subseteq Range(\mathcal{K}): p(\mathcal{K}(D_1) \in S) \leq \exp(\epsilon) \cdot p(\mathcal{K}(D_2) \in S) + \delta \]

%\todo[inline]{A bit more explanation. It was hard to find $\epsilon$, how would a second parameter make this easier? What is suggest by Dwork?}
%\todo[inline,color=red!20]{relaxes the privacy guarantee and thus allows releases/query responses with higher utility.
%abadi: $\delta \in [10^{-5},1]$, there is a trade-off between epsilon and delta, such that a given mechanism can satisfy diff.priv. for an unlimited number of (epsilon,delta) combinations.
%dwork/roth: values of $\delta$ that are less than the inverse of any polynomial in the size of the database. In particular, values of $\delta$ on the order of $1/||x||_1$ are very dangerous: they permit "preserving privacy" by publishing the complete records of a small number of database participants ($||x||_1$ is the size of the database (number of records))
%}

\subsubsection{Distributed Differential Privacy}
\label{metric:distributed-differential-privacy}

Distributed differential privacy extends approximate differential privacy to a setting where distributed entities contribute data to a central data aggregator~\cite{shi_privacy-preserving_2011}.
The data aggregator can be untrusted and possibly colludes with a subset of the participants.
This extension can be useful in smart metering, where users may not trust the energy provider (who acts as data aggregator).
Each user applies randomness to their own values before sending them to the data aggregator.
Distributed differential privacy allows a subset of users $\widehat{U} \subset U$ to collude with the aggregator, while still providing privacy guarantees for the remaining honest users.
To achieve this, distributed differential privacy ensures that the privacy mechanism's probability is taken over the randomness provided by honest users, or in other words, the probability is conditional on the randomness $r_{\widehat{U}}$ provided by compromised users.
%In distributed differential privacy, the output random variables (query responses) are conditioned on the randomness $r_K$ from compromised participants.
%This ensures that differential privacy is achieved using only randomness provided by honest participants.
\[priv_{\text{DDP}} \equiv \forall S \subseteq Range(\mathcal{K}), \forall \widehat{U} \subset U: p(\mathcal{K}(D_1) \in S|r_{\widehat{U}}) \leq \exp(\epsilon) \cdot p(\mathcal{K}(D_2) \in S|r_{\widehat{U}}) + \delta \]

%\todo[inline]{Is there a clearer way to say 'conditioned on the randomness $r_K$. What does that mean?}

\subsubsection{Distributional Privacy}
\label{metric:distributional-privacy}

Distributional privacy extends differential privacy to a setting in which the data sets themselves do not need to be protected, but instead the parameters governing the generation of data.
In a smart metering scenario, for example, these parameters can be user habits, behavioral patterns, or sets of appliances in a home~\cite{jelasity_distributional_2014}.
Distributional privacy assumes a distributed setting in which smart meters apply noise to their local data, limiting the energy provider to querying this distributed database.
Formally, distributional privacy uses two parameter sets $\theta_1$ and $\theta_2$ which govern the creation of two data sets and differ in at most one element.
The privacy mechanism $\mathcal{K}$ is distributionally $\epsilon$-differentially private if 
the probability that query response $\mathcal{K}_j$ is generated is roughly the same, regardless of whether the underlying parameter set is $\theta_1$ or $\theta_2$.
%the series of query responses $\mathcal{K}_j$ for the two parameter sets differs by at most $exp(\epsilon)$.
\[priv_{\text{DSP}} \equiv p(\theta_1|\mathcal{K}_j) \leq \exp(\epsilon) \cdot p(\theta_2|\mathcal{K}_j) \]

%\todo[inline]{Diff privacy doesn't protect the datasets, but the responses. Notation should have the same order than before. We should back-reference to $\mathcal{K}$ and not write as if it was something new. I can't quite put my finger on the notation $P(\theta | \mathcal{K})$. probably need to read the original paper}
%\todo[inline,color=red!20]{not true. the non-interactive setting mentioned above encompasses the setting where noisy data are being released, and thus diff.priv protects the data release, not just the responses. Maybe need to make this clearer.}

% \subsubsection{Positive Membership Privacy}
% \todo{kann raus, wuesste auch nicht wo man das mitreinnehmen koennte um einen erzaehlfaden aurecht zu erhalten}
% Positive membership privacy states that an adversary cannot significantly improve his confidence that an entity is in the data set, specifically that his belief can only increase by a constant factor $\gamma$.
% Many variants of differential privacy can be derived from the formulation of positive membership privacy~\cite{li_membership_2013}.
% \[priv_{\text{PMP}} \equiv Pr[r|Y] \leq \gamma Pr[r] \wedge Pr[r|Y] \geq \frac{Pr[r]}{\gamma} \]

\subsubsection{Geo-Indistinguishability}
\label{metric:geo-indistinguishability}

Geo-indistinguishability extends differential privacy to location privacy scenarios.
The idea is to apply two-dimensional (planar) noise to the user's geographical location so that the differential privacy requirements are met, ensuring that the user enjoys $\epsilon d$-differential privacy within any distance $d > 0$.
Importantly, this definition implies that the user's protection level depends on the distance $d$.
This could mean, for example, that a location-based service provider would be able to distinguish which city the user is in, but not the location within the city.
To achieve geo-indistinguishability, the privacy mechanism $\mathcal{K}$ generates randomized location observations so that the distance between any two locations $d(l_1,l_2)$ is roughly the same as the distance between the distributions of randomized location observations $d_\mathcal{P}(\mathcal{K}(y_1),\mathcal{K}(y_2))$ \cite{andres_geo-indistinguishability:_2013}. 
%any two locations $x_1$, $x_2$ within a given Euclidean distance $d(\cdotp,\cdotp)$ produce observations with distributions that are similar within a small multiplicative factor $\epsilon$~\cite{andres_geo-indistinguishability:_2013}.

\[priv_{\text{GI}} \equiv d_\mathcal{P}(\mathcal{K}(y_1),\mathcal{K}(y_2)) \leq \epsilon d(l_1,l_2) \]

%\todo[inline]{I cannot find this definition in the original paper, nor do i understand it. We should mention this sentence from the paper "`Thus, $\epsilon$-geo-indistinguishability can be though [sic] as differential privacy under the Euclidean metric. I recommend we use something like this notation (based on the paper plus my changes):
%
%$P(x_1 \in S) \leq e^{\epsilon d(x_1,x_2)} \cdot P(x_2 \in S)$ $\forall S \subseteq Range(\mathcal{K})$
%
%$Range(\mathcal{K})$ now includes spatial points instead of arbitrary data.
%The paper uses $P(S|x)$ instead of $P(x \in S)$, it should be the same, but i can't put my finger on it.}

\subsubsection{d-$\chi$-Privacy}
\label{metric:d-chi-privacy}

d-$\chi$-privacy is a generalization of differential privacy that uses distinguishability metrics $d_{\chi}$ to characterize the distance between two datasets instead of the Hamming distance used in standard differential privacy \cite{chatzikokolakis2013broadening}.
In standard differential privacy, the distinguishability level between two datasets of distance 1 is $\epsilon$. In d-$\chi$-privacy, the distinguishability level between datasets of arbitrary distance is given by the distinguishability metric $d_\chi$.
\[ priv_{\text{DX}} \equiv d_\mathcal{P}(\mathcal{K}(D_1), \mathcal{K}(D_2)) \leq d_\chi(D_1,D_2) \]

Depending on the choice of metric, d-$\chi$-privacy can represent different notions of privacy.
For example, the Euclidean distance is suitable for location privacy and results in geo-indistinguishability described above.
In smart metering, the maximum metric (or Chebyshev distance) can be used to distort the accuracy of meter readings while leaving general trends intact.

d-$\chi$-privacy can also be used to construct \textit{elastic} metrics that adapt to the characteristics of the application domain.
For example, in location privacy, the point-of-interest density may influence the level of privacy we expect from geo-indistinguishability: in a rural area with few points of interest, we may need a larger radius compared to an urban area to achieve the same level of privacy \cite{chatzikokolakis2015constructing}.

\subsubsection{Joint Differential Privacy}
\label{metric:joint-differential-privacy}

The idea of joint differential privacy \cite{kearns2014mechanism} is that an individual's private data can be disclosed to the individual him/herself, but not to other individuals.
Applied to a game theoretic problem and focusing on player $u$, for example, joint differential privacy requires that the joint distribution on outputs given to other players, i.e. $\mathcal{K}(D)_{-u}$, is differentially private in player $u$'s input \cite{hsu2014private}.

% joint diffpriv in matching and allocation problems \cite{hsu2014private} (clear definition)
% joint diffpriv in game theory \cite{kearns2014mechanism} (less clear definition)
% from dwork/roth book: ``the solution concept of Joint Differential Privacy requires that for every player i, the joint distribution on messages sent to other players j!=i be differentially private in i's report.''

\[priv_{\text{JDP}} \equiv \forall S \subseteq Range(\mathcal{K}): p(\mathcal{K}(D_1)_{-u} \in S) \leq \exp(\epsilon) \cdot p(\mathcal{K}(D_2)_{-u} \in S) + \delta \]

%\todo[inline]{Not a fan of "`it is ok.."'. $i$ is defined twice. We have $i$ players, and player $i$'s input. I don't understand what $\mathcal{K}_{-i}$ does.} 

% \todo[inline]{What's with N. Li, W. Qardaji, and D. Su, “On Sampling, Anonymization, and
% Differential Privacy or, K-anonymization Meets Differential Privacy,”
% in Proceedings of the 7th ACM Symposium on Information, Computer
% and Communications Security, ser. ASIACCS ’12. ACM, 2012, pp.
% 32–33. It's more of a PET (composable differential privacy), but so is differential privacy. make something $3\epsilon$ by applying the method three times.}
% \todo[inline,color=red!20]{they show how k-anonymity can be done to guarantee differential privacy, but they don't really introduce a new metric?
% I've included it in the k-anonymity description}

\subsubsection{Computational Differential Privacy}
\label{metric:computational-differential-privacy}

Computational differential privacy replaces the unrestricted adversary used in differential privacy with a computationally bounded adversary.
By using a weaker adversary model, computationally differentially private mechanisms can give more accurate query responses.
Informally, computational differential privacy requires that the outputs produced by the privacy mechanism ``look'' differentially private to every adversary.
Depending on how ``look'' is formalized, the definitions of computational differential privacy can be different~\cite{mironov_computational_2009}.
% ; they all depend on a security parameter $\kappa$
For example, a definition based on indistinguishability replaces the unrestricted adversary with a computationally bounded adversary, and 
% \[ priv_\text{IND\_CDP} \equiv p[A_{\kappa}(\mathcal{K}_{\kappa}(D_1)) = 1] \leq \exp(\epsilon_{\kappa}) \cdot p[A_{\kappa}(\mathcal{K}_{\kappa}(D_2)) = 1] + \text{negl}(\kappa) \]
% 
% \todo[inline]{no idea what's going on here. The Adversary $A_k$ (no idea why index $k$) suddenly becomes a function that can be $1$. The $\text{negl}(\kappa)$ is similar to the $\delta$ we used before all the time. maybe we can use $\delta_\kappa$. The last sentence in the description repeats the first one.}
% 
a definition based on simulation requires that the outputs from randomized functions are computationally indistinguishable from the outputs from $\epsilon$-differentially private mechanisms $\mathcal{K}$.
% \[ priv_\text{SIM\_CDP} \equiv |p[A_{\kappa}(\mathcal{K}_{\kappa}(D)) = 1] - p[A_{\kappa}(R_{\kappa}(D)) = 1]| \leq \text{negl}(\kappa) \]

% \todo[inline]{same comment from above applies here}
% \todo[inline,color=red!20]{we could leave out the two equations in this section. if anyone is interested in CDP, they need to go to mironov's paper anyway (unless they are cryptographers, they would probably understand the notation), and the two sentences that explain the two versions should be good enough for our survey}

\subsubsection{Information Privacy}
\label{metric:information-privacy}

Information privacy captures the notion that the prior and posterior probabilities of inferring sensitive data $x^*$ do not change significantly, given query outputs $y$.
$\epsilon$-information privacy implies $2\epsilon$-differential privacy, but additionally bounds the maximum information leakage (Section~\ref{metric:maximum-information-leakage}) to at most $\epsilon / \ln 2$ bits~\cite{du_pin_calmon_privacy_2012}.
% Formally, a privacy preserving mapping from the user's data $s$ to a privacy-preserving output $u$, $p_{S|U}(\cdotp|\cdotp)$, provides $\epsilon$-information privacy if for all sensitive values $s$, the ratio of posterior probability $p_{S|U}(s|u)$ to prior probability $p_S(s)$ is very close to $1$.
Formally, a privacy-preserving query output $y$ provides $\epsilon$-information privacy if for all sensitive values $x^*$, the ratio of posterior probability $p(x^*|y)$ to prior probability $p(x^*)$ is very close to $1$.

\[priv_{\text{IP}} \equiv \exp(-\epsilon) \leq \frac{p(x^*|y)}{p(x^*)} \leq \exp(\epsilon) \text{, } \forall y \in Y: p(y) > 0 \]

In the context of wireless sensor networks, information privacy indicates that event sources cannot be observed by an adversary.
Event source unobservability requires that for all possible observations of events in a system, the adversary's prior probability equals the posterior~\cite{yang_towards_2008}.

%\todo[inline]{
%%Query outputs have never been $u$. Notation is different from the rest in this section (especially $p_{S|U}(s|u)$). 
%what is the probability space? The query output $U$ should again be $Range(\mathcal{K})$ something, and $u$ is actually $S \subseteq Range(\mathcal{K})$ if I understood correctly, but i don't know how to express s here.)
%
%\[priv_{\text{IP}} \equiv e^{-\epsilon} \cdot {p(s)} \leq p(s|S) \leq e^\epsilon \cdot {p(s)}\]
%
%This is also similar to other metrics we have discussed before}
%\todo[inline,color=red!20]{I wouldn't do this, because the interesting thing about this metric is that they arrive at a diff.priv guarantee by relating posterior and prior probabilities}
%\todo[inline, color=green]{This is up to you. If you leave it, make sure it doesn't conflict with the notation}

\subsubsection{Observational Equivalence}
\label{metric:observational-equivalence}

Observational equivalence is a formal property that states that the adversary cannot distinguish between two situations, for example which user sent a given message~\cite{hughes_information_2004}.
To use this metric, privacy protocols are modeled using a formal process calculus such as the applied $\pi$-calculus.\footnote{A process calculus is a formal method to model and reason about concurrent systems.
The applied $\pi$-calculus is a process calculus that includes cryptographic primitives and has thus been used extensively to check properties of cryptographic protocols.
To verify privacy properties of a protocol, the protocol is modeled in the applied $\pi$-calculus, and an automated tool such as ProVerif can verify whether the privacy properties hold for all possible executions of the protocol.}
% In the applied $\pi$-calculus, privacy properties can be defined and verified using 
Observational equivalence is fulfilled if the observable outputs from protocol runs in two situations are equivalent.
This has been used, e.g., in voting privacy~\cite{delaune_verifying_2009}, mobile telephony~\cite{arapinis_new_2012} and webs of trust~\cite{backes_anonymous_2010}.

% \todo[inline,color=blue!20]{Reviewer 2: page 18, line 49: Is there a more intuitive explanation that does refer to a "process calculus" and a "pi-calculus"?}
% \todo[inline,color=red!20]{protocol gets written in process calculus; tool support for crypto primitives; tool (e.g., ProVerif) can then verify whether privacy properties such as anonymity are fulfilled or not; I really don't know how to say this in a more intuitive way. maybe mention of the ProVerif tool would help? I think sample process calculus notation would just confuse everyone even more.}
% \subsubsection{Event Source Unobservability}
% \todo{ist das nicht Information Privacy oder Increase in Adversary’s Belief oder Privacy Breach Level. Klingt fuer mich alles ziemlich gleich}
% This metric states that event sources in a wireless sensor network are unobservable by an adversary if for all possible observations of events in a system, the adversary's prior probability equals the posterior~\cite{yang_towards_2008}.
% \[priv_{\text{ESU}} \equiv \text{true, if } \forall Y \forall X: P(X) = P(X|Y) \]
 
\end{subsection}

%%%%%%%%%%%%%%%%%%%%%%%%%%%%%%%%%%%%%%%%%%%%%%%%%%
% Adversary's Success Probability
%%%%%%%%%%%%%%%%%%%%%%%%%%%%%%%%%%%%%%%%%%%%%%%%%%
\begin{subsection}{Adversary's Success Probability}

Metrics based on the adversary's success probability can be seen as general-purpose metrics that subsume many other aspects of privacy.
They depend strongly on the adversary model (see Section \ref{sec:adversaries}) and on how exactly success is defined.
Even though the metrics in this section mostly originate from the communication and database domains, they can be applied in every domain and setting where an adversary can be defined.
In addition to the adversary's success (cases where the adversary successfully identifies the correct individual, or the true positive rate), metrics in these section should also consider the false positive and false negative rates, i.e. cases where the adversary identifies an incorrect individual, and cases where the adversary fails to identify the correct individual.

%\newcolumntype{V}{>{\centering\arraybackslash}p{2.5cm}}
%\begin{table}[htp]
 %\tbl{Success Metrics by Domain\label{tab:success-metricsbydomain}}{
%%  \resizebox{.7\textwidth}{!}{%
  %\resizebox*{\textwidth}{!}{%
  	%\rowcolors{2}{gray!20}{white}
 %\begin{tabular}{>{\raggedleft\slshape}p{3.3cm}VVVVVV}
 %\upshape\bfseries Metric & \bfseries Communication & \bfseries Databases & \bfseries Location & \bfseries Smart Metering & \bfseries Social Networks & \bfseries Genome Privacy \\
%\midrule
     %Adversary's success rate & Wright 2003 & Narayanan 2008 & & & & \\
     %(d,$\gamma$)-privacy & & Rastogi 2007& & & &  \\
     %Degrees of Anonymity & Reiter 1998 & & & & & \\
     %$\delta$-presence & & Nergiz 2007 & & & &  \\
     %Hiding property & Toth 2004 & & & & & \\
 %\end{tabular}}}
%\end{table}

\begin{table}[htp]
 \caption{Metrics and references in the success category and the domains they originated in}
 \label{tab:success-metricsbydomain}
%  \resizebox{.7\textwidth}{!}{%
%   \resizebox*{\textwidth}{!}{%
  	\rowcolors{2}{gray!20}{white}
			\resizebox{.85\textwidth}{!}{%
 \begin{tabular}{p{1.2cm}p{6cm}p{2.8cm}p{2cm}}
 \upshape\bfseries Section & \bfseries Metric & \bfseries Original Domain & \bfseries Reference \\
\midrule
 \ref{metric:adversarys-success-rate} & Adversary's success rate & Communication & \cite{wright_defending_2003} \\
 \ref{metric:degrees-of-anonymity} & Degrees of anonymity & Communication & \cite{reiter_crowds:_1998} \\
 \ref{metric:privacy-breach-level} & Privacy breach level & Databases & \cite{evfimievski_privacy_2004} \\
 \ref{metric:d-gamma-privacy} & (d,$\gamma$)-privacy & Databases & \cite{rastogi_boundary_2007} \\
 \ref{metric:delta-presence} & $\delta$-presence & Databases & \cite{nergiz_hiding_2007} \\
 \ref{metric:hiding-property} & Hiding property & Communication & \cite{toth_measuring_2004} \\
 \end{tabular}%
}
\end{table}

\subsubsection{Adversary's Success Rate}
\label{metric:adversarys-success-rate}

This metric measures the probability that the adversary is successful, or the percentage of successes in a large number of attempts~\cite{wright_defending_2003}.
Depending on the application scenario, success can be defined in different ways: in databases, for example, the adversary is successful when he can find a record $s'$ that is similar to the target record $s$ with a similarity threshold of $\tau_s$ and an error threshold of $\tau_e$~\cite{narayanan_robust_2008}.
\[priv_{\text{SRD}} \equiv p(Sim(s,s') \geq \tau_s) \geq \tau_e \]

% \todo[inline]{can we get rid of the sim and introduce a generic distance function $P(d(x,x') \leq \theta) \geq \omega$}
% \todo[inline,color=red!20]{not sure. similarity seems to be a somewhat fuzzier notion than distance. also, we already use Pop and Vis above, so Sim doesn't stand out.}

\label{metric:probability-of-path-compromise}
In communication systems, the adversary is successful when he can identify the sender of a message~\cite{shmatikov_probabilistic_2002},
or when he can compromise a communication path with a given amount of resources (e.g., number of nodes and bandwidth)~\cite{murdoch_metrics_2008}.
% or the probability distribution on the percentage of adversary successes for a given target user and time period (e.g., for path compromises in Tor~\cite{johnson_users_2013}).

%\subsubsection{Number of Successes in a Time Period}
%
%For a given target user and a given time period, this metric gives the probability distribution on the number or percentage of adversary successes (e.g., path compromises in Tor) ~\cite{johnson_users_2013}.

\subsubsection{Degrees of Anonymity}
\label{metric:degrees-of-anonymity}

%this metric can be shortened if we need space as it lists everything twice

\citeANP{reiter_crowds:_1998}~\cite{reiter_crowds:_1998} define six degrees of anonymity for communication systems, which depend on how likely the adversary's success is.
In communication systems, for example, $p(x)$ indicates the adversary's probability to identify the sender (or receiver) of a message.
`Absolute privacy' states that the communication produced no observable effects.
`Beyond suspicion' indicates that the sender is equally as likely as all other potential senders.
`Probable innocence' means that the sender is as likely as not to be the originator of a message.
`Possible innocence' states that there is a nontrivial probability $\delta$ that the sender is someone else.
`Exposed' indicates that the adversary's probability is above a threshold $\tau$.
Lastly, `provably exposed' says that the adversary can prove who the sender is.
\[priv_{\text{DOA}} \equiv 
 \begin{cases}
  \text{absolute privacy,} &\text{if } p(x) = 0 \\
  \text{beyond suspicion,} &\text{if } p(x) = \frac{1}{|X|} \\ 
  \text{probable innocence,} &\text{if } p(x) \leq 0.5 \\
  \text{possible innocence,} &\text{if } p(x) < 1-\delta \\
  \text{exposed,} &\text{if } p(x) \geq \tau \\
  \text{provably exposed,} &\text{if } p(x) = 1
 \end{cases}
\]
However, it has been noted that the degree of anonymity does not reflect the adversary's real probability of success, because it ignores the cardinality of the anonymity set~\cite{murdoch_quantifying_2014}.

% \todo[inline]{what is between 0 and min? What exactly is min and max? what is Pr? Just "`probability"' is a bit tricky.}
% \todo[inline,color=red!20]{min actually reads like equiprobability. }

User-specified innocence~\cite{chen_measuring_2012} merges two degrees of anonymity, probable and possible innocence, by introducing a parameter $\alpha$ that represents the probability of the most likely user in the anonymity set.
% \todo[inline]{this is basically a similar concept as Min Entropy, right?}
% \todo[inline,color=red!20]{except that we don't have a logarithm here}

%\todo[inline,color=blue!20]{Reviewer 2: page 19, lines 21-43: It may be good to distinguish between type I/type
%II error (a.k.a. soundness/completeness, significance/power, false
%positive/false negative).}
%\todo[inline, color=green]{I don't think we should. This will be a lot of text and explanation and would be inconsistent in the level of detail compared to other metrics. These are more general terms that are not specific to the discussed metric. If you can do it in 1-2 sentence, by all means, lets do it, but I don't think its possible}

\subsubsection{Privacy Breach Level}
\label{metric:privacy-breach-level}

A privacy breach occurs if the posterior probability of a property, given its prior probability, is higher than the threshold $\tau$. In a data mining scenario, for example, a server (e.g., a recommender system) mines association rules between items (e.g., books) based on their occurrence in user transactions, and users can randomize their transactions to hide which user has which items.
The privacy breach level then uses the probability that an item $s$ is contained in a transaction $\mathcal{T}_{x^*}$, given the probability that the item is part of an item set $S$, which is a subset of the randomized transaction $\mathcal{T}_y$ that was transmitted to the server~\cite{evfimievski_privacy_2004}.

\[priv_{\text{PBL}} \equiv \tau \text{, where } \exists s \in S \text{ so that } p(s \in \mathcal{T}_{x^*} | S \subseteq \mathcal{T}_y) \geq \tau \]
The privacy breach level can also measure privacy in networking, where the metric refers to the conditional probability that a node generated a message with specific characteristics, given that another node received such a message~\cite{seys_arm:_2009}.

\subsubsection{$(d,\gamma)$-Privacy}
\label{metric:d-gamma-privacy}

An extension of the privacy breach level is $d,\gamma$-privacy, which introduces additional bounds on the prior and posterior probabilities ($d$ and $\gamma$, respectively) so that the ratio between posterior and prior probability cannot drop by more than a factor of $d/\gamma$~\cite{rastogi_boundary_2007}.
This metric is similar to Information Privacy (Section \ref{metric:information-privacy}), but uses more detailed bounds.
\[priv_{\text{DG}} \equiv \frac{d}{\gamma} \leq \frac{p(s|S)}{p(s)} \text{, where } p(s) \leq d \text{ and } p(s|S) \leq \gamma \]

% \todo[inline]{this one of the few metrics that do not have their own subsubsection but are part of another metric. why? Also this reads like some other metrics we already discussed. Information privacy is similar, it bounds it to $\pm e^\epsilon$ and also similar to increase in adversary's belief. The latter could be moved to this section actually.}
% \todo[inline,color=red!20]{I don't remember why no subsubsection, we could add it again.
% I agree that it's very similar to information privacy, but the bounds here are a bit more detailed. increase in adv. belief uses a subtraction instead of ratio.
% maybe mention the similarities here instead of moving metrics around?

% we don't really need the post and prior suffixes, do we?
% }

\subsubsection{$\delta$-Presence}
\label{metric:delta-presence}

%In the database domain, $\delta$-presence bounds the adversary's probability of inferring that an individual $y$ is in a database $D_{X^*}$.
%  between $\delta_{min}$ and $\delta_{max}$
In databases, $\delta$-presence bounds the adversary's probability of inferring that an individual $u$ is part of some published data $D_Y$, assuming that the adversary has access to external database tables $D_Z$ so that all individuals in $D_Y$ are also in $D_Z$~\cite{nergiz_hiding_2007}.
\[priv_{\text{DLP}} \equiv (\delta_{min}, \delta_{max}) \text{, where }  \forall u \in U_Z : \delta_{min} \leq p(u \in U_Y) \leq \delta_{max} \]

The adversary's probability can be based on comparing the number of users in the data table (e.g., $p(u \in U_Y) = \frac{|U_Y|}{|U_Z|}$), or on elimating rows based on other attributes.
However, this model assumes that the adversary and the data publisher who assesses whether $\delta$-presence is satisfied have access to the same external tables.
This assumption may not hold in practice~\cite{fung_privacy-preserving_2010}.
% \todo[inline,color=red!20]{this sounds very similar to the privacy breach level, just with two bounds instead of one}
% \todo[inline,color=blue!20]{Reviewer 2: page 20, line 15: What is the probability space?}

% \todo[inline]{I do not understand why the individiual is element of the Prior Knowledge. How can prior knowledge be modelled here?}
% \todo[inline,color=red!20]{for all individuals that are part of the prior knowledge (external database tables), we look at the probability that this individual is part of the published (sanitized) database}

\subsubsection{Hiding Property}
\label{metric:hiding-property}

%\todo[inline, color=blue!20]{The relationship of the measures with statistical disclosure and
%anonymity can be better defined. Many of the measure follow the classic
%statistical disclosure classification of identity hiding (anonymity) or
%property hiding. In many of the measures, there is some assumption of
%anonymity or property hiding, but this assumption is not made explicit
%in the text.}

In communication systems, the source (or destination) hiding property measures the adversary's maximum probability $p(x_{(m,u)})$ for any user $u$ to be sender (or recipient) of a given message $m$.
The source (or destination) is assumed to be hidden if this probability is smaller than a threshold $\tau$~\cite{toth_measuring_2004}.
\[priv_{\text{HP}} \equiv \tau \text{, where } \forall m, \forall u: p(x_{(m,u)}) \leq \tau \]

% \todo[inline]{notation: we've never used $p_{subscript}$ before. Needs letters $m$ and $u$ in the text. Could this be $P(u|m)$?}
% \todo[inline,color=red!20]{it's a joint probability, not a conditional}

\end{subsection}

%%%%%%%%%%%%%%%%%%%%%%%%%%%%%%%%%%%%%%%%%%%%%%%%%%
% Error
%%%%%%%%%%%%%%%%%%%%%%%%%%%%%%%%%%%%%%%%%%%%%%%%%%
\begin{subsection}{Error}

Error-based metrics quantify the error an adversary makes in creating his estimate.
Because information about the true outcome is needed to compute these metrics, they cannot be computed by the adversary.
Similar to the adversary's success probability category, metrics in the error category are applicable to all domains.

\begin{table}[htp]
 \caption{Metrics and references in the error category and the domains they originated in}
 \label{tab:error-metricsbydomain}
%  \resizebox{.7\textwidth}{!}{%
%   \resizebox*{\textwidth}{!}{%
  	\rowcolors{2}{gray!20}{white}
			\resizebox{.85\textwidth}{!}{%
 \begin{tabular}{p{1.2cm}p{6cm}p{2.8cm}p{2cm}}
 \upshape\bfseries Section & \bfseries Metric & \bfseries Original Domain & \bfseries Reference \\
\midrule
    \ref{metric:expected-estimation-error} & Adversary's expected estimation error & Location & \cite{shokri_quantifying_2011}\\
%     \ref{metric:expected-estimation-error} & Adversary's Expected Estimation Error & Genome Privacy & \cite{humbert_addressing_2013} \\
    \ref{metric:expectation-of-distance-error} & Expectation of distance error & Location & \cite{hoh_protecting_2005} \\
    \ref{metric:mean-squared-error} & Mean squared error & Communication & \cite{oya_dummies_2014} \\
%     \ref{metric:percentage-incorrectly-classified} & Percentage incorrectly classified & Smart Metering & \cite{lisovich_inferring_2010} \\
    \ref{metric:percentage-incorrectly-classified} & Percentage incorrectly classified & Social networks & \cite{narayanan_-anonymizing_2009} \\
    \ref{metric:health-privacy} & Health privacy & Genome privacy & \cite{humbert_addressing_2013} \\
 \end{tabular}%
}
\end{table}

\subsubsection{Adversary's Expected Estimation Error}
\label{metric:expected-estimation-error}

In location privacy, the adversary's expected estimation error measures the adversary's correctness by computing the expected distance between the true location $x^*$ and the estimated location $x$ using a distance metric $d()$, for example the Euclidean distance or a metric that yields either $0$ or $1$ (in this case, the metric reduces to the adversary's probability of error).
The expectation is computed over the posterior probability of the adversary's estimated locations $x$ based on his observations $y$~\cite{shokri_quantifying_2011}.
%Privacy is defined as the adversary's incorrectness.
\[priv_{\text{AEE}} \equiv \sum_{x \in X}{p(x|y)d(x,x^*)} \]
The metric can also be used in other domains if an appropriate distance metric is available.
In genomic privacy, for example, the distance metric depends on how the values of genetic variations are encoded~\cite{humbert_addressing_2013}.

\subsubsection{Expectation of Distance Error}
\label{metric:expectation-of-distance-error}

Similar to the adversary's expected estimation error, the expectation of distance error measures the expected distance error of an adversary, but over multiple timesteps $T$ and location assignment hypotheses $\mathcal{H}$~\cite{hoh_protecting_2005}.
Each hypothesis $h$ assigns a user to a location with probability $p_{h,t}(x)$, and the distance $d_{h,t}(x,x^*)$ indicates the distance between the correct user location and the location in hypothesis $h$ at timestep $t$.
\[priv_{\text{EDE}} \equiv \frac{1}{|U|T} \sum_{t \in T} \sum_{h \in \mathcal{H}} p_{h,t}(x) d_{h,t}(x,x^*) \]

\subsubsection{Mean Squared Error}
\label{metric:mean-squared-error}
In statistical parameter estimations, a common goal is to minimize the mean squared error.
As a privacy metric, the mean squared error describes the error between observations $y$ by the adversary and the true outcome $x^*$, for example the error in the assignment of communication relationships~\cite{oya_dummies_2014}, or the error in reconstructing user data in participatory sensing~\cite{ganti_poolview:_2008}.
\[priv_{\text{MSE}} \equiv \frac{1}{|X^*|}\sum_{x^* \in X^*}{ \|x^*-y\|^2} \]

%\todo[inline]{observations could be $z \in Z$ again. Notation is different again. Should have a set of observations and a set real states. Given $n$ observations it is then $\frac{1}{n} \sum_{i}^{n}{ \|x_i-o_i\|^2}$
%or just 
%
%\[ \frac{1}{|Z|}\sum_{x \in X, z \in Z}{ \|x-z\|^2} \]
%
%is this clear, or can this imply that subtract every z from every possible x? not sure
%}

\subsubsection{Percentage Incorrectly Classified}
\label{metric:percentage-incorrectly-classified}

This metric measures the percentage of incorrectly classified users or events $U'$ within the set of all users or events $U$, for example users that were incorrectly de-anonymized by the adversary~\cite{narayanan_-anonymizing_2009}, or events that were incorrectly classified in a smart metering scenario~\cite{lisovich_inferring_2010}.
\[priv_\text{PIC} \equiv \frac{U'}{U} \]

\subsubsection{Health Privacy}
\label{metric:health-privacy}

Health privacy is a metric from genome privacy that captures privacy with regard to a specific disease~\cite{humbert_addressing_2013}.
The metric assumes that a set of genetic variations $V$ contributes to the disease risk, where each variation contributes to a varying extent $\omega_v$.
The better an adversary can predict the individual genetic variations, the better he is able to infer the user's disease risk.
The metric is computed as the weighted, normalized sum over a base metric $B_v$ which measures the privacy of each genetic variation.
Base metrics can be normalized entropy (Section \ref{metric:normalized-entropy}), normalized mutual information (Section \ref{metric:normalized-mutual-information}), or expected estimation error (Section \ref{metric:expected-estimation-error})~\cite{humbert_addressing_2013}.
Depending on the base metric, health privacy measures a different kind of output; in the case of expected estimation error, health privacy measures the adversary's weighted average error.
\[priv_{\text{HLP}} \equiv \frac{1}{\sum_{v \in V} \omega_v} \sum_{v \in V} \omega_v B_v \]

%\todo[inline]{Example would help "`It assumes that the better an adversary can guess the single genetic variations (as indicated by the base metric), the better he is able to conclude a disease risk for the user'. (not sure if that actually makes sense. Variations are $S$, not sure whether this is inline with $S$ is in the rest of the paper. Could be $V$}

\end{subsection}

%%%%%%%%%%%%%%%%%%%%%%%%%%%%%%%%%%%%%%%%%%%%%%%%%%
% 
%%%%%%%%%%%%%%%%%%%%%%%%%%%%%%%%%%%%%%%%%%%%%%%%%%
\begin{subsection}{Time}

Time-based metrics focus on time as a resource that the adversary needs to spend to compromise users' privacy.
Some time-based metrics measure the time until the adversary succeeds, assuming \acp{PET} will fail eventually, while others measure the time until the adversary's confusion, assuming \acp{PET} will succeed eventually.
These metrics originate (and are usually applied) in the communication and location domains, but have also found application in smart metering.

%
%\newcolumntype{V}{>{\centering\arraybackslash}p{2.5cm}}
%\begin{table}[htp]
 %\tbl{Time Metrics by Domain\label{tab:time-metricsbydomain}}{
%%  \resizebox{.7\textwidth}{!}{%
  %\resizebox*{\textwidth}{!}{%
  	%\rowcolors{2}{gray!20}{white}
 %\begin{tabular}{>{\raggedleft\slshape}p{3.3cm}VVVVVV}
 %\upshape\bfseries Metric & \bfseries Communication & \bfseries Databases & \bfseries Location & \bfseries Smart Metering & \bfseries Social Networks & \bfseries Genome Privacy \\
%\midrule
     %Maximum tracking time & & & Sampigethaya 2005 & Lisovich 2010 & & \\
     %Mean time to confusion & & & Hoh 2007 & & & \\
     %Time until adversary's success & Wright 2002, Agrawal 2003 & & & & & \\
 %\end{tabular}}}
%\end{table}

\begin{table}[htp]
 \caption{Metrics and references in the time category and the domains they originated in}
 \label{tab:time-metricsbydomain}
%  \resizebox{.7\textwidth}{!}{%
%   \resizebox*{\textwidth}{!}{%
  	\rowcolors{2}{gray!20}{white}
			\resizebox{.85\textwidth}{!}{%
 \begin{tabular}{p{1.2cm}p{6cm}p{2.8cm}p{2cm}}
 \upshape\bfseries Section & \bfseries Metric & \bfseries Original Domain & \bfseries Reference \\
\midrule
 \ref{metric:time-until-success} & Time until adversary's success & Communication & \cite{wright_analysis_2002} \\
 \ref{metric:maximum-tracking-time} & Maximum tracking time & Location & \cite{sampigethaya_caravan:_2005} \\
 \ref{metric:mean-time-to-confusion} & Mean time to confusion & Location & \cite{hoh_preserving_2007} \\
 \end{tabular}%
}
\end{table}

\subsubsection{Time until Adversary's Success}
\label{metric:time-until-success}

The most general time-based metric measures the time until the adversary's success~\cite{wright_analysis_2002}.
It assumes that the adversary will succeed eventually, and is therefore an example of a pessimistic metric.
This metric relies on a definition of success, and varies depending on how success is defined in a scenario.
For example, success in a communication system can be if the adversary identifies $n$ out of $N$ of the target's possible communication partners~\cite{agrawal_measuring_2003}.
%In an anonymization system that shuffles $b$ messages in every round (a so-called batch mix), in which a single sender communicates with $m$ recipients, and the security parameter is $l$, the metric computes the expected number of rounds until the adversary succeeds.
%% This metric is defined as the number of observations, or rounds, that are needed to identify a subset of a sender's recipients.
%\[priv_{\text{TAI}} \equiv \left[m \cdot l\left(\sqrt{\frac{N-1}{N}(b-1)}+\sqrt{\frac{N-1}{N^2}(b-1)+\frac{m-1}{m}}\right)\right]^2 \]
%
%\todo[inline]{check paper for a simpler equation. I don't really know whats happening, and I'm surprised the reviewer did}
% \subsubsection{Time to First Path Compromise}

Success can also be when the adversary first compromises a communication path~\cite{johnson_users_2013,vratonjic_how_2013}.
In an onion routing system such as Tor~\cite{dingledine_tor:_2004}, path compromise happens when the adversary controls all relays on a user's onion routing path.

\subsubsection{Maximum Tracking Time}
\label{metric:maximum-tracking-time}
In location privacy, the adversary often aims to not only break privacy at a single point in time, but to track a target's location over time.
The adversary's tracking ability is measured by the maximum tracking time, defined as the cumulative time that the size of the target $u$'s anonymity set remains~1~\cite{sampigethaya_caravan:_2005}.
\[priv_{\text{MTT}} \equiv \text{Cumulative time when } |AS_u| = 1 \]
This metric tends to overestimate a target's privacy because it assumes that the adversary has to be completely certain, i.e., the anonymity set has to be of size 1, to be successful.
In reality, however, an adversary may be capable to continue tracking despite a small number of users in the target's anonymity set.

% This can be seen as a best case metric for the user as tracking fails when user has a very low probability to be the target.
In a smart metering scenario, the maximum tracking time describes the percentage of a time interval during which the adversary can correctly classify the user's load transitions~\cite{lisovich_inferring_2010}.

\subsubsection{Mean Time to Confusion}
\label{metric:mean-time-to-confusion}

To avoid the maximum tracking time's overestimation of privacy, the mean time to confusion measures the time during which the adversary's uncertainty stays below a confusion threshold $\tau$~\cite{hoh_preserving_2007}.
The adversary's uncertainty is measured using the entropy $H(X)$ (Section~\ref{metric:entropy}), with the random variable $X$ indicating the adversary's estimated probabilities for each member of the anonymity set.
\[priv_{\text{MTC}} \equiv \text{Time during which } H(X) < \tau \]
Instead of time to confusion, the metric can also measure the distance to confusion, i.e., the travel distance until the adversary's tracking uncertainty rises above the threshold.
\end{subsection}

%%%%%%%%%%%%%%%%%%%%%%%%%%%%%%%%%%%%%%%%%%%%%%%%%%
% Accuracy / Precision
%%%%%%%%%%%%%%%%%%%%%%%%%%%%%%%%%%%%%%%%%%%%%%%%%%
\begin{subsection}{Accuracy / Precision}

Accuracy metrics quantify the accuracy of the adversary's estimate.
Although it can be argued that the accuracy of an estimate is not correlated with privacy because it does not allow to draw conclusions about the adversary's correctness or certainty~\cite{shokri_quantifying_2011}, inaccurate estimates can lead to higher privacy and are thus an important aspect of privacy.
Most metrics in this category originate from the domain of location-based services and measure geographic precision, but others are applicable more widely, including databases and communication systems.

%\newcolumntype{V}{>{\centering\arraybackslash}p{2.5cm}}
%\begin{table}[htp]
 %\tbl{Accuracy / Precision Metrics by Domain\label{tab:accuracy-metricsbydomain}}{
%%  \resizebox{.7\textwidth}{!}{%
  %\resizebox*{\textwidth}{!}{%
  	%\rowcolors{2}{gray!20}{white}
 %\begin{tabular}{>{\raggedleft\slshape}p{3.3cm}VVVVVV}
 %\upshape\bfseries Metric & \bfseries Communication & \bfseries Databases & \bfseries Location & \bfseries Smart Metering & \bfseries Social Networks & \bfseries Genome Privacy \\
%\midrule
     %Accuracy of obfuscated region & & & Ardagna 2007 & & & \\
     %Confidence interval width & & Agrawal 2000 & & & & \\
     %Coverage of sensitive region & & & Cheng 2006  & & & \\
     %Size of uncertainty region & & & Cheng 2006 & & & \\
     %Statistically strong event unobservability & Shao 2008 (WSNs) & & & & & \\
     %(t,$\delta$) privacy violation & & Kantarcioglu 2004 & & & & \\
 %\end{tabular}}}
%\end{table}

\begin{table}[htp]
 \caption{Metrics and references in the accuracy/precision category and the domains they originated in}
 \label{tab:accuracy-metricsbydomain}
%  \resizebox{.7\textwidth}{!}{%
%   \resizebox*{\textwidth}{!}{%
  	\rowcolors{2}{gray!20}{white}
			\resizebox{.85\textwidth}{!}{%
 \begin{tabular}{p{1.2cm}p{6cm}p{2.8cm}p{2cm}}
 \upshape\bfseries Section & \bfseries Metric & \bfseries Original Domain & \bfseries Reference \\
\midrule
 \ref{metric:confidence-interval-width} & Confidence interval width & Databases & \cite{agrawal_privacy-preserving_2000} \\
 \ref{metric:td-privacy-violation} & $(t,\delta)$ privacy violation & Databases & \cite{kantarcioglu_when_2004} \\
 \ref{metric:statistically-strong-event-unobservability} & Statistically strong event unobservability & Communication & \cite{shao_towards_2008} \\
 \ref{metric:size-of-uncertainty-region} & Size of uncertainty region & Location & \cite{cheng_preserving_2006} \\
 \ref{metric:accuracy-of-obfuscated-region} & Accuracy of obfuscated region & Location & \cite{ardagna_location_2007} \\
 \ref{metric:coverage-of-sensitive-region} & Coverage of sensitive region & Location & \cite{cheng_preserving_2006} \\
 \end{tabular}%
}
\end{table}

\subsubsection{Confidence Interval Width}
\label{metric:confidence-interval-width}

According to the confidence interval width, the amount of privacy at $\tau$\% confidence is given by the width of the confidence interval for the adversary's estimate $x \in [x_2, x_1]$ in which the true outcome $x^*$ lies~\cite{agrawal_privacy-preserving_2000}.
\[priv_{\text{CIW}} \equiv |x_2 - x_1| \text{ where } p(x_1 \leq x < x_2) = \tau/100 \]
However, when publishing perturbed data, knowledge of the confidence interval width may allow reconstruction of the original distribution~\cite{agrawal_design_2001}.

% \todo[inline]{how can the adversary's estimate have 2 values? usually he guesses 1 value}

\subsubsection{$(t,\delta)$ Privacy Violation}
\label{metric:td-privacy-violation}

In data mining, $(t,\delta)$ privacy violation gives information whether the release of a classifier for public data is a privacy threat, depending on how many training samples $t$ are available to the adversary.
Training samples link public data $D$ to sensitive data $S$ for some individuals, and privacy is violated when an adversary can infer sensitive information from public data for individuals who are not in the training samples.
% by building a classifier $C$ with Bayes error $\rho$.
The metric compares the Bayes errors $\rho$ for the cases when the adversary builds a classifier based on training samples alone ($\rho(t)$), or based on training samples and a given classifier for public data ($\rho(t, C(D))$).
The classifier $C(D)$ is $(t,\delta)$ privacy violating if it reduces the adversary's Bayes error by more than the privacy parameter $\delta$~\cite{kantarcioglu_when_2004}.
%is based on available training samples $t$, and its prediction accuracy is bounded by a privacy parameter~$p$~\cite{kantarcioglu_when_2004}.
%\[priv_{\text{TPP}} \equiv \rho(t;C) \leq \rho(t) - p \text{, with } \rho(t) = \rho_{\{t;D,S\}} \text{ and } \rho(t;C) = \rho_{\{t;D,C,S\}} \]
\[priv_{\text{TPP}} \equiv \rho(t;C(D)) \leq \rho(t) - \delta \]
%\todo[inline,color=red!20]{what's the interpretation for this metric? what do its values mean?}

%\todo[inline]{the privacy parameter should not be p. I don't understand what $\rho(x)$ does, as $\rho$ is the error and not a function. We need to explain this metric better}
%\todo[inline,color=red!20]{errors can be functions -- just says that the error depends on something}

\subsubsection{Statistically Strong Event Unobservability}
\label{metric:statistically-strong-event-unobservability}

In wireless sensor networks, a privacy goal is to hide where in the network an event has occurred. 
Statistically strong event unobservability compares the message patterns in all parts of the network so that event locations are not revealed by a sudden burst of messages.
For example, the event sources in a wireless sensor network are unobservable if the distributions of inter-message delays are roughly the same in all parts of the network.
% This metric has been used in wireless sensor networks to measure the unobservability of events by computing how close the distributions of inter-message delays from different cells in the network are.
Specifically, the metric requires that the distance between distributions $d(F_1,F_2)$ is smaller than $\tau$, and that the difference between the distribution parameters $f$ is smaller than $\epsilon$~\cite{shao_towards_2008}. 
However, the metric is limited to distributions that have a single parameter, such as the exponential distribution.
\[priv_{\text{SEU}} \equiv (\tau,\epsilon) \text{, where } d(F_1,F_2) \leq \tau \wedge (1-\epsilon)f_1 \leq f_2 \leq (1+\epsilon)f_1 \]
%\todo[inline,color=red!20]{what's the interpretation for this metric? what do its values mean?}
%\todo[inline]{This needs to be better explained. What are the distributions? What is $F_1$ and $F_2$? now d() is the area, and not the distance like in the rest of the paper. It used to be $Area$ before, but here it's more like an integral? privacy parameter should not be $p$, to avoid confusing it with probability. Why are there so many $\leq$ in the equation. Last sentence should mention 1 or 2 distributions that can be used}

\subsubsection{Size of Uncertainty Region}
\label{metric:size-of-uncertainty-region}
%this metric can be shortened if we need space, the equation is no contribution

In location privacy, the size of the uncertainty region denotes the minimal size of the region $R_U$ to which an adversary can narrow down the position of a target user $u$~\cite{cheng_preserving_2006}.
%The adversary cannot determine the user's position with a finer resolution than this region.
\[priv_{\text{SUR}} \equiv Area(R_U) \]

\subsubsection{Accuracy of Obfuscated Region}
\label{metric:accuracy-of-obfuscated-region}

In location-based services, users may report a certain region back to a service provider, e.g. to ask for local services in that region.
To protect their location privacy, users can obfuscate this region before submitting it by enlarging it to a point where it satisfies a chosen minimum user requirement $r_\text{min}$ (assuming circular areas).
The accuracy of the obfuscated region then indicates how relevant to a service provider the reported area is, a value of 0 representing the lowest relevance, or highest level of privacy respectively.
The metric can be computed based on the optimal accuracy provided by the used sensing technology $r_\text{opt}$ and the user-specified minimum $r_\text{min}$~\cite{ardagna_location_2007}.
\[priv_{\text{AOR}} \equiv \frac{r_\text{opt}^2}{r_\text{min}^2} \]

\subsubsection{Coverage of Sensitive Region}
\label{metric:coverage-of-sensitive-region}

The coverage of the sensitive region evaluates how a user's sensitive regions $R_S$ overlap with the adversary's uncertainty region $R_U$ (see Section \ref{metric:size-of-uncertainty-region})~\cite{cheng_preserving_2006}.
A sensitive region can be, for example, a hospital or a nightclub.
The uncertainty region indicates the smallest region of which the adversary is certain that it includes the user.
If the two regions overlap, the adversary succeeds in linking the user to the sensitive region.

The metric is normalized to the area of the uncertainty region, so that it becomes $1$ when $R_U$ equals or is fully contained in $R_S$, in which case the adversary can indubitably associate a user with the sensitive region.

\[priv_{\text{CSR}} \equiv \frac{Area(R_S \cap R_U)}{Area(R_U)} \]

% \todo[inline]{the uncertainty region here is basically the metric from in 5.8.4, maybe backreference more clearly, or even use the symbol. the symbols in this metric HAVE to be the same as in 5.8.4}
\end{subsection}

\begin{printonly}
\section{Supplementary Materials} 	 
\label{sec:recommendations} 	 
\label{sec:futurework} 	 
The online version of this paper includes two additional sections: first, a guide on how to select suitable privacy metrics, and second, a discussion of promising future research directions. 	 
It also includes additional references, as well as tables \ref{tab:privacymetrics1} and \ref{tab:privacymetrics2} that summarize how each metric can be classified according to the characteristics introduced in \ref{sec:characteristics}.

\manuallabel{tab:privacymetrics1}{10} 	 
\manuallabel{tab:privacymetrics2}{11} 	 
	  	 
\end{printonly}

\begin{screenonly}

% WATCH OUT!
% The lines in this table have been auto-generated from metrics-table.xslx
% using http://ctan.org/tex-archive/support/excel2latex
% DO NOT MODIFY!
\begin{table}[htp]
 \caption{Privacy Metrics (1): Uncertainty, Information Gain/Loss, and Similarity/Diversity Outputs}
 \label{tab:privacymetrics1}
 \resizebox*{\textwidth}{!}{%
 \begin{tabular}{p{0.3cm}p{5.5cm}lllllllll} % p{0.15cm}p{0.15cm}p{0.15cm}p{0.15cm}p{0.15cm}
%  \toprule
  & & &  & &  & \multicolumn{5}{c}{Inputs} \\
  % make parbox values bigger to move text down, smaller to move text up
  % these two headings have to be in the first row because they are taller than any of the input headings
  \cmidrule(l){7-11}
  \rotatebox[origin=l]{90}{Output} & \rotatebox[origin=l]{90}{Metric} & \rotatebox[origin=l]{90}{Value range} & \rotatebox[origin=l]{90}{\parbox{2.6cm}{high (H) or low (L)\\ values indicate\\ high privacy}} & \rotatebox[origin=l]{90}{\parbox{2cm}{Primary\\data source}} & \rotatebox[origin=l]{90}{\parbox{2.6cm}{(I)dentity/(P)roperty}} & \rotatebox[origin=l]{90}{Adv. estimate} & \rotatebox[origin=l]{90}{Adv. resources} & \rotatebox[origin=l]{90}{True outcome} & \rotatebox[origin=l]{90}{Prior knowledge} & \rotatebox[origin=l]{90}{Parameters} \\
  \midrule
%   \hline
    \multirow{17}{*}{\rotatebox[origin=c]{90}{Uncertainty}} & Anonymity set size & $[0,|X|]$ & H     & obs   & IP & \multicolumn{1}{l}{x} &       &       &       &  \\
    & Asymmetric entropy & $[0,1]$ & H     & obs, pub  & IP & \multicolumn{1}{l}{x} &       &       & x     &  \\
    & Conditional entropy & $[0, \infty]$ & H     & obs, pub  & IP & \multicolumn{1}{l}{x} &       &       & x     &  \\
    & Conditional privacy & $[1, \infty]$ & H     & obs, pub  & IP & \multicolumn{1}{l}{x} &       &       & x     &  \\
    & Cross-entropy & $[0, \infty]$ & H     & pub    & IP & \multicolumn{1}{l}{x} &       & x     &       &  \\
    & Cumulative entropy & $[0,\infty]$ & H     & obs    & IP & \multicolumn{1}{l}{x} &       &       &       &  \\
    & Degree of unlinkability & $[0,\infty]$ & H     & obs, pub  & P & \multicolumn{1}{l}{x} &       &       & (x)   &  \\
    & Entropy & $[0,H_0(X)]$ & H     & obs, pub  & IP & \multicolumn{1}{l}{x} &       &       &       &  \\
%     & Entropy + Bayes & $[0,\infty]$ & H     & obs    &  & \multicolumn{1}{l}{x} &       &       & x     &  \\
    & Genomic privacy & $[0,\infty]$ & H     & pub    	& P & \multicolumn{1}{l}{x} &       &       &       & x \\
    & Inherent privacy & $[1,|X|]$ & H     & obs, pub  & IP & \multicolumn{1}{l}{x} &       &       &       &  \\
    & Max-entropy (Hartley) & $[0,\infty]$ & H     & obs, pub  & IP & \multicolumn{1}{l}{x} &       &       &       &  \\
    & Min-entropy & $[0,\infty]$ & H     & obs, pub  & IP & \multicolumn{1}{l}{x} &       &       &       &  \\
%     & Normalized conditional entropy & $[0,1]$ & H     & obs, pub  &  & \multicolumn{1}{l}{x} &       &       & x     &  \\
    & Normalized entropy & $[0,1]$ & H     & obs, pub  & IP & \multicolumn{1}{l}{x} &       &       &       &  \\
    & Protection level & $[0, \infty]$ & H     & obs    & P & \multicolumn{1}{l}{x} &       &       &       & x \\
    & Quantiles on entropy & $[0,H_0(X)]$ & H     & obs, pub  & IP & \multicolumn{1}{l}{x} &       &       &       & x \\
    & R\'{e}nyi entropy & $[0, \infty]$ & H     & obs, pub  & IP & \multicolumn{1}{l}{x} &       &       &       & x \\
    & User-centric privacy & $[0, H_0(U)]$ & H     & obs    & IP & \multicolumn{1}{l}{x} &       &       &       & x \\
    \midrule
    \multirow{16}{*}{\rotatebox[origin=c]{90}{Information Gain}} & Amount of leaked information & $[0,\infty]$ & L     & pub, oth  & IP & \multicolumn{1}{l}{} &       & x     &       &  \\
    & Conditional mutual information & $[0,\infty]$ & L     & obs, pub  & IP & \multicolumn{1}{l}{x} &       & x     & x     &  \\
    & Conditional privacy loss & $[0,1]$ & L     & obs, pub  & IP & \multicolumn{1}{l}{x} &       & x     &       &  \\
    & Full/partial disclosure & $[0,1]$ & L     & obs, pub  & IP & \multicolumn{1}{l}{x} & & & & x \\
    & Increase in adversary's belief & true, false, $\delta$: $[0,1]$ & L     & obs, pub  & IP & \multicolumn{1}{l}{x} &       &       & x     & x \\
    & Information surprisal & $]0,\infty]$ & L     & pub    & P & \multicolumn{1}{l}{x} &       & x     &       &  \\
    & Maximum information leakage & $[0,\infty]$ & L     & obs, pub  & IP & \multicolumn{1}{l}{x} &       &       &       &  \\
    & Mutual information & $[0, \infty]$ & L     & obs, pub  & IP & \multicolumn{1}{l}{x} &       & x     &       &  \\
    & Normalized mutual information & $[0,1]$ & H     & obs, pub  & IP & \multicolumn{1}{l}{x} &       & x     &       &  \\
    & Pearson's correlation coefficient & $[0,1]$ & L     & obs, rep  & IP & \multicolumn{1}{l}{} &       & x     &       &  \\
    & Positive information disclosure & $[0,1]$ & L     & obs  & IP & \multicolumn{1}{l}{x} &       &      &       &\\
    & Privacy score & $[0,\infty]$ & L     & pub    & P & \multicolumn{1}{l}{} &       &       &       & x \\
    & Reduction in observable features & $[0,1]$ & L     & obs, rep  & P & \multicolumn{1}{l}{} &       & x     &       &  \\
    & Relative entropy & $[0, \infty]$ & H     & obs, pub  & IP & \multicolumn{1}{l}{x} &       & x     &       &  \\
    & (Relative) Loss of anonymity & $[0,H(X)]$ & L     & obs    & IP & \multicolumn{1}{l}{x} &       & x     & (x)   &  \\
    & System anonymity level & $[0, \infty]$ & H     & obs    & I & \multicolumn{1}{l}{x} &       & x     &       &  \\
    \midrule
    \multirow{16}{*}{\rotatebox[origin=c]{90}{Similarity}} & ($\alpha$,k)-anonymity & k: $[0,\infty]$, $\alpha$: $[0,1]$ & k: H, $\alpha$: L & pub    & IP & \multicolumn{1}{l}{} &       &       &       & x \\
    & (c,t)-isolation & $[0,\infty]$ & H     & pub    & IP & \multicolumn{1}{l}{x} &       & x     &       & x \\
    & Cluster similarity & $[0,1]$ & L     & obs, rep  & P & \multicolumn{1}{l}{} &       & x     &       &  \\
    & Coefficient of determination $R^2$ & $[0,1]$ & L     & obs, rep  & P & \multicolumn{1}{l}{} &       & x     &       &  \\
    & ($\epsilon$,m)-anonymity & $\epsilon: [0,1]$, m: $[1,\infty]$ & $\epsilon$: H, m: H & pub    & IP & \multicolumn{1}{l}{} &       &       &       & x \\
%     & Haplotype-SNP-test & true, false & H     & pub    &  & \multicolumn{1}{l}{} &       &       &       & x \\
    & Historical $k$-anonymity & $[0,\infty]$ & H     & obs    & IP & \multicolumn{1}{l}{} &       & x     &       & x \\
    & $k$-anonymity & $[1, |D|]$ & H     & pub    & I & \multicolumn{1}{l}{} &       &       &       & x \\
    & (k,e)-anonymity & $[0,\infty]$ & H     & pub    & IP & \multicolumn{1}{l}{} &       &       &       & x \\
    & $\ell$-diversity & $[0, \infty]$ & H     & pub    & IP & \multicolumn{1}{l}{} &       &       &       & x \\
    & $m$-invariance & $[0,\infty]$ & H     & pub    & IP & \multicolumn{1}{l}{} &       &       &       & x \\
    & Multirelational $k$-anonymity & $[0,\infty]$ & H     & pub    & I & \multicolumn{1}{l}{} &       & x     &       & x \\
    & Normalized variance & $[0,1]$ & H     & pub    & IP & \multicolumn{1}{l}{} &       & x     &       &  \\
    & Stochastic $t$-closeness & t: $[0, \infty]$, $\epsilon$: $[0, \infty]$ & L & pub  & IP & \multicolumn{1}{l}{} &       & x     &       & x \\
    & $t$-closeness & $[0, \infty]$ & L     & pub    & IP & \multicolumn{1}{l}{} &       & x     &       & x \\
    & (X,Y)-privacy & $]0,1]$ & L     & pub    & IP & \multicolumn{1}{l}{} &       & x     &       & x \\
%     \bottomrule
 \end{tabular}}%
\end{table}

\begin{table}[htp]
 \caption{Privacy Metrics (2): Indistinguishability, Adversary's Success Probability, Error, Accuracy/Precision, and Time Outputs}
 \label{tab:privacymetrics2}
 \resizebox*{\textwidth}{!}{%
 \begin{tabular}{p{0.3cm}p{5.5cm}lllllllll} % p{0.15cm}p{0.15cm}p{0.15cm}p{0.15cm}p{0.15cm}
%  \toprule
  & & &  & &  & \multicolumn{5}{c}{Inputs} \\
  % make parbox values bigger to move text down, smaller to move text up
  % these two headings have to be in the first row because they are taller than any of the input headings
  \cmidrule(l){7-11}
  \rotatebox[origin=l]{90}{Output} & \rotatebox[origin=l]{90}{Metric} & \rotatebox[origin=l]{90}{Value range} & \rotatebox[origin=l]{90}{\parbox{2.6cm}{high (H) or low (L)\\ values indicate\\ high privacy}} & \rotatebox[origin=l]{90}{\parbox{2cm}{Primary\\data source}} & \rotatebox[origin=l]{90}{\parbox{2.6cm}{(I)dentity/(P)roperty}} & \rotatebox[origin=l]{90}{Adv. estimate} & \rotatebox[origin=l]{90}{Adv. resources} & \rotatebox[origin=l]{90}{True outcome} & \rotatebox[origin=l]{90}{Prior knowledge} & \rotatebox[origin=l]{90}{Parameters} \\
  \midrule
%   \hline
    \multirow{11}{*}{\rotatebox[origin=c]{90}{Indistinguishability}} & Approximate differential privacy & $\epsilon$: $[0,\infty]$, $\delta$: $[0,\infty]$ & $\epsilon$: L, $\delta$: L & pub   &  IP & \multicolumn{1}{l}{} &       & x     &       & x \\
    & Computational differential privacy & $[0,\infty]$ & L     & pub    & IP & \multicolumn{1}{l}{x} & x     & x     &       & x \\
    & Crypto. game / semantic security & true, false & H     & obs    & IP & \multicolumn{1}{l}{x} &       & x     &       & x \\
    & d-$\chi$-privacy & $[0,\infty]$ & L     & pub    & IP & \multicolumn{1}{l}{} &       & x     &       & x \\
    & Differential privacy & $[0,\infty]$ & L     & pub    & IP & \multicolumn{1}{l}{} &       & x     &       & x \\
    & Distributed differential privacy & $\epsilon$: $[0,\infty]$, $\delta$: $[0,\infty]$ & $\epsilon$: L, $\delta$: L & pub, rep  & IP & \multicolumn{1}{l}{} &       & x     &       & x \\
    & Distributional privacy & $[0,\infty]$ & L     & pub, rep  & P & \multicolumn{1}{l}{} &       & x     &       & x \\
    & Geo-indistinguishability & $[0,\infty]$ & L     & obs    & P & \multicolumn{1}{l}{} &       & x     &       & x \\
    & Information privacy & true, false & H     & obs    & IP & \multicolumn{1}{l}{x} &       &       &       & x \\
    & Joint differential privacy & $\epsilon$: $[0,\infty]$, $\delta$: $[0,\infty]$ & $\epsilon$: L, $\delta$: L & pub    & IP & \multicolumn{1}{l}{} &       & x     &       & x \\
    & Observational equivalence & true, false & H     & obs    & IP & \multicolumn{1}{l}{x} &       & x     &       &  \\
%     & Unconditional / computational privacy & true, false & L     & obs    &  & \multicolumn{1}{l}{x} &       & x     &       & x \\
    \midrule
    \multirow{6}{*}{\rotatebox[origin=c]{90}{Success}} & Adversary's success rate & $[0,1]$ & L     & obs    & IP & \multicolumn{1}{l}{x} &       & x     &       & (x) \\
    & (d,$\gamma$)-privacy & $[0,1]$ & L     & obs    & IP & \multicolumn{1}{l}{x} &       &       & x     & x \\
    & Degrees of anonymity & $[0,1]$ & L     & obs    & IP & \multicolumn{1}{l}{x} &       & x     &       & x \\
    & $\delta$-presence & $[0,1]$ & L     & pub    & I & \multicolumn{1}{l}{x} &       &       & x     & x \\
    & Hiding property & $[0,1]$ & L     & obs    & I & \multicolumn{1}{l}{x} &       &       &       & x \\
    & Privacy breach level & $[0,1]$ & L     & obs   &  IP & \multicolumn{1}{l}{x} &       &       & x     & x \\
%     & Probability of path compromise & $[0,1]$ & L     & obs    &  & \multicolumn{1}{l}{x} & x     & x     &       &  \\
    \midrule
    \multirow{4}{*}{\rotatebox[origin=c]{90}{Error}} & Adv.'s expected estimation error & $[0,1]$ & L     & obs    & IP & \multicolumn{1}{l}{x} &       & x     &       &  \\
    & Expectation of distance error & $[0,\infty]$ & H     & obs    & P & \multicolumn{1}{l}{x} &       & x     &       &  \\
%     & Health privacy & depends & depends & obs, pub  &  & \multicolumn{4}{c}{depends on base metric} & x \\
    & Mean squared error & $[0,\infty]$ & H     & obs    & IP & \multicolumn{1}{l}{x} &       & x     &       &  \\
    & Percentage incorrectly classified & $[0,1]$ & H     & obs, rep  & IP & \multicolumn{1}{l}{x} &       & x     &       &  \\
    \midrule
    \multirow{6}{*}{\rotatebox[origin=c]{90}{Accuracy}} & Accuracy of obfuscated region & $[0,1]$ & L     & obs    & P & \multicolumn{1}{l}{} &       &       &       & x \\
    & Confidence interval width & $[0,\infty]$ & H     & pub, obs  & IP & \multicolumn{1}{l}{x} &       &       &       & x \\
    & Coverage of sensitive region & $[0,1]$ & L     & obs    & P & \multicolumn{1}{l}{x} &       &       &       & x \\
    & Size of uncertainty region & $[0,\infty]$ & H     & obs    & P & \multicolumn{1}{l}{x} &       &       &       &  \\
    & Stat. strong event unobservability & $[0,\infty]$ & L     & obs    & P & \multicolumn{1}{l}{x} &       &       &       & x \\
    & (t,$\delta$) privacy violation & $[0,1]$ & L     & pub    & P & \multicolumn{1}{l}{x} &       & x     & x     & x \\
     \midrule
    \multirow{3}{*}{\rotatebox[origin=c]{90}{Time}} & Maximum tracking time & $[0,\infty]$ & L     & obs    & I & \multicolumn{1}{l}{x} &       &       &       &  \\
    & Mean time to confusion & $[0,\infty]$ & L     & obs    & I & \multicolumn{1}{l}{x} &       &       &       & x \\
    & Time until adversary's success & $[0,\infty]$ & H     & obs    & IP & \multicolumn{1}{l}{x} &       & x     &       & (x) \\
 \end{tabular}}
\end{table}

\section{How to Select Suitable Privacy Metrics}
\label{sec:recommendations}

% \todo[inline,color=blue!20]{Lastly, Section 6 is very interesting, and it could extend the audience
% of the survey to practitioners. To this aim, it would be useful to have
% few examples of practical usage of these metrics, and some discussions
% on the problem of parameter setting.}

Given the number and diversity of privacy metrics, selecting metrics for a given scenario can be difficult.
We suggest a series of nine questions to guide the selection process.
Answering each of the questions makes sure that all aspects of metric selection are considered.
Where possible and appropriate, we point to metrics or groups of metrics that we associate with particular answers.

The first two questions ask about which aspects of privacy should be quantified (question \ref{question:output}), and which adversary types we need to protect against (question \ref{question:adversary}).
Next, we suggest to consider which data sources need to be protected (question \ref{question:data}), and which input data are available to compute the metrics (question \ref{question:input}).
We then move on to consider the requirements of the target audience (question \ref{question:audience}) and which metrics have been used in related work (question \ref{question:related-work}).
We also suggest to check whether any of the selected metrics have flaws (question \ref{question:quality}), and whether validated implementations for the metrics are available (question \ref{question:implementations}).
Finally, we consider strategies to choose parameter settings for the selected metrics (question \ref{question:parameters}).

We have already succesfully applied this selection strategy in a case study for genomic privacy~\cite{wagner2017evaluating}, and found the following questions useful to support the selection process.

\subsection{Suitable Output Measures?}
\label{question:output}
% what aspects do we want to measure?
\textit{Which aspects of privacy do we want to quantify?
Do we want to give privacy guarantees, or is some loss of privacy acceptable?}

The pool of potential metrics can be narrowed down by deciding which outputs we want to measure.
In Section \ref{sec:taxonomy}, we classify the output measures of privacy metrics into eight categories.
Figure~\ref{fig_outcomes} and the \emph{Output} column in Tables~\ref{tab:privacymetrics1} and \ref{tab:privacymetrics2} list the output measure for each metric.
% to narrow down metric pool: are there metrics that measure the same or similar things? That are shown to have the same shape?
% how do the properties that we measure relate to a user's privacy level?

If the application scenario requires privacy guarantees in the sense that privacy properties can be proven to hold, the only viable choices for metrics are in the indistinguishability category.
If the application instead calls for a quantification of privacy levels, metrics from the other categories are more suitable.

Instead of fixing a single output measure for a scenario, we recommend to measure several different outputs.
Because none of the metrics measures `privacy' directly, but only quantities assumed to be related to privacy, each additional output category gives information about an additional aspect of privacy.
% The development of \acp{PET} can only benefit from measuring several different aspects of privacy.

For example, a study about location privacy by \citeANP{shokri_quantifying_2011}~\cite{shokri_quantifying_2011} used metrics from three different categories to measure the adversary's accuracy (confidence interval width, Section~\ref{metric:confidence-interval-width}), uncertainty (entropy, Section~\ref{metric:entropy}), and error (expected estimation error, Section~\ref{metric:expected-estimation-error}).
Following our recommendation, this selection could be extended with a success metric that quantifies how likely it is for the adversary to succeed, or with a time metric that measures the time until the adversary's success.
We might also add a second uncertainty metric that indicates the size of the crowd into which an individual can blend.

%For a given scenario, we therefore recommend to select metrics from as many different categories from our taxonomy in Section~\ref{sec:taxonomy} as appropriate.
% Our taxonomy of outputs in Section~\ref{sec:taxonomy} shows that it is possible to quantify other aspects of privacy as well.
% We therefore suggest to add more metrics as appropriate.

Besides including metrics from different categories, we recommend to select metrics that reflect the average case, the distribution of privacy values, and the worst case.

% In addition to privacy metrics, orthogonal metrics that measure cost or utility may be needed.
% For example, in a data publishing scenario a utility metric can be used to measure how useful the published data are when a \ac{PET} has been applied.
% Utility metrics are out of scope for this paper; as starting points for further investigation, we refer interested readers to~\cite{bertino_survey_2008,fung_privacy-preserving_2010,ghosh_universally_2012}.

\subsection{Adversary Models?}
\label{question:adversary}
\textit{What are the characteristics of the adversary we consider? 
How do we incorporate the adversary's goals and their knowledge?}

% paper presents attack? -> adversary-based metric
We observed that papers presenting attacks against privacy tend to use metrics based on time, error, or the adversary's success probability, whereas papers presenting new \acp{PET} tend towards accuracy, similarity, and indistinguishability metrics.
In both cases, this is a convenient choice: most metrics in the first group have a stronger focus on the adversary, while the metrics in the second group emphasize the efficacy of the presented \ac{PET}.
%allow \ac{PET} papers to focus more on the actual system that they try to protect.
However, as we have argued before, the measurement of privacy benefits when more aspects of privacy are measured.
We therefore believe that both the `attack' and `defense' perspective can benefit from selecting metrics from the other side.

We also observed that different privacy domains make different assumptions about the adversary.
For example, time-based metrics in communication systems measure the time until the adversary's success, whereas time-based metrics in location privacy measures the time until the adversary's confusion.
This is a fundamental difference, and it is not obvious which flavor of the assumption holds in other privacy domains.

% metrics that are independent of adversary are risky - k-anonymity is a prime example
Care must be taken when choosing metrics that do not consider an adversary model.
For example, most data similarity metrics such as $k$-anonymity (Section~\ref{sec:metrics:similarity}) compute the level of privacy depending only on properties of the data.
However, if the adversary happens to have relevant prior knowledge, the privacy level indicated by $k$ is no longer accurate.

%\todo[inline, color=green!20]{ I am not fully convinced about its
%applicability in real world scenarios. E.g., the questions proposed to
%decide the value of k for k-anonymity are rather generic (e.g., level of
%privacy acceptable).  In any case, this is a very difficult problem in
%practice, and the proposed questions could be an useful guide.}

We found few metrics that explicitly consider the resources an adversary has to expend in order to succeed.
Aside from time-based metrics, the only other metric considering resources is probability of compromising a communication path (a variant of the adversary's success rate, see Section~\ref{metric:probability-of-path-compromise}), where bandwidth and the number of nodes are the constrained resources.
%\todo[inline]{Isa: The reference here points to 5.5.1 and I can't find this metric anywhere in the paper? If you refer to the second paragraph of the adversary's success, then we should mention the metric name verbatim}
Resource-based metrics are an interesting area for future research, which means that if we consider a resource-constrained adversary, we will have to create new metrics.

Lastly, it is important to consider which type of sensitive information the adversary aims to reveal, i.e. either user identities or properties, and to select metrics that are able to measure the relevant aspect.

\subsection{Data Source?}
\label{question:data}
\textit{Which data sources do we aim to protect?}

We introduced four data sources in Section~\ref{sec:sources} -- published, observable, re-purposed, or all other data.
Depending on which data source needs protecting, different metrics apply.
We summarize the primary data sources for each of the metrics in the \textit{Primary data source} column in Tables~\ref{tab:privacymetrics1} and \ref{tab:privacymetrics2}.

Although in many scenarios one data source will be the main cause of concern, considering all four data sources reduces the likelihood that unforeseen events compromise the entire system. 
It also enables informed decisions about which privacy risks should be mitigated or accepted.
In addition, considering all four data sources can emphasize the need for data minimization, because data that is not there does not need protection.

% \todo{It is also matter of perspective: e.g. from a user's adversary’s perspective (what are the implications of the data that is public) or a providers perspective (which data can/do we make public, or what happens if data that is not supposed to be public becomes public), or even a protocol designer (what can be or must not be in the observable information)}

\subsection{Availability of Input Data?}
\label{question:input}
% what input data do we want to consider?
% or, related, what input data are available in our scenario?
\textit{Which types of input data do we want to consider, and which are available in our scenario?}

Input data refers to the information that is needed to compute a metric, such as the adversary's estimate, resources, and prior knowledge, the true outcome, or parameter values.
If a certain kind of input data is not available or applicable in a scenario, we can disregard all metrics that need this input type.
Similarly, if we explicitly want to consider a certain input, we can disregard metrics that do not use this input type.
We describe different kinds of input data in Section \ref{sec:inputs} and show the kinds of input data for each metric in the \emph{Inputs} column of Tables~\ref{tab:privacymetrics1} and \ref{tab:privacymetrics2}.

\subsection{Target Audience?}
\label{question:audience}
\textit{What is the intended audience for our study?
What are their expectations regarding the presentation of results, and do they understand the interpretations of our metrics?}
% what is the target audience for our study, and what are their expectations regarding presentation of results?
% do the metrics have an interpretation that is easy to understand?

An important consideration for the selection of metrics is the intended audience, especially with regard to laypeople and researchers in other academic disciplines.

Whenever results need to be communicated to laypeople, it is important to select metrics that can be understood easily.
This does not mean that the formal definition of the metric has to be simplistic; rather, it means that the metric should have an intuitive interpretation, even if it simplifies the underlying technical details.
However, we are not aware of user studies that evaluate how easily different metrics are understood by laypeople, or which interpretations help understanding.
%this is an interesting topic for future research.

% need metrics that are usable/understandable for researchers in other areas, e.g, social science (see daries_privacy_2014 about MOOCs) or biology (genomic privacy)
% (I suspect biologists will have less trouble than social scientists)
Whenever metrics are intended to be used by researchers in other academic disciplines, it may be beneficial to use methods and terminology common in the respective discipline.
Consider genome privacy as an example: in many areas of biology it is common to conduct statistical analyses; for non-privacy researchers in this field, metrics based on accuracy, error, or success will therefore be easier to understand and adopt than, say, metrics based on indistinguishability.

\subsection{Related Work?}
\label{question:related-work}
% what metrics are used by the papers in our related work section?
\textit{Which metrics are used by work that is related to ours, and would those metrics be suitable in our work as well?
Which mathematical concepts or formalisms are used by others in our field?
Which of these are already available in the tools we use?}

To enable comparisons between different studies in the same privacy domain, it is useful to select metrics that have already been used by related work, even if those metrics would otherwise not be the first choice.
In addition, well-known metrics are likely to be more easily understood by other researchers in the same field.

A related consideration is expertise.
Some metrics are conceptually difficult, and hard to use correctly.
To reduce the risk of invalidating the results of an entire study, we recommend to select both comparatively simple metrics and more complex ones.

% \subsubsection{Parameter Values?}
% % what are appropriate values for parameters?
% If candidate metrics rely on parameters, what are appropriate parameter values?
% How can we determine missing parameter values?

\subsection{Quality of Metrics?}
\label{question:quality}
% have any candidate metrics been shown to be defective, or particularly good or useful?
% if there is no data, can we conduct a study to verify that the metrics behave as we intend?
\textit{Do any of the candidate metrics have known flaws?
%If there is no information about the goodness of a metric:
Is it feasible to conduct a study that verifies that candidate metrics indeed behave as we intend?}

Even though it is desirable to work with high-quality metrics, few studies systematically evaluate the quality of privacy metrics.
This means that information about metric quality is not readily available at the time of this writing.
Even so, some metrics do have known weaknesses (which we have pointed out throughout Section~\ref{sec:metrics}) and should only be used with caution.
If selecting known weak metrics, we recommend to use them in combination with other metrics to help offset the weaknesses.

If results about metric quality are not available for a particular privacy domain, it may be possible to conduct a small study to evaluate how candidate metrics perform.

\subsection{Metric Implementations?}
\label{question:implementations}
\textit{Are there implementations of the candidate metrics that we can use, or compare our implementation with?}

Even when metrics are easy to understand, implementing them in a particular scenario can be difficult, and challenges can arise with unexpected aspects of a metric.
For example, when implementing the entropy of an anonymity set, the challenge may not be entropy itself, but the propagation of anonymity set probabilities over multiple timesteps.
% The authors have experienced this while creating an implementation for the entropy of an anonymity set in a location privacy scenario.
% The challenge in this case was not entropy, but the propagation of anonymity set probabilities over multiple timesteps.
% 
Common challenges like this are likely to be solved to different degrees in different implementations.
The more research groups use and validate an implementation, the higher the chance of detecting implementation errors.
We therefore recommend to consider selecting metrics for which a validated implementation exists.
Ultimately, only implementations that have been thoroughly validated can lead to consistent results across studies.

% [14:22:45] D. Eckh.: aber prinzipiell sollte man natuerlich seine auswahl nicht darauf beschraenken ob das schon mal jemand implementiert hat
% [14:22:55] D. Eckh.: aber wenn es schon eine gibt, sollte man seine implementierung vergleichen zumindest
% [14:23:00] D. Eckh.: oder halt die andere nehmen

\subsection{Metric Parameters?}
\label{question:parameters}
\textit{How should we choose the parameter values for the candidate metrics?}

Many metrics use parameters to adapt to the privacy requirements of specific scenarios (see the \textit{Parameters} column in Tables \ref{tab:privacymetrics1} and \ref{tab:privacymetrics2}).
For example, $k$-anonymity (Section \ref{metric:k-anonymity}) uses the parameter $k$ to indicate how many individuals in a database should be indistinguishable from each other, user-centric privacy (Section \ref{metric:user-centric-privacy}) uses a parameter to indicate how fast (in the user's opinion) their privacy decays over time, and health privacy (Section \ref{metric:health-privacy}) uses weights to indicate the contribution of genetic varations to a disease.
However, it is often difficult to decide how these parameters should be set.
For example, studies using differential privacy (Section \ref{metric:differential-privacy}) have used values for $\epsilon$ that span five orders of magnitude (from 0.01 to 100), and aside from \citeANP{lee2011how}~\cite{lee2011how}, there is not much literature on parameter setting for differential privacy.
For $k$-anonymity (Section \ref{metric:k-anonymity}), some authors argue that $k=3$ satisfies US regulations for the release of educational data \cite{daries2014privacy}, and some have used $k=5$ for the release of medical data \cite{rynkiewicz2015private}.

% in a nutshell:
% first step: get clear about requirements
% then: ask users, consider utility, use real-world data, add metrics without parameters, and evaluate several parameters settings
There are a number of strategies that can help determine parameter settings or mitigate suboptimal parameter settings.
Most important is to clearly state the requirements of the application scenario.
Then, we recommend five strategies:
(1) Ask users what levels of privacy they would deem acceptable.
However, care must be taken to present privacy levels and the influence of parameter settings in an accessible way so that users do not need extensive technical knowledge to participate.
(2) Consider the required utility, especially when there is concern that higher privacy will result in lower utility.
(3) Use real-world data to determine parameter settings for case studies.
(4) Evaluate several parameter settings to analyze how the parameter values influence privacy.
% this seems a bit circular
(5) Finally, we recommend to also include metrics that do not have parameters.

\section{Future Research Directions}
\label{sec:futurework}

Despite the substantial body of research into privacy metrics presented in the previous sections, there are a number of questions that merit further research.

\subsection{Interdependent Privacy}

% metrics that measure the impact and severity of privacy breaches
% effects of privacy breaches analogous to Richter vs. Mercalli scale
% Returning to our earlier analogy of seismology, we observe that most of the privacy metrics that have been proposed focus on measuring the equivalent of released energy. 
% Not much work has been done on measuring effects, such as the impact and severity of privacy breaches. 
% We admit that this would be extremely difficult to quantify; however, analogous to the Mercalli scale, it may be helpful to have a natural language description of the scale in order to rate privacy incidents after they have occurred.
% Such a scale could encompass the number of individuals concerned and the nature of the damages (e.g., embarrassment, financial, or physical).

% Among the privacy metrics we presented in our review, there were no metrics suitable for measuring interdependent privacy.
% 
Interdependent privacy refers to scenarios in which actions of one user affect the privacy of other users, for example in social networks~\cite{thomas_unfriendly_2010}, location privacy~\cite{vratonjic_how_2013}, or genome privacy~\cite{humbert_addressing_2013}. 
There are two options for measuring interdependent privacy.
The first option is to measure how the value of an existing privacy metric changes when the degree of interdependency increases.
The effect of interdependency can then be shown by comparing absolute values~\cite{bloessl_scrambler_2015}, or by computing a difference~\cite{olteanu_quantifying_2014}.

% For the first option, we need to consider measurement scales (see Section~\ref{sec:scales}).
% The scale indicates how privacy values under varying levels of interdependency can be compared: by stating which value is greater/smaller (ordinal scale), by stating the difference between two values (interval scale), or by stating the ratio between two values (ratio scale).

The second option is to create new metrics that explicitly consider interdependency.
In this case, it can be beneficial to make use of metrics that measure the consequences that one user's actions have on the privacy of another user.
For example, this is done in game theory, where the widely used Helly metric~\cite{vorob1977infinite} assesses players' strategies in terms of their consequences which are the payoffs for each player.
We believe further research is needed to investigate the capabilities of these two options.

\subsection{Privacy Attitudes and Behaviors}
\label{sec:future-work-attitudes}
In this survey, we focused on technical privacy metrics and did not consider metrics that measure users' privacy attitudes, behaviors, or perception \cite{preibusch_guide_2013}.
User-assigned privacy or privacy risk scores vary greatly in how information is collected from the user.
For example, some studies measure users' perception of privacy risks or privacy attitudes on Likert scales~\cite{acquisti_economics_2003,achara_short_2014}. 
Others require users to label sensitive data~\cite{zhang_sedic:_2011}, assign privacy scores to their credentials~\cite{yao_private_2008}, or configure existing mechanisms according to their privacy needs~\cite{xiao_personalized_2006}. 
Some studies work with risk attitudes that are inferred from user actions via machine learning~\cite{akcora_privacy_2012}.

Some metrics in our survey combine a technical metric with parameters that are specified by users to reflect their preferences, for example user-centric privacy (Section~\ref{metric:user-centric-privacy}), coverage of sensitive region (Section~\ref{metric:coverage-of-sensitive-region}), or privacy score (Section~\ref{metric:privacy-score}).
In general, however, it is an open question how best to integrate user attitudes, behaviors, or perceptions with technical metrics.
In addition, it is unclear whether this integration is generally useful, and which scenarios would benefit most.

% Useful starting points are: 
% \cite{preibusch_guide_2013} % * privacy attitude measurement and behavioral metrics 
% \cite{albert_measuring_2013} % * user experience measurement book/papers 

% user risk attitudes that are inferred by machine learning (akcora2012privacy)
% perception of privacy risk on scale of 1 (low risk) to 10 (high risk) (achara2014wifileaks)
% users label sensitive data (zhang2011sedic)
% users value their credentials with privacy scores, indicating how reluctant they are to reveal them; scores may also be classified in different types (addresses, financial, etc.) (yao2008private)
% privacy attitudes on 7-point Likert scale in different domains: privacy policy, importance in daily life, and concern (acquisti2006imagined)

\subsection{Aggregating Metrics}

% \todo[inline,color=blue!20]{Reviewer 2: page 26, line 46, section 7.3: This is sometimes called "composing"
% rather than "aggregating" metrics.}

In scenarios with a large number of entities, such as thousands of genomic variations or users in a communication system, it can be beneficial to aggregate (or \textit{compose}), metrics.
Some metrics in our survey attempt to do this, for example cumulative entropy (Section~\ref{metric:cumulative-entropy}), genomic privacy (Section~\ref{metric:genomic-privacy}), health privacy (Section~\ref{metric:health-privacy}), or expected estimation error (Section~\ref{metric:expected-estimation-error}).
All of these metrics are based on an addition of privacy values.
Their results are a sum (cumulative entropy, genomic privacy), a weighted arithmetic mean (health privacy), or an expected value (expected estimation error).
% However, sums or arithmetic means are not allowed on all measurement scales (see Section~\ref{sec:scales}).
% As an example, consider the situation in benchmarking, i.e. the comparison of computer performance measurements.
% Research in this area argues that 
% On the ordinal scale, only the median and mode (most frequent value) are allowed.
% The interval scale additionally allows the arithmetic mean, and only the ratio scale allows the geometric and harmonic means.
However, depending on the distribution of the underlying population, the arithmetic mean may lead to biased results~\cite{mashey_war_2004}.
In some situations, a geometric mean is preferable because it assumes a log-normal, rather than normal, distribution, and is less biased by outlier values~\cite{citron_harmonic_2006}.
% Benchmark suites like SPEC use the geometric mean, because it assumes a log-normal, rather than normal, distribution, and is less biased by outlier values~\cite{citron_harmonic_2006}.
% In the field of privacy measurement, however, a similar discussion has not yet taken place.
However, in the field of privacy measurement it is not clear what these situations are.
We therefore believe that privacy research would benefit from a rigorous study of ways to aggregate metrics. %, especially considering the different ways to compute means, and the situations in which they can be applied.

Another option to aggregate privacy values is visualization.
When metrics are visualized, a common option is to display averages -- the same strategy as with aggregate metrics.
However, more sophisticated plot types can highlight issues such as fairness that are hidden when averages are used.
For example, box plots display the smallest and largest privacy values as well as the first, second, and third quartile; violin plots add kernel density plots to visualize the distribution of privacy values.
These plots give more information than aggregate metrics; however, it is unclear how aggregate metrics can be designed so that the benefits of these plots are preserved. 

% simple addition may not lead to useful insight (e.g., 5 cars in one mix zone vs. 100 in another, but has been done: cumulative entropy), so at least normalization may be needed in addition to aggregation

\subsection{Combining Metrics}

Whereas the aggregation of metrics considers values of the same privacy metric for many entities, the combination of metrics considers values of different privacy metrics for one entity.
Combining different metrics can be useful if the combination retains the strengths of each metric while reducing their weaknesses.
It can also simplify interpretation to express the performance of a \ac{PET} with a single number.
Metrics in our survey use three methods to combine metrics: adding sensitivity scores, normalizing metrics, and extending metrics to new contexts.

Metrics that combine a sensitivity score with a technical metric are user-centric privacy (using a linear combination, Section~\ref{metric:user-centric-privacy}) and privacy score (using sensitivity as a weighting factor, Section~\ref{metric:privacy-score}).
As mentioned in Section~\ref{sec:future-work-attitudes} above, it is not clear how sensitivity scores and technical metrics can best be combined.
In addition, it is not clear whether the resulting values have a meaningful interpretation.

Metrics that combine two technical metrics typically use one metric to normalize another, for example normalized entropy (Section~\ref{metric:normalized-entropy}), normalized mutual information (Section~\ref{metric:normalized-mutual-information}), or reduction in observable features (Section~\ref{metric:reduction-in-observable-features}).
Normalization can make it easier to interpret privacy measurements, but for some metrics, is is not clear if and how they can be normalized, or which normalization method works best.

Metrics that adapt a privacy metric so that it can be used in a new context are computational differential privacy (Section~\ref{metric:computational-differential-privacy}) which adapts differential privacy to a new adversary type, and entropy combined with Bayesian belief tables to apply entropy across multiple time-steps (Section~\ref{metric:entropy}).
These innovative metrics raise two questions: first, whether their mechanisms can extend the range of use for other metrics as well, and second, whether there are other mechanisms that can be used in a similar way to adapt existing metrics to new use cases.

% The metrics in the combination category all produce one integrated number by mathematically combining their respective source metrics using one of two approaches:
% the first approach uses general-purpose solutions, such as fractions~\cite{ardagna_location_2007,cheng_preserving_2006}, weighted sums~\cite{humbert_addressing_2013}, or linear combinations~\cite{freudiger_non-cooperative_2009}.
% The second approach uses metric-specific mechanisms to merge the two metrics, for example by updating the probabilities to compute entropy using Bayesian belief tables~\cite{ma_measuring_2010}, or by replacing a privacy mechanism with a computational version~\cite{mironov_computational_2009,bernhard_measuring_2012}.
% bertino_survey_2008 propose a weighted mean between a privacy metric and a utility metric
% These approaches simplify interpretation because they sum up the performance of a \ac{PET} with a single number.
%In addition, especially for combinations with indistinguishability-based metrics, this merging of two metrics is sometimes the only viable alternative.
%For example,~\cite{bertino_survey_2008} propose to combine a privacy metric with a utility metric using a weighted mean.

\subsection{Quality of Metrics}

We presented a number of quality indicators for privacy metrics in Section~\ref{sec:requirements}.
While there is a general consensus that high-quality metrics should be used, there is no consensus what exactly constitutes high quality and how it should be measured.
As a result, there are few studies investigating the quality of privacy metrics.
For example, in a previous study, we systematically compared 22 metrics for genome privacy and found that metrics varied greatly with regard to consistency and monotonicity~\cite{wagner2015genomic,wagner2017evaluating}. 
Although our study yielded good results for a selection of privacy metrics in one specific scenario, it was limited in terms of the scenario, quality indicators, and number of privacy metrics.
It is unclear whether the results of our study would hold in general, and therefore we believe that more studies are needed that rigorously evaluate the quality and thus the meaningfulness of privacy metrics.

\end{screenonly}

\section{Conclusion}
\label{sec:conclusion}
In this survey we presented a comprehensive review of privacy metrics.
We described and discussed a selection of over eighty privacy metrics using examples from six different privacy domains.

To structure the complex landscape of privacy metrics, we introduced categorizations based on the aspect of privacy they measure, their required inputs, and the type of data that needs protection.
In addition, we highlighted topics where we believe additional work on privacy metrics is needed.
This includes research toward the combination and aggregation of privacy metrics as well as the field of interdependent privacy.

Finally, we presented a method on how to choose privacy metrics based on nine questions that help identify the right privacy metrics for a given scenario.
Most importantly, we argue for the selection of multiple metrics to cover multiple aspects of privacy.
We believe that our systematization will serve as a reference guide for privacy metrics that allows informed choices of suitable privacy metrics and thus serves as a useful toolbox for privacy researchers.

\section*{Acknowledgements}
David Eckhoff is financially supported by the Singapore National Research Foundation under its Campus for Research Excellence And Technological Enterprise (CREATE) programme.



\bibliographystyle{ACM-Reference-Format}
\bibliography{sok-privacymetrics,temp}


\begin{thebibliography}{162}

%%% the Web address.

\ifx \showCODEN    \undefined \def \showCODEN     #1{\unskip}     \fi
\ifx \showDOI      \undefined \def \showDOI       #1{#1}\fi
\ifx \showISBNx    \undefined \def \showISBNx     #1{\unskip}     \fi
\ifx \showISBNxiii \undefined \def \showISBNxiii  #1{\unskip}     \fi
\ifx \showISSN     \undefined \def \showISSN      #1{\unskip}     \fi
\ifx \showLCCN     \undefined \def \showLCCN      #1{\unskip}     \fi
\ifx \shownote     \undefined \def \shownote      #1{#1}          \fi
\ifx \showarticletitle \undefined \def \showarticletitle #1{#1}   \fi
\ifx \showURL      \undefined \def \showURL       {\relax}        \fi
% invisible to TeX
\providecommand\bibfield[2]{#2}
\providecommand\bibinfo[2]{#2}
\providecommand\natexlab[1]{#1}
\providecommand\showeprint[2][]{arXiv:#2}

\bibitem[\protect\citeauthoryear{Abadi, Chu, Goodfellow, McMahan, Mironov,
  Talwar, and Zhang}{Abadi et~al\mbox{.}}{2016}]%
        {abadi_deep_2016}
\bibfield{author}{\bibinfo{person}{Martin Abadi}, \bibinfo{person}{Andy Chu},
  \bibinfo{person}{Ian Goodfellow}, \bibinfo{person}{H.~Brendan McMahan},
  \bibinfo{person}{Ilya Mironov}, \bibinfo{person}{Kunal Talwar}, {and}
  \bibinfo{person}{Li Zhang}.} \bibinfo{year}{2016}\natexlab{}.
\newblock \showarticletitle{Deep {Learning} with {Differential} {Privacy}}. In
  \bibinfo{booktitle}{\emph{Proceedings of the 2016 {ACM} {SIGSAC} {Conference}
  on {Computer} and {Communications} {Security}}}. \bibinfo{publisher}{ACM},
  \bibinfo{address}{Vienna, Austria}, \bibinfo{pages}{308--318}.
\newblock
\showISBNx{978-1-4503-4139-4}


\bibitem[\protect\citeauthoryear{Achara, Cunche, Roca, and Francillon}{Achara
  et~al\mbox{.}}{2014}]%
        {achara_short_2014}
\bibfield{author}{\bibinfo{person}{Jagdish~Prasad Achara},
  \bibinfo{person}{Mathieu Cunche}, \bibinfo{person}{Vincent Roca}, {and}
  \bibinfo{person}{Aurélien Francillon}.} \bibinfo{year}{2014}\natexlab{}.
\newblock \showarticletitle{{WifiLeaks: Underestimated Privacy Implications of
  the ACCESS\_WIFI\_STATE Android Permission}}. In
  \bibinfo{booktitle}{\emph{Proc. 7th ACM Conf. on Security and Privacy in
  Wireless and Mobile Networks (WiSec 2014)}}. \bibinfo{publisher}{{ACM}},
  \bibinfo{address}{Oxford, UK}, \bibinfo{pages}{231--236}.
\newblock
\showISBNx{978-1-4503-2972-9}


\bibitem[\protect\citeauthoryear{Acquisti, Dingledine, and Syverson}{Acquisti
  et~al\mbox{.}}{2003}]%
        {acquisti_economics_2003}
\bibfield{author}{\bibinfo{person}{Alessandro Acquisti}, \bibinfo{person}{Roger
  Dingledine}, {and} \bibinfo{person}{Paul Syverson}.}
  \bibinfo{year}{2003}\natexlab{}.
\newblock \showarticletitle{{On the Economics of Anonymity}}. In
  \bibinfo{booktitle}{\emph{Proc. 7th Int. Financial Cryptography Conf
  (FC03)}}. \bibinfo{publisher}{Springer}, \bibinfo{address}{Gosier,
  Guadeloupe}, \bibinfo{pages}{84--102}.
\newblock


\bibitem[\protect\citeauthoryear{Aggarwal}{Aggarwal}{2005}]%
        {aggarwal_k-anonymity_2005}
\bibfield{author}{\bibinfo{person}{Charu~C. Aggarwal}.}
  \bibinfo{year}{2005}\natexlab{}.
\newblock \showarticletitle{{On k-Anonymity and the Curse of Dimensionality}}.
  In \bibinfo{booktitle}{\emph{Proc. 31st Int. Conf. on Very Large Data Bases
  (VLDB 2005)}}. \bibinfo{publisher}{{VLDB} Endowment},
  \bibinfo{address}{Trondheim, Norway}, \bibinfo{pages}{901--909}.
\newblock


\bibitem[\protect\citeauthoryear{Agrawal and Aggarwal}{Agrawal and
  Aggarwal}{2001}]%
        {agrawal_design_2001}
\bibfield{author}{\bibinfo{person}{Dakshi Agrawal} {and}
  \bibinfo{person}{Charu~C. Aggarwal}.} \bibinfo{year}{2001}\natexlab{}.
\newblock \showarticletitle{{On the Design and Quantification of Privacy
  Preserving Data Mining Algorithms}}. In \bibinfo{booktitle}{\emph{Proc. 20th
  {ACM} {SIGMOD}-{SIGACT}-{SIGART} Symp. on Principles of Database Systems
  (PODS 2001)}}. \bibinfo{publisher}{{ACM}}, \bibinfo{address}{Santa Barbara,
  CA, USA}, \bibinfo{pages}{247--255}.
\newblock


\bibitem[\protect\citeauthoryear{Agrawal and Kesdogan}{Agrawal and
  Kesdogan}{2003}]%
        {agrawal_measuring_2003}
\bibfield{author}{\bibinfo{person}{Dakshi Agrawal} {and} \bibinfo{person}{Dogan
  Kesdogan}.} \bibinfo{year}{2003}\natexlab{}.
\newblock \showarticletitle{{Measuring Anonymity: The Disclosure Attack}}.
\newblock \bibinfo{journal}{\emph{IEEE Security \& Privacy}}
  \bibinfo{volume}{1}, \bibinfo{number}{6} (\bibinfo{date}{November}
  \bibinfo{year}{2003}), \bibinfo{pages}{27--34}.
\newblock
\showISSN{1540-7993}


\bibitem[\protect\citeauthoryear{Agrawal and Srikant}{Agrawal and
  Srikant}{2000}]%
        {agrawal_privacy-preserving_2000}
\bibfield{author}{\bibinfo{person}{Rakesh Agrawal} {and}
  \bibinfo{person}{Ramakrishnan Srikant}.} \bibinfo{year}{2000}\natexlab{}.
\newblock \showarticletitle{{Privacy-preserving Data Mining}}. In
  \bibinfo{booktitle}{\emph{Proc. 2000 ACM SIGMOD Int. Conf. on Management of
  Data (SIGMOD'00)}}. \bibinfo{publisher}{{ACM}}, \bibinfo{address}{Dallas, TX,
  USA}, \bibinfo{pages}{439--450}.
\newblock
\showISBNx{1-58113-217-4}


\bibitem[\protect\citeauthoryear{Akcora, Carminati, and Ferrari}{Akcora
  et~al\mbox{.}}{2012}]%
        {akcora_privacy_2012}
\bibfield{author}{\bibinfo{person}{Cuneyt~Gurcan Akcora},
  \bibinfo{person}{Barbara Carminati}, {and} \bibinfo{person}{Elena Ferrari}.}
  \bibinfo{year}{2012}\natexlab{}.
\newblock \showarticletitle{{Privacy in Social Networks: How Risky is Your
  Social Graph?}}. In \bibinfo{booktitle}{\emph{Proc {IEEE} 28th Int. Conf. on
  Data Engineering (ICDE'12)}}. \bibinfo{publisher}{{IEEE}},
  \bibinfo{address}{Washington, DC, USA}, \bibinfo{pages}{9--19}.
\newblock


\bibitem[\protect\citeauthoryear{Alexander and Smith}{Alexander and
  Smith}{2003}]%
        {alexander_engineering_2003}
\bibfield{author}{\bibinfo{person}{James Alexander} {and}
  \bibinfo{person}{Jonathan Smith}.} \bibinfo{year}{2003}\natexlab{}.
\newblock \showarticletitle{{Engineering Privacy in Public: Confounding Face
  Recognition}}. In \bibinfo{booktitle}{\emph{Proc. 3rd Int. Workshop on
  Privacy Enhancing Technologies (PET 2003)}} \emph{(\bibinfo{series}{LNCS
  2760})}. \bibinfo{publisher}{Springer}, \bibinfo{address}{Dresden, Germany},
  \bibinfo{pages}{88--106}.
\newblock
\showISBNx{978-3-540-20610-1, 978-3-540-40956-4}


\bibitem[\protect\citeauthoryear{Andersson and Lundin}{Andersson and
  Lundin}{2008}]%
        {andersson_fundamentals_2008}
\bibfield{author}{\bibinfo{person}{Christer Andersson} {and}
  \bibinfo{person}{Reine Lundin}.} \bibinfo{year}{2008}\natexlab{}.
\newblock \showarticletitle{{On the Fundamentals of Anonymity Metrics}}. In
  \bibinfo{booktitle}{\emph{{Proc. 3rd IFIP Int. Summer School on The Future of
  Identity in the Information Society}}}. \bibinfo{publisher}{Springer},
  \bibinfo{address}{Karlstad, Sweden}, \bibinfo{pages}{325--341}.
\newblock


\bibitem[\protect\citeauthoryear{Andrés, Bordenabe, Chatzikokolakis, and
  Palamidessi}{Andrés et~al\mbox{.}}{2013}]%
        {andres_geo-indistinguishability:_2013}
\bibfield{author}{\bibinfo{person}{Miguel~E. Andrés},
  \bibinfo{person}{Nicolás~E. Bordenabe}, \bibinfo{person}{Konstantinos
  Chatzikokolakis}, {and} \bibinfo{person}{Catuscia Palamidessi}.}
  \bibinfo{year}{2013}\natexlab{}.
\newblock \showarticletitle{{Geo-Indistinguishability: Differential Privacy for
  Location-Based Systems}}. In \bibinfo{booktitle}{\emph{Proc. 20th ACM Conf.
  on Computer and Communications Security (CCS'13)}}.
  \bibinfo{publisher}{{ACM}}, \bibinfo{address}{Berlin, Germany},
  \bibinfo{pages}{901--914}.
\newblock
\showISBNx{978-1-4503-2477-9}


\bibitem[\protect\citeauthoryear{Arapinis, Mancini, Ritter, Ryan, Golde, Redon,
  and Borgaonkar}{Arapinis et~al\mbox{.}}{2012}]%
        {arapinis_new_2012}
\bibfield{author}{\bibinfo{person}{Myrto Arapinis}, \bibinfo{person}{Loretta
  Mancini}, \bibinfo{person}{Eike Ritter}, \bibinfo{person}{Mark Ryan},
  \bibinfo{person}{Nico Golde}, \bibinfo{person}{Kevin Redon}, {and}
  \bibinfo{person}{Ravishankar Borgaonkar}.} \bibinfo{year}{2012}\natexlab{}.
\newblock \showarticletitle{{New Privacy Issues in Mobile Telephony: Fix and
  Verification}}. In \bibinfo{booktitle}{\emph{Proc. 19th {ACM} Conf. on
  Computer and Communications Security (CCS'12)}}. \bibinfo{publisher}{{ACM}},
  \bibinfo{address}{Raleigh, NC, USA}, \bibinfo{pages}{205--216}.
\newblock
\showISBNx{978-1-4503-1651-4}


\bibitem[\protect\citeauthoryear{Ardagna, Cremonini, Damiani, Di~Vimercati, and
  Samarati}{Ardagna et~al\mbox{.}}{2007}]%
        {ardagna_location_2007}
\bibfield{author}{\bibinfo{person}{Claudio~Agostino Ardagna},
  \bibinfo{person}{Marco Cremonini}, \bibinfo{person}{Ernesto Damiani},
  \bibinfo{person}{S.~De~Capitani Di~Vimercati}, {and}
  \bibinfo{person}{Pierangela Samarati}.} \bibinfo{year}{2007}\natexlab{}.
\newblock \showarticletitle{{Location Privacy Protection Through
  Obfuscation-based Techniques}}. In \bibinfo{booktitle}{\emph{Data and
  Applications Security XXI: 21st Annu. IFIP Working Conf. on Data and
  Applications Security}}. \bibinfo{publisher}{Springer},
  \bibinfo{address}{Redondo Beach, CA, USA}, \bibinfo{pages}{47--60}.
\newblock


\bibitem[\protect\citeauthoryear{Ayday, Raisaro, and Hubaux}{Ayday
  et~al\mbox{.}}{2013a}]%
        {ayday_personal_2013}
\bibfield{author}{\bibinfo{person}{Erman Ayday}, \bibinfo{person}{Jean~Louis
  Raisaro}, {and} \bibinfo{person}{Jean-Pierre Hubaux}.}
  \bibinfo{year}{2013}\natexlab{a}.
\newblock \showarticletitle{{Personal Use of the Genomic Data: Privacy vs.
  Storage Cost}}. In \bibinfo{booktitle}{\emph{Proc. {IEEE} Global
  Communications Conf. (GLOBECOM 2013)}}. \bibinfo{publisher}{IEEE},
  \bibinfo{address}{Atlanta, GA, USA}, \bibinfo{pages}{2723--2729}.
\newblock


\bibitem[\protect\citeauthoryear{Ayday, Raisaro, Hubaux, and Rougemont}{Ayday
  et~al\mbox{.}}{2013b}]%
        {ayday_protecting_2013}
\bibfield{author}{\bibinfo{person}{Erman Ayday}, \bibinfo{person}{Jean~Louis
  Raisaro}, \bibinfo{person}{Jean-Pierre Hubaux}, {and}
  \bibinfo{person}{Jacques Rougemont}.} \bibinfo{year}{2013}\natexlab{b}.
\newblock \showarticletitle{{Protecting and Evaluating Genomic Privacy in
  Medical Tests and Personalized Medicine}}. In \bibinfo{booktitle}{\emph{Proc.
  12th {ACM} Workshop on Workshop on Privacy in the Electronic Society
  (WPES'13)}}. \bibinfo{publisher}{{ACM}}, \bibinfo{address}{Berlin, Germany},
  \bibinfo{pages}{95--106}.
\newblock
\showISBNx{978-1-4503-2485-4}


\bibitem[\protect\citeauthoryear{Backes, Lorenz, Maffei, and Pecina}{Backes
  et~al\mbox{.}}{2010}]%
        {backes_anonymous_2010}
\bibfield{author}{\bibinfo{person}{Michael Backes}, \bibinfo{person}{Stefan
  Lorenz}, \bibinfo{person}{Matteo Maffei}, {and} \bibinfo{person}{Kim
  Pecina}.} \bibinfo{year}{2010}\natexlab{}.
\newblock \showarticletitle{{Anonymous Webs of Trust}}. In
  \bibinfo{booktitle}{\emph{Proc. 10th Int. Symp. on Privacy Enhancing
  Technologies (PETS 2010)}} \emph{(\bibinfo{series}{LNCS 6205})}.
  \bibinfo{publisher}{Springer}, \bibinfo{address}{Berlin, Germany},
  \bibinfo{pages}{130--148}.
\newblock
\showISBNx{978-3-642-14526-1, 978-3-642-14527-8}


\bibitem[\protect\citeauthoryear{Backstrom, Dwork, and Kleinberg}{Backstrom
  et~al\mbox{.}}{2007}]%
        {backstrom_wherefore_2007}
\bibfield{author}{\bibinfo{person}{Lars Backstrom}, \bibinfo{person}{Cynthia
  Dwork}, {and} \bibinfo{person}{Jon Kleinberg}.}
  \bibinfo{year}{2007}\natexlab{}.
\newblock \showarticletitle{{Wherefore Art Thou R3579x?: Anonymized Social
  Networks, Hidden Patterns, and Structural Steganography}}. In
  \bibinfo{booktitle}{\emph{16th Int. Conf. on World Wide Web}}.
  \bibinfo{publisher}{ACM}, \bibinfo{address}{Banff, Canada},
  \bibinfo{pages}{181--190}.
\newblock


\bibitem[\protect\citeauthoryear{Beimel, Nissim, and Stemmer}{Beimel
  et~al\mbox{.}}{2013}]%
        {beimel2013private}
\bibfield{author}{\bibinfo{person}{Amos Beimel}, \bibinfo{person}{Kobbi
  Nissim}, {and} \bibinfo{person}{Uri Stemmer}.}
  \bibinfo{year}{2013}\natexlab{}.
\newblock \showarticletitle{Private {{Learning}} and {{Sanitization}}: {{Pure}}
  vs. {{Approximate Differential Privacy}}}.
\newblock In \bibinfo{booktitle}{\emph{Approximation, {{Randomization}}, and
  {{Combinatorial Optimization}}. {{Algorithms}} and {{Techniques}}}},
  \bibfield{editor}{\bibinfo{person}{Prasad Raghavendra},
  \bibinfo{person}{Sofya Raskhodnikova}, \bibinfo{person}{Klaus Jansen}, {and}
  \bibinfo{person}{Jos{\'e} D.~P. Rolim}} (Eds.). Number 8096 in
  \bibinfo{series}{Lecture Notes in Computer Science}.
  \bibinfo{publisher}{{Springer Berlin Heidelberg}}, \bibinfo{pages}{363--378}.
\newblock


\bibitem[\protect\citeauthoryear{Bertino, Lin, and Jiang}{Bertino
  et~al\mbox{.}}{2008}]%
        {bertino_survey_2008}
\bibfield{author}{\bibinfo{person}{Elisa Bertino}, \bibinfo{person}{Dan Lin},
  {and} \bibinfo{person}{Wei Jiang}.} \bibinfo{year}{2008}\natexlab{}.
\newblock \showarticletitle{{A Survey of Quantification of Privacy Preserving
  Data Mining Algorithms}}.
\newblock In \bibinfo{booktitle}{\emph{Privacy-Preserving Data Mining: Models
  and Algorithms}}. Number~34 in \bibinfo{series}{Advances in Database
  Systems}. \bibinfo{publisher}{Springer}, Chapter~8,
  \bibinfo{pages}{183--205}.
\newblock
\showISBNx{978-0-387-70991-8, 978-0-387-70992-5}


\bibitem[\protect\citeauthoryear{Bettini, Wang, and Jajodia}{Bettini
  et~al\mbox{.}}{2005}]%
        {bettini_protecting_2005}
\bibfield{author}{\bibinfo{person}{Claudio Bettini}, \bibinfo{person}{X.~Sean
  Wang}, {and} \bibinfo{person}{Sushil Jajodia}.}
  \bibinfo{year}{2005}\natexlab{}.
\newblock \showarticletitle{{Protecting Privacy Against Location-based Personal
  Identification}}. In \bibinfo{booktitle}{\emph{2nd VLDB Works. on Secure Data
  Management}} \emph{(\bibinfo{series}{LNCS 3674})}.
  \bibinfo{publisher}{Springer}, \bibinfo{address}{Trondheim, Norway},
  \bibinfo{pages}{185--199}.
\newblock


\bibitem[\protect\citeauthoryear{Bezzi}{Bezzi}{2010}]%
        {bezzi2010information}
\bibfield{author}{\bibinfo{person}{Michele Bezzi}.}
  \bibinfo{year}{2010}\natexlab{}.
\newblock \showarticletitle{An {{Information Theoretic Approach}} for {{Privacy
  Metrics}}}.
\newblock \bibinfo{journal}{\emph{Trans. Data Privacy}} \bibinfo{volume}{3},
  \bibinfo{number}{3} (\bibinfo{year}{2010}), \bibinfo{pages}{199--215}.
\newblock
\showISSN{1888-5063}


\bibitem[\protect\citeauthoryear{Bloessl, Sommer, Dressler, and
  Eckhoff}{Bloessl et~al\mbox{.}}{2015}]%
        {bloessl_scrambler_2015}
\bibfield{author}{\bibinfo{person}{Bastian Bloessl}, \bibinfo{person}{Christoph
  Sommer}, \bibinfo{person}{Falko Dressler}, {and} \bibinfo{person}{David
  Eckhoff}.} \bibinfo{year}{2015}\natexlab{}.
\newblock \showarticletitle{The scrambler attack: {A} robust physical layer
  attack on location privacy in vehicular networks}. In
  \bibinfo{booktitle}{\emph{2015 {International} {Conference} on {Computing},
  {Networking} and {Communications} ({ICNC})}}. \bibinfo{pages}{395--400}.
\newblock


\bibitem[\protect\citeauthoryear{Blum, Ligett, and Roth}{Blum
  et~al\mbox{.}}{2013}]%
        {blum_learning_2013}
\bibfield{author}{\bibinfo{person}{Avrim Blum}, \bibinfo{person}{Katrina
  Ligett}, {and} \bibinfo{person}{Aaron Roth}.}
  \bibinfo{year}{2013}\natexlab{}.
\newblock \showarticletitle{A learning theory approach to noninteractive
  database privacy}.
\newblock \bibinfo{journal}{\emph{Journal of the ACM (JACM)}}
  \bibinfo{volume}{60}, \bibinfo{number}{2} (\bibinfo{year}{2013}),
  \bibinfo{pages}{12}.
\newblock


\bibitem[\protect\citeauthoryear{Chatzikokolakis, Andr{\'e}s, Bordenabe, and
  Palamidessi}{Chatzikokolakis et~al\mbox{.}}{2013}]%
        {chatzikokolakis2013broadening}
\bibfield{author}{\bibinfo{person}{Konstantinos Chatzikokolakis},
  \bibinfo{person}{Miguel~E. Andr{\'e}s}, \bibinfo{person}{Nicol{\'a}s~Emilio
  Bordenabe}, {and} \bibinfo{person}{Catuscia Palamidessi}.}
  \bibinfo{year}{2013}\natexlab{}.
\newblock \showarticletitle{Broadening the {{Scope}} of {{Differential Privacy
  Using Metrics}}}.
\newblock In \bibinfo{booktitle}{\emph{Privacy {{Enhancing Technologies}}}},
  \bibfield{editor}{\bibinfo{person}{Emiliano~De Cristofaro} {and}
  \bibinfo{person}{Matthew Wright}} (Eds.). Number 7981 in
  \bibinfo{series}{Lecture Notes in Computer Science}.
  \bibinfo{publisher}{{Springer Berlin Heidelberg}}, \bibinfo{pages}{82--102}.
\newblock
\showISBNx{978-3-642-39076-0 978-3-642-39077-7}


\bibitem[\protect\citeauthoryear{Chatzikokolakis, Palamidessi, and
  Panangaden}{Chatzikokolakis et~al\mbox{.}}{2007}]%
        {chatzikokolakis_anonymity_2007}
\bibfield{author}{\bibinfo{person}{Konstantinos Chatzikokolakis},
  \bibinfo{person}{Catuscia Palamidessi}, {and} \bibinfo{person}{Prakash
  Panangaden}.} \bibinfo{year}{2007}\natexlab{}.
\newblock \showarticletitle{{Anonymity Protocols as Noisy Channels}}. In
  \bibinfo{booktitle}{\emph{Proc. 3rd Int. Symp. Trustworthy Global Computing
  (TGC'2007)}}. \bibinfo{address}{Sophia-Antipolis, France},
  \bibinfo{pages}{281--300}.
\newblock


\bibitem[\protect\citeauthoryear{Chatzikokolakis, Palamidessi, and
  Stronati}{Chatzikokolakis et~al\mbox{.}}{2015}]%
        {chatzikokolakis2015constructing}
\bibfield{author}{\bibinfo{person}{Konstantinos Chatzikokolakis},
  \bibinfo{person}{Catuscia Palamidessi}, {and} \bibinfo{person}{Marco
  Stronati}.} \bibinfo{year}{2015}\natexlab{}.
\newblock \showarticletitle{Constructing Elastic Distinguishability Metrics for
  Location Privacy}.
\newblock \bibinfo{journal}{\emph{Proceedings on Privacy Enhancing
  Technologies}} \bibinfo{volume}{2015}, \bibinfo{number}{2}
  (\bibinfo{year}{2015}), \bibinfo{pages}{156--170}.
\newblock
\showISSN{2299-0984}


\bibitem[\protect\citeauthoryear{Chaum}{Chaum}{1988}]%
        {chaum_dining_1988}
\bibfield{author}{\bibinfo{person}{David Chaum}.}
  \bibinfo{year}{1988}\natexlab{}.
\newblock \showarticletitle{{The Dining Cryptographers Problem: Unconditional
  Sender and Recipient Untraceability}}.
\newblock \bibinfo{journal}{\emph{Journal of Cryptology}} \bibinfo{volume}{1},
  \bibinfo{number}{1} (\bibinfo{date}{January} \bibinfo{year}{1988}),
  \bibinfo{pages}{65--75}.
\newblock
\showISSN{0933-2790, 1432-1378}


\bibitem[\protect\citeauthoryear{Chawla, Dwork, McSherry, Smith, and
  Wee}{Chawla et~al\mbox{.}}{2005}]%
        {chawla_toward_2005}
\bibfield{author}{\bibinfo{person}{Shuchi Chawla}, \bibinfo{person}{Cynthia
  Dwork}, \bibinfo{person}{Frank McSherry}, \bibinfo{person}{Adam Smith}, {and}
  \bibinfo{person}{Hoeteck Wee}.} \bibinfo{year}{2005}\natexlab{}.
\newblock \showarticletitle{{Toward Privacy in Public Databases}}. In
  \bibinfo{booktitle}{\emph{Proc. 2nd Int. Conf. on Theory of Cryptography
  (TCC'05)}}. \bibinfo{publisher}{Springer}, \bibinfo{address}{Cambridge, MA,
  USA}, \bibinfo{pages}{363--385}.
\newblock


\bibitem[\protect\citeauthoryear{Chen, Chaabane, Tournoux, Kaafar, and
  Boreli}{Chen et~al\mbox{.}}{2013}]%
        {chen_how_2013}
\bibfield{author}{\bibinfo{person}{Terence Chen}, \bibinfo{person}{Abdelberi
  Chaabane}, \bibinfo{person}{Pierre~Ugo Tournoux},
  \bibinfo{person}{Mohamed-Ali Kaafar}, {and} \bibinfo{person}{Roksana
  Boreli}.} \bibinfo{year}{2013}\natexlab{}.
\newblock \showarticletitle{{How Much Is Too Much? Leveraging Ads Audience
  Estimation to Evaluate Public Profile Uniqueness}}. In
  \bibinfo{booktitle}{\emph{Proc. 13th Int. Symp. on Privacy Enhancing
  Technologies (PETS 2013)}} \emph{(\bibinfo{series}{LNCS 7981})}.
  \bibinfo{publisher}{Springer}, \bibinfo{address}{Bloomington, IN, USA},
  \bibinfo{pages}{225--244}.
\newblock
\showISBNx{978-3-642-39076-0, 978-3-642-39077-7}


\bibitem[\protect\citeauthoryear{Chen and Pang}{Chen and Pang}{2012}]%
        {chen_measuring_2012}
\bibfield{author}{\bibinfo{person}{Xihui Chen} {and} \bibinfo{person}{Jun
  Pang}.} \bibinfo{year}{2012}\natexlab{}.
\newblock \showarticletitle{{Measuring Query Privacy in Location-based
  Services}}. In \bibinfo{booktitle}{\emph{Proc. 2nd {ACM} Conf. on Data and
  Application Security and Privacy (CODASPY'12)}}. \bibinfo{publisher}{{ACM}},
  \bibinfo{address}{San Antonio, TX, USA}, \bibinfo{pages}{49--60}.
\newblock
\showISBNx{978-1-4503-1091-8}


\bibitem[\protect\citeauthoryear{Cheng, Zhang, Bertino, and Prabhakar}{Cheng
  et~al\mbox{.}}{2006}]%
        {cheng_preserving_2006}
\bibfield{author}{\bibinfo{person}{Reynold Cheng}, \bibinfo{person}{Yu Zhang},
  \bibinfo{person}{Elisa Bertino}, {and} \bibinfo{person}{Sunil Prabhakar}.}
  \bibinfo{year}{2006}\natexlab{}.
\newblock \showarticletitle{{Preserving User Location Privacy in Mobile Data
  Management Infrastructures}}. In \bibinfo{booktitle}{\emph{Proc. 6th Int.
  Workshop on Privacy Enhancing Technologies (PET 2006)}}
  \emph{(\bibinfo{series}{LNCS 4258})}. \bibinfo{publisher}{Springer},
  \bibinfo{address}{Cambridge, UK}, \bibinfo{pages}{393--412}.
\newblock
\showISBNx{978-3-540-68790-0, 978-3-540-68793-1}


\bibitem[\protect\citeauthoryear{Citron, Hurani, and Gnadrey}{Citron
  et~al\mbox{.}}{2006}]%
        {citron_harmonic_2006}
\bibfield{author}{\bibinfo{person}{Daniel Citron}, \bibinfo{person}{Adham
  Hurani}, {and} \bibinfo{person}{Alaa Gnadrey}.}
  \bibinfo{year}{2006}\natexlab{}.
\newblock \showarticletitle{The {Harmonic} or {Geometric} {Mean}: {Does} {It}
  {Really} {Matter}?}
\newblock \bibinfo{journal}{\emph{SIGARCH Comput. Archit. News}}
  \bibinfo{volume}{34}, \bibinfo{number}{4} (\bibinfo{date}{Sept.}
  \bibinfo{year}{2006}), \bibinfo{pages}{18--25}.
\newblock
\showISSN{0163-5964}


\bibitem[\protect\citeauthoryear{Clau{\ss} and Schiffner}{Clau{\ss} and
  Schiffner}{2006}]%
        {claus_structuring_2006}
\bibfield{author}{\bibinfo{person}{Sebastian Clau{\ss}} {and}
  \bibinfo{person}{Stefan Schiffner}.} \bibinfo{year}{2006}\natexlab{}.
\newblock \showarticletitle{{Structuring Anonymity Metrics}}. In
  \bibinfo{booktitle}{\emph{{Proc. 13th ACM Conf. on Computer and
  Communications Security 2006 (CCS'06): 2nd ACM Workshop on Digital Identity
  Management (DIM'06)}}}. \bibinfo{publisher}{{ACM}},
  \bibinfo{address}{Alexandria, VA, USA}, \bibinfo{pages}{55--62}.
\newblock
\showISBNx{1-59593-547-9}


\bibitem[\protect\citeauthoryear{Coble}{Coble}{2008}]%
        {coble_formalized_2008}
\bibfield{author}{\bibinfo{person}{Aaron~R. Coble}.}
  \bibinfo{year}{2008}\natexlab{}.
\newblock \showarticletitle{{Formalized Information-Theoretic Proofs of Privacy
  Using the {HOL}4 Theorem-Prover}}. In \bibinfo{booktitle}{\emph{Proc 8th Int.
  Symp. on Privacy Enhancing Technologies (PETS 2008)}}.
  \bibinfo{publisher}{Springer}, \bibinfo{address}{Leuven, Belgium},
  \bibinfo{pages}{77--98}.
\newblock


\bibitem[\protect\citeauthoryear{Daries, Reich, Waldo, Young, Whittinghill, Ho,
  Seaton, and Chuang}{Daries et~al\mbox{.}}{2014}]%
        {daries2014privacy}
\bibfield{author}{\bibinfo{person}{Jon~P. Daries}, \bibinfo{person}{Justin
  Reich}, \bibinfo{person}{Jim Waldo}, \bibinfo{person}{Elise~M. Young},
  \bibinfo{person}{Jonathan Whittinghill}, \bibinfo{person}{Andrew~Dean Ho},
  \bibinfo{person}{Daniel~Thomas Seaton}, {and} \bibinfo{person}{Isaac
  Chuang}.} \bibinfo{year}{2014}\natexlab{}.
\newblock \showarticletitle{Privacy, {{Anonymity}}, and {{Big Data}} in the
  {{Social Sciences}}}.
\newblock \bibinfo{journal}{\emph{Commun. ACM}} \bibinfo{volume}{57},
  \bibinfo{number}{9} (\bibinfo{date}{Sept.} \bibinfo{year}{2014}),
  \bibinfo{pages}{56--63}.
\newblock
\showISSN{0001-0782}


\bibitem[\protect\citeauthoryear{Delaune, Kremer, and Ryan}{Delaune
  et~al\mbox{.}}{2009}]%
        {delaune_verifying_2009}
\bibfield{author}{\bibinfo{person}{Stéphanie Delaune}, \bibinfo{person}{Steve
  Kremer}, {and} \bibinfo{person}{Mark Ryan}.} \bibinfo{year}{2009}\natexlab{}.
\newblock \showarticletitle{{Verifying Privacy-type properties of electronic
  voting protocols}}.
\newblock \bibinfo{journal}{\emph{Journal of Computer Security}}
  \bibinfo{volume}{17}, \bibinfo{number}{4} (\bibinfo{date}{December}
  \bibinfo{year}{2009}), \bibinfo{pages}{435--487}.
\newblock


\bibitem[\protect\citeauthoryear{Deng, Pang, and Wu}{Deng
  et~al\mbox{.}}{2007}]%
        {deng_measuring_2007}
\bibfield{author}{\bibinfo{person}{Yuxin Deng}, \bibinfo{person}{Jun Pang},
  {and} \bibinfo{person}{Peng Wu}.} \bibinfo{year}{2007}\natexlab{}.
\newblock \showarticletitle{{Measuring Anonymity with Relative Entropy}}. In
  \bibinfo{booktitle}{\emph{Proc. 8th Int. Workshop on Formal Aspects in
  Security and Trust (FAST 2011)}}. \bibinfo{publisher}{Springer},
  \bibinfo{address}{Leuven, Belgium}, \bibinfo{pages}{65--79}.
\newblock


\bibitem[\protect\citeauthoryear{Diaz}{Diaz}{2006}]%
        {diaz2006anonymity}
\bibfield{author}{\bibinfo{person}{Claudia Diaz}.}
  \bibinfo{year}{2006}\natexlab{}.
\newblock \showarticletitle{{Anonymity Metrics Revisited}}. In
  \bibinfo{booktitle}{\emph{Anonymous Communication and its Applications}}
  \emph{(\bibinfo{series}{Dagstuhl Seminar Proceedings})},
  \bibfield{editor}{\bibinfo{person}{Shlomi Dolev}, \bibinfo{person}{Rafail
  Ostrovsky}, {and} \bibinfo{person}{Andreas Pfitzmann}} (Eds.).
  \bibinfo{publisher}{Internationales Begegnungs- und Forschungszentrum f{\"u}r
  Informatik (IBFI), Schloss Dagstuhl, Germany}, \bibinfo{address}{Dagstuhl,
  Germany}.
\newblock
\showISSN{1862-4405}


\bibitem[\protect\citeauthoryear{Diaz, Seys, Claessens, and Preneel}{Diaz
  et~al\mbox{.}}{2003}]%
        {diaz_towards_2003}
\bibfield{author}{\bibinfo{person}{Claudia Diaz}, \bibinfo{person}{Stefaan
  Seys}, \bibinfo{person}{Joris Claessens}, {and} \bibinfo{person}{Bart
  Preneel}.} \bibinfo{year}{2003}\natexlab{}.
\newblock \showarticletitle{{Towards Measuring Anonymity}}. In
  \bibinfo{booktitle}{\emph{Proc. 3rd Int. Workshop on Privacy Enhancing
  Technologies (PET 2003)}} \emph{(\bibinfo{series}{LNCS 2482})}.
  \bibinfo{publisher}{Springer}, \bibinfo{address}{Dresden, Germany},
  \bibinfo{pages}{54--68}.
\newblock
\showISBNx{978-3-540-00565-0, 978-3-540-36467-2}


\bibitem[\protect\citeauthoryear{Diaz, Troncoso, and Danezis}{Diaz
  et~al\mbox{.}}{2007}]%
        {diaz_does_2007}
\bibfield{author}{\bibinfo{person}{Claudia Diaz}, \bibinfo{person}{Carmela
  Troncoso}, {and} \bibinfo{person}{George Danezis}.}
  \bibinfo{year}{2007}\natexlab{}.
\newblock \showarticletitle{{Does Additional Information Always Reduce
  Anonymity?}}. In \bibinfo{booktitle}{\emph{Proc. 6th {ACM} Workshop on
  Privacy in Electronic Society (WPES '07)}}. \bibinfo{publisher}{{ACM}},
  \bibinfo{address}{Alexandria, VA, USA}, \bibinfo{pages}{72--75}.
\newblock
\showISBNx{978-1-59593-883-1}


\bibitem[\protect\citeauthoryear{Dingledine, Mathewson, and
  Syverson}{Dingledine et~al\mbox{.}}{2004}]%
        {dingledine_tor:_2004}
\bibfield{author}{\bibinfo{person}{Roger Dingledine}, \bibinfo{person}{Nick
  Mathewson}, {and} \bibinfo{person}{Paul Syverson}.}
  \bibinfo{year}{2004}\natexlab{}.
\newblock \showarticletitle{{Tor: The Second-Generation Onion Router}}. In
  \bibinfo{booktitle}{\emph{Proc. 13th USENIX Security Symp. (Security'04)}}.
  \bibinfo{publisher}{USENIX}, \bibinfo{address}{San Diego, CA, USA},
  \bibinfo{pages}{1--17}.
\newblock


\bibitem[\protect\citeauthoryear{Domingo-Ferrer and Soria-Comas}{Domingo-Ferrer
  and Soria-Comas}{2015}]%
        {domingoferrer2015t-closeness}
\bibfield{author}{\bibinfo{person}{Josep Domingo-Ferrer} {and}
  \bibinfo{person}{Jordi Soria-Comas}.} \bibinfo{year}{2015}\natexlab{}.
\newblock \showarticletitle{From T-Closeness to Differential Privacy and Vice
  Versa in Data Anonymization}.
\newblock \bibinfo{journal}{\emph{Knowledge-Based Systems}}
  \bibinfo{volume}{74} (\bibinfo{date}{Jan.} \bibinfo{year}{2015}),
  \bibinfo{pages}{151--158}.
\newblock
\showISSN{0950-7051}


\bibitem[\protect\citeauthoryear{du~Pin~Calmon and Fawaz}{du~Pin~Calmon and
  Fawaz}{2012}]%
        {du_pin_calmon_privacy_2012}
\bibfield{author}{\bibinfo{person}{Flávio du Pin~Calmon} {and}
  \bibinfo{person}{Nadia Fawaz}.} \bibinfo{year}{2012}\natexlab{}.
\newblock \showarticletitle{{Privacy Against Statistical Inference}}. In
  \bibinfo{booktitle}{\emph{Proc. 50th Annu. Allerton Conf. on Communication,
  Control, and Computing (Allerton 2012)}}. \bibinfo{publisher}{{IEEE}},
  \bibinfo{address}{Monticello, IL, USA}, \bibinfo{pages}{1401--1408}.
\newblock


\bibitem[\protect\citeauthoryear{Duckham and Kulik}{Duckham and Kulik}{2005}]%
        {duckham_formal_2005}
\bibfield{author}{\bibinfo{person}{Matt Duckham} {and} \bibinfo{person}{Lars
  Kulik}.} \bibinfo{year}{2005}\natexlab{}.
\newblock \showarticletitle{A formal model of obfuscation and negotiation for
  location privacy}.
\newblock In \bibinfo{booktitle}{\emph{Pervasive computing}}.
  \bibinfo{publisher}{Springer}, \bibinfo{pages}{152--170}.
\newblock


\bibitem[\protect\citeauthoryear{Dwork}{Dwork}{2006}]%
        {dwork_differential_2006}
\bibfield{author}{\bibinfo{person}{Cynthia Dwork}.}
  \bibinfo{year}{2006}\natexlab{}.
\newblock \showarticletitle{{Differential Privacy}}. In
  \bibinfo{booktitle}{\emph{Proc. 33rd Int. Colloq. on Automata, Languages and
  Programming (ICALP 2006)}} \emph{(\bibinfo{series}{LNCS 4052})}.
  \bibinfo{publisher}{Springer}, \bibinfo{address}{Venice, Italy},
  \bibinfo{pages}{1--12}.
\newblock


\bibitem[\protect\citeauthoryear{Dwork, Kenthapadi, McSherry, Mironov, and
  Naor}{Dwork et~al\mbox{.}}{2006}]%
        {dwork_our_2006}
\bibfield{author}{\bibinfo{person}{Cynthia Dwork}, \bibinfo{person}{Krishnaram
  Kenthapadi}, \bibinfo{person}{Frank McSherry}, \bibinfo{person}{Ilya
  Mironov}, {and} \bibinfo{person}{Moni Naor}.}
  \bibinfo{year}{2006}\natexlab{}.
\newblock \showarticletitle{{Our Data, Ourselves: Privacy via Distributed Noise
  Generation}}. In \bibinfo{booktitle}{\emph{Proc. 25th Int. Cryptology Conf.
  (EUROCRYPT 2006)}}. \bibinfo{publisher}{Springer}, \bibinfo{address}{St.
  Petersburg, Russia}, \bibinfo{pages}{486--503}.
\newblock


\bibitem[\protect\citeauthoryear{Dwork, Naor, Reingold, Rothblum, and
  Vadhan}{Dwork et~al\mbox{.}}{2009}]%
        {dwork_complexity_2009}
\bibfield{author}{\bibinfo{person}{Cynthia Dwork}, \bibinfo{person}{Moni Naor},
  \bibinfo{person}{Omer Reingold}, \bibinfo{person}{Guy~N. Rothblum}, {and}
  \bibinfo{person}{Salil Vadhan}.} \bibinfo{year}{2009}\natexlab{}.
\newblock \showarticletitle{{On the Complexity of Differentially Private Data
  Release: Efficient Algorithms and Hardness Results}}. In
  \bibinfo{booktitle}{\emph{Proc. 41st Annu. ACM Symp. on Theory of Computing
  (STOC'09)}}. \bibinfo{publisher}{{ACM}}, \bibinfo{address}{Bethesda, MD,
  USA}, \bibinfo{pages}{381--390}.
\newblock


\bibitem[\protect\citeauthoryear{Dwork and Roth}{Dwork and Roth}{2014}]%
        {dwork_algorithmic_2014}
\bibfield{author}{\bibinfo{person}{Cynthia Dwork} {and} \bibinfo{person}{Aaron
  Roth}.} \bibinfo{year}{2014}\natexlab{}.
\newblock \bibinfo{booktitle}{\emph{The {Algorithmic} {Foundations} of
  {Differential} {Privacy}}}.
\newblock \bibinfo{publisher}{Now Publishers}.
\newblock
\showISBNx{978-1-60198-818-8}


\bibitem[\protect\citeauthoryear{Eckhoff and Wagner}{Eckhoff and
  Wagner}{2018}]%
        {eckhoff2017privacy}
\bibfield{author}{\bibinfo{person}{David Eckhoff} {and} \bibinfo{person}{Isabel
  Wagner}.} \bibinfo{year}{First Quarter 2018}\natexlab{}.
\newblock \showarticletitle{Privacy in the {{Smart City}} -- {{Applications}},
  {{Technologies}}, {{Challenges}} and {{Solutions}}}.
\newblock \bibinfo{journal}{\emph{IEEE Communications Surveys \& Tutorials}}
  \bibinfo{volume}{20}, \bibinfo{number}{1} (\bibinfo{year}{First Quarter
  2018}), \bibinfo{pages}{489--516}.
\newblock


\bibitem[\protect\citeauthoryear{Erlingsson, Pihur, and Korolova}{Erlingsson
  et~al\mbox{.}}{2014}]%
        {erlingsson2014rappor}
\bibfield{author}{\bibinfo{person}{{\'U}lfar Erlingsson},
  \bibinfo{person}{Vasyl Pihur}, {and} \bibinfo{person}{Aleksandra Korolova}.}
  \bibinfo{year}{2014}\natexlab{}.
\newblock \showarticletitle{{{RAPPOR}}: {{Randomized Aggregatable
  Privacy}}-{{Preserving Ordinal Response}}}. In
  \bibinfo{booktitle}{\emph{Proc. 21st {{ACM Conf}}. on {{Computer}} and
  {{Communications Security}} ({{CCS}} '14)}}. \bibinfo{publisher}{{ACM}},
  \bibinfo{address}{Scottsdale, Arizona, US}, \bibinfo{pages}{1054--1067}.
\newblock
\showISBNx{978-1-4503-2957-6}


\bibitem[\protect\citeauthoryear{{European Parliament \& Council}}{{European
  Parliament \& Council}}{2016}]%
        {european2016general}
\bibfield{author}{\bibinfo{person}{{European Parliament \& Council}}.}
  \bibinfo{year}{2016}\natexlab{}.
\newblock \showarticletitle{General Data Protection Regulation, Regulation (EU)
  2016/679}.
\newblock \bibinfo{journal}{\emph{Official Journal of the European Union}}
  \bibinfo{volume}{L 119} (\bibinfo{date}{April} \bibinfo{year}{2016}),
  \bibinfo{pages}{1--88}.
\newblock


\bibitem[\protect\citeauthoryear{Evfimievski, Srikant, Agrawal, and
  Gehrke}{Evfimievski et~al\mbox{.}}{2004}]%
        {evfimievski_privacy_2004}
\bibfield{author}{\bibinfo{person}{Alexandre Evfimievski},
  \bibinfo{person}{Ramakrishnan Srikant}, \bibinfo{person}{Rakesh Agrawal},
  {and} \bibinfo{person}{Johannes Gehrke}.} \bibinfo{year}{2004}\natexlab{}.
\newblock \showarticletitle{{Privacy Preserving Mining of Association Rules}}.
\newblock \bibinfo{journal}{\emph{Information Systems}} \bibinfo{volume}{29},
  \bibinfo{number}{4} (\bibinfo{date}{June} \bibinfo{year}{2004}),
  \bibinfo{pages}{343--364}.
\newblock


\bibitem[\protect\citeauthoryear{Fanti, Pihur, and Erlingsson}{Fanti
  et~al\mbox{.}}{2016}]%
        {fanti2016building}
\bibfield{author}{\bibinfo{person}{Giulia Fanti}, \bibinfo{person}{Vasyl
  Pihur}, {and} \bibinfo{person}{{\'U}lfar Erlingsson}.}
  \bibinfo{year}{2016}\natexlab{}.
\newblock \showarticletitle{Building a {{RAPPOR}} with the {{Unknown}}:
  {{Privacy}}-{{Preserving Learning}} of {{Associations}} and {{Data
  Dictionaries}}}.
\newblock \bibinfo{journal}{\emph{Proceedings on Privacy Enhancing
  Technologies}} \bibinfo{volume}{2016}, \bibinfo{number}{3}
  (\bibinfo{year}{2016}), \bibinfo{pages}{41--61}.
\newblock
\showISSN{2299-0984}


\bibitem[\protect\citeauthoryear{Fawaz, Kim, and Shin}{Fawaz
  et~al\mbox{.}}{2016}]%
        {fawaz2016privacy}
\bibfield{author}{\bibinfo{person}{Kassem Fawaz}, \bibinfo{person}{Kyu-Han
  Kim}, {and} \bibinfo{person}{Kang~G. Shin}.} \bibinfo{year}{2016}\natexlab{}.
\newblock \showarticletitle{Privacy vs. {{Reward}} in {{Indoor Location-Based
  Services}}}.
\newblock \bibinfo{journal}{\emph{Proceedings on Privacy Enhancing
  Technologies}} \bibinfo{volume}{2016}, \bibinfo{number}{4}
  (\bibinfo{date}{Jan.} \bibinfo{year}{2016}), \bibinfo{pages}{102--122}.
\newblock
\showISSN{2299-0984}


\bibitem[\protect\citeauthoryear{Franz, Meyer, and Pashalidis}{Franz
  et~al\mbox{.}}{2007}]%
        {Franz2007}
\bibfield{author}{\bibinfo{person}{Matthias Franz}, \bibinfo{person}{Bernd
  Meyer}, {and} \bibinfo{person}{Andreas Pashalidis}.}
  \bibinfo{year}{2007}\natexlab{}.
\newblock \showarticletitle{{Attacking Unlinkability: The Importance of
  Context}}. In \bibinfo{booktitle}{\emph{Proc. 7th Int. Symp. on Privacy
  Enhancing Technologies (PETS 2007)}} \emph{(\bibinfo{series}{LNCS 4776})}.
  \bibinfo{publisher}{Springer}, \bibinfo{address}{Ottawa, Canada},
  \bibinfo{pages}{1--16}.
\newblock
\showISBNx{978-3-540-75550-0, 978-3-540-75551-7}


\bibitem[\protect\citeauthoryear{Freudiger, Manshaei, Hubaux, and
  Parkes}{Freudiger et~al\mbox{.}}{2009}]%
        {freudiger_non-cooperative_2009}
\bibfield{author}{\bibinfo{person}{Julien Freudiger},
  \bibinfo{person}{Mohammad~Hossein Manshaei}, \bibinfo{person}{Jean-Pierre
  Hubaux}, {and} \bibinfo{person}{David~C. Parkes}.}
  \bibinfo{year}{2009}\natexlab{}.
\newblock \showarticletitle{{On Non-cooperative Location Privacy: A
  Game-theoretic Analysis}}. In \bibinfo{booktitle}{\emph{Proc. 16th {ACM}
  Conf. on Computer and Communications Security (CCS'09)}}.
  \bibinfo{publisher}{{ACM}}, \bibinfo{address}{Chicago, IL, USA},
  \bibinfo{pages}{324--337}.
\newblock
\showISBNx{978-1-60558-894-0}


\bibitem[\protect\citeauthoryear{Freudiger, Raya, Félegyházi, Papadimitratos,
  and Hubaux}{Freudiger et~al\mbox{.}}{2007}]%
        {freudiger_mix-zones_2007}
\bibfield{author}{\bibinfo{person}{Julien Freudiger}, \bibinfo{person}{Maxim
  Raya}, \bibinfo{person}{Márk Félegyházi}, \bibinfo{person}{Panos
  Papadimitratos}, {and} \bibinfo{person}{Jean-Pierre Hubaux}.}
  \bibinfo{year}{2007}\natexlab{}.
\newblock \showarticletitle{{Mix-Zones for Location Privacy in Vehicular
  Networks}}. In \bibinfo{booktitle}{\emph{Proc. 1st Int. Workshop on Wireless
  Networking for Intelligent Transportation Systems (WiN-ITS 2007)}}.
  \bibinfo{publisher}{ICST}, \bibinfo{address}{Vancouver, Canada}.
\newblock


\bibitem[\protect\citeauthoryear{Fung, Wang, Chen, and Yu}{Fung
  et~al\mbox{.}}{2010}]%
        {fung_privacy-preserving_2010}
\bibfield{author}{\bibinfo{person}{Benjamin Fung}, \bibinfo{person}{Ke Wang},
  \bibinfo{person}{Rui Chen}, {and} \bibinfo{person}{Philip~S. Yu}.}
  \bibinfo{year}{2010}\natexlab{}.
\newblock \showarticletitle{{Privacy-Preserving Data Publishing: A Survey of
  Recent Developments}}.
\newblock \bibinfo{journal}{\emph{ACM Computing Surveys (CSUR)}}
  \bibinfo{volume}{42}, \bibinfo{number}{4} (\bibinfo{date}{June}
  \bibinfo{year}{2010}), \bibinfo{pages}{14}.
\newblock


\bibitem[\protect\citeauthoryear{Ganti, Pham, Tsai, and Abdelzaher}{Ganti
  et~al\mbox{.}}{2008}]%
        {ganti_poolview:_2008}
\bibfield{author}{\bibinfo{person}{Raghu~K. Ganti}, \bibinfo{person}{Nam Pham},
  \bibinfo{person}{Yu-En Tsai}, {and} \bibinfo{person}{Tarek~F. Abdelzaher}.}
  \bibinfo{year}{2008}\natexlab{}.
\newblock \showarticletitle{{PoolView}: stream privacy for grassroots
  participatory sensing}. In \bibinfo{booktitle}{\emph{Proceedings of the 6th
  {ACM} conference on {Embedded} network sensor systems}}.
  \bibinfo{publisher}{ACM}, \bibinfo{pages}{281--294}.
\newblock


\bibitem[\protect\citeauthoryear{Gierlichs, Troncoso, Diaz, Preneel, and
  Verbauwhede}{Gierlichs et~al\mbox{.}}{2008}]%
        {gierlichs_revisiting_2008}
\bibfield{author}{\bibinfo{person}{Benedikt Gierlichs},
  \bibinfo{person}{Carmela Troncoso}, \bibinfo{person}{Claudia Diaz},
  \bibinfo{person}{Bart Preneel}, {and} \bibinfo{person}{Ingrid Verbauwhede}.}
  \bibinfo{year}{2008}\natexlab{}.
\newblock \showarticletitle{{Revisiting a Combinatorial Approach Toward
  Measuring Anonymity}}. In \bibinfo{booktitle}{\emph{Proc. 7th {ACM} Workshop
  on Privacy in the Electronic Society (WPES'08)}}. \bibinfo{publisher}{{ACM}},
  \bibinfo{address}{Alexandria, VA, USA}, \bibinfo{pages}{111--116}.
\newblock
\showISBNx{978-1-60558-289-4}


\bibitem[\protect\citeauthoryear{Golle and Partridge}{Golle and
  Partridge}{2009}]%
        {golle_anonymity_2009}
\bibfield{author}{\bibinfo{person}{Philippe Golle} {and} \bibinfo{person}{Kurt
  Partridge}.} \bibinfo{year}{2009}\natexlab{}.
\newblock \showarticletitle{On the anonymity of home/work location pairs}.
\newblock In \bibinfo{booktitle}{\emph{Pervasive {Computing}}}.
  \bibinfo{publisher}{Springer}, \bibinfo{pages}{390--397}.
\newblock


\bibitem[\protect\citeauthoryear{Hamel, Grégoire, and Goldberg}{Hamel
  et~al\mbox{.}}{2011}]%
        {hamel_misentropists:_2011}
\bibfield{author}{\bibinfo{person}{Angèle Hamel},
  \bibinfo{person}{Jean-Charles Grégoire}, {and} \bibinfo{person}{Ian
  Goldberg}.} \bibinfo{year}{2011}\natexlab{}.
\newblock \bibinfo{booktitle}{\emph{{The Misentropists: New Approaches to
  Measures in Tor}}}.
\newblock \bibinfo{type}{{T}echnical {R}eport}. \bibinfo{institution}{Technical
  Report 2011-18, Cheriton School of Computer Science, University of Waterloo}.
\newblock
\urldef\tempurl%
\url{http://cacr.uwaterloo.ca/techreports/2011/cacr2011-18.pdf}
\showURL{%
\tempurl}


\bibitem[\protect\citeauthoryear{Hermans, Pashalidis, Vercauteren, and
  Preneel}{Hermans et~al\mbox{.}}{2011}]%
        {hermans_new_2011}
\bibfield{author}{\bibinfo{person}{Jens Hermans}, \bibinfo{person}{Andreas
  Pashalidis}, \bibinfo{person}{Frederik Vercauteren}, {and}
  \bibinfo{person}{Bart Preneel}.} \bibinfo{year}{2011}\natexlab{}.
\newblock \showarticletitle{{A new {RFID} Privacy Model}}. In
  \bibinfo{booktitle}{\emph{Proc. 16th Symp. on Research in Computer Security
  (ESORICS 2011)}} \emph{(\bibinfo{series}{LNCS 6879})}.
  \bibinfo{publisher}{Springer}, \bibinfo{address}{Leuven, Belgium},
  \bibinfo{pages}{568--587}.
\newblock


\bibitem[\protect\citeauthoryear{Heurix, Zimmermann, Neubauer, and Fenz}{Heurix
  et~al\mbox{.}}{2015}]%
        {heurix2015taxonomy}
\bibfield{author}{\bibinfo{person}{Johannes Heurix}, \bibinfo{person}{Peter
  Zimmermann}, \bibinfo{person}{Thomas Neubauer}, {and} \bibinfo{person}{Stefan
  Fenz}.} \bibinfo{year}{2015}\natexlab{}.
\newblock \showarticletitle{A Taxonomy for Privacy Enhancing Technologies}.
\newblock \bibinfo{journal}{\emph{Computers \& Security}}  \bibinfo{volume}{53}
  (\bibinfo{date}{Sept.} \bibinfo{year}{2015}), \bibinfo{pages}{1--17}.
\newblock
\showISSN{0167-4048}


\bibitem[\protect\citeauthoryear{Heydt-Benjamin, Chae, Defend, and
  Fu}{Heydt-Benjamin et~al\mbox{.}}{2006}]%
        {heydt-benjamin_privacy_2006}
\bibfield{author}{\bibinfo{person}{Thomas~S. Heydt-Benjamin},
  \bibinfo{person}{Hee-Jin Chae}, \bibinfo{person}{Benessa Defend}, {and}
  \bibinfo{person}{Kevin Fu}.} \bibinfo{year}{2006}\natexlab{}.
\newblock \showarticletitle{Privacy for {Public} {Transportation}}.
\newblock In \bibinfo{booktitle}{\emph{Privacy {Enhancing} {Technologies}}},
  \bibfield{editor}{\bibinfo{person}{George Danezis} {and}
  \bibinfo{person}{Philippe Golle}} (Eds.). Number 4258 in
  \bibinfo{series}{Lecture {Notes} in {Computer} {Science}}.
  \bibinfo{publisher}{Springer Berlin Heidelberg}, \bibinfo{pages}{1--19}.
\newblock
\showISBNx{978-3-540-68790-0 978-3-540-68793-1}


\bibitem[\protect\citeauthoryear{Hoh and Gruteser}{Hoh and Gruteser}{2005}]%
        {hoh_protecting_2005}
\bibfield{author}{\bibinfo{person}{Baik Hoh} {and} \bibinfo{person}{Marco
  Gruteser}.} \bibinfo{year}{2005}\natexlab{}.
\newblock \showarticletitle{{Protecting Location Privacy Through Path
  Confusion}}. In \bibinfo{booktitle}{\emph{Proc. 1st Int. Conf. on Security
  and Privacy for Emerging Areas in Communications Networks, (SecureComm
  2005)}}. \bibinfo{address}{Athens, Greece}, \bibinfo{pages}{194--205}.
\newblock


\bibitem[\protect\citeauthoryear{Hoh, Gruteser, Xiong, and Alrabady}{Hoh
  et~al\mbox{.}}{2007}]%
        {hoh_preserving_2007}
\bibfield{author}{\bibinfo{person}{Baik Hoh}, \bibinfo{person}{Marco Gruteser},
  \bibinfo{person}{Hui Xiong}, {and} \bibinfo{person}{Ansaf Alrabady}.}
  \bibinfo{year}{2007}\natexlab{}.
\newblock \showarticletitle{{Preserving Privacy in GPS Traces via
  Uncertainty-Aware Path Cloaking}}. In \bibinfo{booktitle}{\emph{14th {ACM}
  Conf. on Computer and Communications Security (CCS)}}.
  \bibinfo{address}{Alexandria, VA}, \bibinfo{pages}{161--171}.
\newblock
\showISBNx{978-1-59593-703-2}


\bibitem[\protect\citeauthoryear{Hsu, Gaboardi, Haeberlen, Khanna, Narayan,
  Pierce, and Roth}{Hsu et~al\mbox{.}}{2014a}]%
        {hsu_differential_2014}
\bibfield{author}{\bibinfo{person}{Justin Hsu}, \bibinfo{person}{Marco
  Gaboardi}, \bibinfo{person}{Andreas Haeberlen}, \bibinfo{person}{Sanjeev
  Khanna}, \bibinfo{person}{Arjun Narayan}, \bibinfo{person}{Benjamin~C.
  Pierce}, {and} \bibinfo{person}{Aaron Roth}.}
  \bibinfo{year}{2014}\natexlab{a}.
\newblock \showarticletitle{Differential {{Privacy}}: {{An Economic Method}}
  for {{Choosing Epsilon}}}. In \bibinfo{booktitle}{\emph{2014 {{IEEE}} 27th
  {{Computer Security Foundations Symposium}}}}. \bibinfo{pages}{398--410}.
\newblock


\bibitem[\protect\citeauthoryear{Hsu, Huang, Roth, Roughgarden, and Wu}{Hsu
  et~al\mbox{.}}{2014b}]%
        {hsu2014private}
\bibfield{author}{\bibinfo{person}{Justin Hsu}, \bibinfo{person}{Zhiyi Huang},
  \bibinfo{person}{Aaron Roth}, \bibinfo{person}{Tim Roughgarden}, {and}
  \bibinfo{person}{Zhiwei~Steven Wu}.} \bibinfo{year}{2014}\natexlab{b}.
\newblock \showarticletitle{Private {{Matchings}} and {{Allocations}}}. In
  \bibinfo{booktitle}{\emph{Proceedings of the 46th {{Annual ACM Symposium}} on
  {{Theory}} of {{Computing}}}} \emph{(\bibinfo{series}{STOC '14})}.
  \bibinfo{publisher}{{ACM}}, \bibinfo{pages}{21--30}.
\newblock
\showISBNx{978-1-4503-2710-7}


\bibitem[\protect\citeauthoryear{Hughes and Shmatikov}{Hughes and
  Shmatikov}{2004}]%
        {hughes_information_2004}
\bibfield{author}{\bibinfo{person}{Dominic Hughes} {and}
  \bibinfo{person}{Vitaly Shmatikov}.} \bibinfo{year}{2004}\natexlab{}.
\newblock \showarticletitle{{Information Hiding, Anonymity and Privacy: A
  Modular Approach}}.
\newblock \bibinfo{journal}{\emph{ACM Journal of Computer Security}}
  \bibinfo{volume}{12}, \bibinfo{number}{1} (\bibinfo{date}{January}
  \bibinfo{year}{2004}), \bibinfo{pages}{3--36}.
\newblock


\bibitem[\protect\citeauthoryear{Humbert, Ayday, Hubaux, and Telenti}{Humbert
  et~al\mbox{.}}{2013}]%
        {humbert_addressing_2013}
\bibfield{author}{\bibinfo{person}{Mathias Humbert}, \bibinfo{person}{Erman
  Ayday}, \bibinfo{person}{Jean-Pierre Hubaux}, {and} \bibinfo{person}{Amalio
  Telenti}.} \bibinfo{year}{2013}\natexlab{}.
\newblock \showarticletitle{{Addressing the Concerns of the Lacks Family:
  Quantification of Kin Genomic Privacy}}. In \bibinfo{booktitle}{\emph{Proc.
  20th ACM Conf. on Computer and Communications Security (CCS'13)}}.
  \bibinfo{publisher}{{ACM}}, \bibinfo{address}{Berlin, Germany},
  \bibinfo{pages}{1141--1152}.
\newblock
\showISBNx{978-1-4503-2477-9}


\bibitem[\protect\citeauthoryear{Jelasity and Birman}{Jelasity and
  Birman}{2014}]%
        {jelasity_distributional_2014}
\bibfield{author}{\bibinfo{person}{Márk Jelasity} {and}
  \bibinfo{person}{Kenneth~P. Birman}.} \bibinfo{year}{2014}\natexlab{}.
\newblock \showarticletitle{{Distributional Differential Privacy for
  Large-scale Smart Metering}}. In \bibinfo{booktitle}{\emph{2nd {ACM} Workshop
  on Information Hiding and Multimedia Security (IH\&MMSec'14)}}.
  \bibinfo{publisher}{ACM}, \bibinfo{address}{Salzburg, Austria},
  \bibinfo{pages}{141--146}.
\newblock


\bibitem[\protect\citeauthoryear{Johnson, Wacek, Jansen, Sherr, and
  Syverson}{Johnson et~al\mbox{.}}{2013}]%
        {johnson_users_2013}
\bibfield{author}{\bibinfo{person}{Aaron Johnson}, \bibinfo{person}{Chris
  Wacek}, \bibinfo{person}{Rob Jansen}, \bibinfo{person}{Micah Sherr}, {and}
  \bibinfo{person}{Paul Syverson}.} \bibinfo{year}{2013}\natexlab{}.
\newblock \showarticletitle{{Users Get Routed: Traffic Correlation on Tor by
  Realistic Adversaries}}. In \bibinfo{booktitle}{\emph{Proc. 20th ACM Conf. on
  Computer and Communications Security (CCS'13)}}. \bibinfo{publisher}{{ACM}},
  \bibinfo{address}{Berlin, Germany}, \bibinfo{pages}{337–348}.
\newblock
\showISBNx{978-1-4503-2477-9}


\bibitem[\protect\citeauthoryear{Juels and Weis}{Juels and Weis}{2009}]%
        {juels_defining_2009}
\bibfield{author}{\bibinfo{person}{Ari Juels} {and} \bibinfo{person}{Stephen~A.
  Weis}.} \bibinfo{year}{2009}\natexlab{}.
\newblock \showarticletitle{Defining {Strong} {Privacy} for {RFID}}.
\newblock \bibinfo{journal}{\emph{ACM Trans. Inf. Syst. Secur.}}
  \bibinfo{volume}{13}, \bibinfo{number}{1} (\bibinfo{date}{Nov.}
  \bibinfo{year}{2009}), \bibinfo{pages}{7:1--7:23}.
\newblock
\showISSN{1094-9224}


\bibitem[\protect\citeauthoryear{Kalogridis, Efthymiou, Denic, Lewis, and
  Cepeda}{Kalogridis et~al\mbox{.}}{2010}]%
        {kalogridis_privacy_2010}
\bibfield{author}{\bibinfo{person}{Georgios Kalogridis},
  \bibinfo{person}{Costas Efthymiou}, \bibinfo{person}{Stojan~Z. Denic},
  \bibinfo{person}{Tim~A. Lewis}, {and} \bibinfo{person}{Rafael Cepeda}.}
  \bibinfo{year}{2010}\natexlab{}.
\newblock \showarticletitle{{Privacy for Smart Meters: Towards Undetectable
  Appliance Load Signatures}}. In \bibinfo{booktitle}{\emph{Proc. 1st Int.
  Conf. on Smart Grid Communications (SmartGridComm 2010)}}.
  \bibinfo{publisher}{IEEE}, \bibinfo{address}{Gaithersburg, MD, USA},
  \bibinfo{pages}{232--237}.
\newblock


\bibitem[\protect\citeauthoryear{Kantarcioğlu, Jin, and Clifton}{Kantarcioğlu
  et~al\mbox{.}}{2004}]%
        {kantarcioglu_when_2004}
\bibfield{author}{\bibinfo{person}{Murat Kantarcioğlu},
  \bibinfo{person}{Jiashun Jin}, {and} \bibinfo{person}{Chris Clifton}.}
  \bibinfo{year}{2004}\natexlab{}.
\newblock \showarticletitle{{When do Data Mining Results Violate Privacy?}}. In
  \bibinfo{booktitle}{\emph{Proc. 10th {ACM} {SIGKDD} Int. Conf. on Knowledge
  Discovery and data Mining (KDD'04)}}. \bibinfo{publisher}{{ACM}},
  \bibinfo{address}{Seatlle, WA, USA}, \bibinfo{pages}{599--604}.
\newblock


\bibitem[\protect\citeauthoryear{Kearns, Pai, Roth, and Ullman}{Kearns
  et~al\mbox{.}}{2014}]%
        {kearns2014mechanism}
\bibfield{author}{\bibinfo{person}{Michael Kearns}, \bibinfo{person}{Mallesh
  Pai}, \bibinfo{person}{Aaron Roth}, {and} \bibinfo{person}{Jonathan Ullman}.}
  \bibinfo{year}{2014}\natexlab{}.
\newblock \showarticletitle{Mechanism {{Design}} in {{Large Games}}:
  {{Incentives}} and {{Privacy}}}. In \bibinfo{booktitle}{\emph{Proc. 5th
  {{Conf.}} on {{Innovations}} in {{Theoretical Computer Science}}}}
  \emph{(\bibinfo{series}{ITCS '14})}. \bibinfo{publisher}{{ACM}},
  \bibinfo{pages}{403--410}.
\newblock
\showISBNx{978-1-4503-2698-8}


\bibitem[\protect\citeauthoryear{Kelly, Raines, Grimaila, Baldwin, and
  Mullins}{Kelly et~al\mbox{.}}{2008}]%
        {kelly_survey_2008}
\bibfield{author}{\bibinfo{person}{Douglas~J. Kelly},
  \bibinfo{person}{Richard~A. Raines}, \bibinfo{person}{Michael~R. Grimaila},
  \bibinfo{person}{Rusty~O. Baldwin}, {and} \bibinfo{person}{Barry~E.
  Mullins}.} \bibinfo{year}{2008}\natexlab{}.
\newblock \showarticletitle{{A Survey of State-of-the-art in Anonymity
  Metrics}}. In \bibinfo{booktitle}{\emph{Proc. 15th ACM Conf. on Computer and
  Communications Security 2008 (CCS'08): 1st Workshop on Network Data
  Anonymization (NDA'08)}}. \bibinfo{publisher}{{ACM}},
  \bibinfo{address}{Alexandria, VA, USA}, \bibinfo{pages}{31--40}.
\newblock
\showISBNx{978-1-60558-301-3}


\bibitem[\protect\citeauthoryear{Kenthapadi, Mishra, and Nissim}{Kenthapadi
  et~al\mbox{.}}{2005}]%
        {kenthapadi2005simulatable}
\bibfield{author}{\bibinfo{person}{Krishnaram Kenthapadi},
  \bibinfo{person}{Nina Mishra}, {and} \bibinfo{person}{Kobbi Nissim}.}
  \bibinfo{year}{2005}\natexlab{}.
\newblock \showarticletitle{Simulatable {{Auditing}}}. In
  \bibinfo{booktitle}{\emph{Proc. 24th {{ACM SIGMOD}}-{{SIGACT}}-{{SIGART
  Symposium}} on {{Principles}} of {{Database Systems}}}}
  \emph{(\bibinfo{series}{PODS '05})}. \bibinfo{publisher}{{ACM}},
  \bibinfo{address}{New York, NY, USA}, \bibinfo{pages}{118--127}.
\newblock
\showISBNx{978-1-59593-062-0}


\bibitem[\protect\citeauthoryear{Kesdogan, Egner, and Büschkes}{Kesdogan
  et~al\mbox{.}}{1998}]%
        {kesdogan_stop-_1998}
\bibfield{author}{\bibinfo{person}{Dogan Kesdogan}, \bibinfo{person}{Jan
  Egner}, {and} \bibinfo{person}{Roland Büschkes}.}
  \bibinfo{year}{1998}\natexlab{}.
\newblock \showarticletitle{{Stop- and- Go-MIXes Providing Probabilistic
  Anonymity in an Open System}}. In \bibinfo{booktitle}{\emph{Proc. 2nd Int.
  Workshop on Information Hiding (IH'98)}} \emph{(\bibinfo{series}{LNCS
  1525})}. \bibinfo{publisher}{Springer}, \bibinfo{address}{Portland, OR},
  \bibinfo{pages}{83--98}.
\newblock
\showISBNx{978-3-540-65386-8, 978-3-540-49380-8}


\bibitem[\protect\citeauthoryear{Kifer and Machanavajjhala}{Kifer and
  Machanavajjhala}{2011}]%
        {kifer_no_2011}
\bibfield{author}{\bibinfo{person}{Daniel Kifer} {and} \bibinfo{person}{Ashwin
  Machanavajjhala}.} \bibinfo{year}{2011}\natexlab{}.
\newblock \showarticletitle{{No Free Lunch in Data Privacy}}. In
  \bibinfo{booktitle}{\emph{Proc. 2011 {ACM} {SIGMOD} Int. Conf. on Management
  of Data (SIGMOD'11)}}. \bibinfo{publisher}{{ACM}}, \bibinfo{address}{Athens,
  Greece}, \bibinfo{pages}{193--204}.
\newblock


\bibitem[\protect\citeauthoryear{Kim, Ngai, and Srivastava}{Kim
  et~al\mbox{.}}{2011}]%
        {kim_cooperative_2011}
\bibfield{author}{\bibinfo{person}{Younghun Kim}, \bibinfo{person}{Edith C.-H.
  Ngai}, {and} \bibinfo{person}{Mani~B. Srivastava}.}
  \bibinfo{year}{2011}\natexlab{}.
\newblock \showarticletitle{{Cooperative State Estimation for Preserving
  Privacy of User Behaviors in Smart Grid}}. In \bibinfo{booktitle}{\emph{Proc.
  2nd IEEE Int. Conf. on Smart Grid Communications (SmartGridComm 2011)}}.
  \bibinfo{publisher}{{IEEE}}, \bibinfo{address}{Brussels, Belgium},
  \bibinfo{pages}{178--183}.
\newblock


\bibitem[\protect\citeauthoryear{Krumm}{Krumm}{2009}]%
        {krumm_survey_2009}
\bibfield{author}{\bibinfo{person}{John Krumm}.}
  \bibinfo{year}{2009}\natexlab{}.
\newblock \showarticletitle{{A Survey of Computational Location Privacy}}.
\newblock \bibinfo{journal}{\emph{Personal and Ubiquitous Computing}}
  \bibinfo{volume}{13}, \bibinfo{number}{6} (\bibinfo{date}{August}
  \bibinfo{year}{2009}), \bibinfo{pages}{391--399}.
\newblock


\bibitem[\protect\citeauthoryear{Lai, Ho, and Poor}{Lai et~al\mbox{.}}{2011}]%
        {lai_privacy_2011}
\bibfield{author}{\bibinfo{person}{Lifeng Lai}, \bibinfo{person}{Siu-Wai Ho},
  {and} \bibinfo{person}{Vincent~H. Poor}.} \bibinfo{year}{2011}\natexlab{}.
\newblock \showarticletitle{{Privacy-Security Trade-Offs in Biometric Security
  Systems---Part I: Single Use Case}}.
\newblock \bibinfo{journal}{\emph{IEEE Trans. on Information Forensics
  Security}} \bibinfo{volume}{6}, \bibinfo{number}{1} (\bibinfo{date}{March}
  \bibinfo{year}{2011}), \bibinfo{pages}{122--139}.
\newblock
\showISSN{1556-6013}


\bibitem[\protect\citeauthoryear{Lee and Clifton}{Lee and Clifton}{2011}]%
        {lee2011how}
\bibfield{author}{\bibinfo{person}{Jaewoo Lee} {and} \bibinfo{person}{Chris
  Clifton}.} \bibinfo{year}{2011}\natexlab{}.
\newblock \showarticletitle{How {{Much Is Enough}}? {{Choosing}} $\epsilon$ for
  {{Differential Privacy}}}. In \bibinfo{booktitle}{\emph{Information
  {{Security}}}} \emph{(\bibinfo{series}{Lecture Notes in Computer Science})}.
  \bibinfo{publisher}{{Springer, Berlin, Heidelberg}},
  \bibinfo{pages}{325--340}.
\newblock
\showISBNx{978-3-642-24860-3 978-3-642-24861-0}


\bibitem[\protect\citeauthoryear{Li, Tao, and Xiao}{Li et~al\mbox{.}}{2008}]%
        {li_preservation_2008}
\bibfield{author}{\bibinfo{person}{Jiexing Li}, \bibinfo{person}{Yufei Tao},
  {and} \bibinfo{person}{Xiaokui Xiao}.} \bibinfo{year}{2008}\natexlab{}.
\newblock \showarticletitle{{Preservation of Proximity Privacy in Publishing
  Numerical Sensitive Data}}. In \bibinfo{booktitle}{\emph{Proc. 2008 {ACM}
  Int. Conf. on Management of Data (SIGMOD 2008)}}. \bibinfo{publisher}{{ACM}},
  \bibinfo{address}{Vancouver, Canada}, \bibinfo{pages}{473--486}.
\newblock


\bibitem[\protect\citeauthoryear{Li, Li, and Venkatasubramanian}{Li
  et~al\mbox{.}}{2007}]%
        {li_t-closeness:_2007}
\bibfield{author}{\bibinfo{person}{Ninghui Li}, \bibinfo{person}{Tiancheng Li},
  {and} \bibinfo{person}{S. Venkatasubramanian}.}
  \bibinfo{year}{2007}\natexlab{}.
\newblock \showarticletitle{{t-Closeness: Privacy Beyond k-Anonymity and
  l-Diversity}}. In \bibinfo{booktitle}{\emph{Proc. IEEE 23rd Int. Conf. on
  Data Engineering (ICDE 2007)}}. \bibinfo{publisher}{IEEE},
  \bibinfo{address}{Istanbul, Turkey}, \bibinfo{pages}{106--115}.
\newblock


\bibitem[\protect\citeauthoryear{Li, Qardaji, and Su}{Li et~al\mbox{.}}{2012}]%
        {li2012sampling}
\bibfield{author}{\bibinfo{person}{Ninghui Li}, \bibinfo{person}{Wahbeh
  Qardaji}, {and} \bibinfo{person}{Dong Su}.} \bibinfo{year}{2012}\natexlab{}.
\newblock \showarticletitle{On {{Sampling}}, {{Anonymization}}, and
  {{Differential Privacy}} Or, {{K}}-Anonymization {{Meets Differential
  Privacy}}}. In \bibinfo{booktitle}{\emph{Proceedings of the 7th {{ACM
  Symposium}} on {{Information}}, {{Computer}} and {{Communications Security}}
  (AsiaCCS)}}. \bibinfo{publisher}{{ACM}}, \bibinfo{address}{Seoul, Republic of
  Korea}, \bibinfo{pages}{32--33}.
\newblock
\showISBNx{978-1-4503-1648-4}


\bibitem[\protect\citeauthoryear{Lin, Hewett, and Altman}{Lin
  et~al\mbox{.}}{2002}]%
        {lin_using_2002}
\bibfield{author}{\bibinfo{person}{Zhen Lin}, \bibinfo{person}{Michael Hewett},
  {and} \bibinfo{person}{Russ~B. Altman}.} \bibinfo{year}{2002}\natexlab{}.
\newblock \showarticletitle{{Using Binning to Maintain Confidentiality of
  Medical Data}}. In \bibinfo{booktitle}{\emph{Proc. AMIA Symp. (American
  Medical Informatics Association 2002)}}. \bibinfo{address}{San Antonio, TX,
  USA}, \bibinfo{pages}{454--458}.
\newblock
\showISSN{1531-605X}


\bibitem[\protect\citeauthoryear{Lisovich, Mulligan, and Wicker}{Lisovich
  et~al\mbox{.}}{2010}]%
        {lisovich_inferring_2010}
\bibfield{author}{\bibinfo{person}{Mikhail~A. Lisovich},
  \bibinfo{person}{Deirdre~K. Mulligan}, {and} \bibinfo{person}{Stephen~B.
  Wicker}.} \bibinfo{year}{2010}\natexlab{}.
\newblock \showarticletitle{Inferring personal information from demand-response
  systems}.
\newblock \bibinfo{journal}{\emph{Security \& Privacy, IEEE}}
  \bibinfo{volume}{8}, \bibinfo{number}{1} (\bibinfo{year}{2010}),
  \bibinfo{pages}{11--20}.
\newblock


\bibitem[\protect\citeauthoryear{Liu and Terzi}{Liu and Terzi}{2010}]%
        {liu_framework_2010}
\bibfield{author}{\bibinfo{person}{Kun Liu} {and} \bibinfo{person}{Evimaria
  Terzi}.} \bibinfo{year}{2010}\natexlab{}.
\newblock \showarticletitle{{A Framework for Computing the Privacy Scores of
  Users in Online Social Networks}}.
\newblock \bibinfo{journal}{\emph{ACM Transactions on Knowledge Discovery from
  Data}} \bibinfo{volume}{5}, \bibinfo{number}{1} (\bibinfo{date}{December}
  \bibinfo{year}{2010}), \bibinfo{pages}{6:1--6:30}.
\newblock
\showISSN{1556-4681}


\bibitem[\protect\citeauthoryear{Ma, Kargl, and Weber}{Ma
  et~al\mbox{.}}{2010}]%
        {ma_measuring_2010}
\bibfield{author}{\bibinfo{person}{Zhendong Ma}, \bibinfo{person}{Frank Kargl},
  {and} \bibinfo{person}{Michael Weber}.} \bibinfo{year}{2010}\natexlab{}.
\newblock \showarticletitle{{Measuring long-term location privacy in vehicular
  communication systems}}.
\newblock \bibinfo{journal}{\emph{Elsevier Computer Communications}}
  \bibinfo{volume}{33}, \bibinfo{number}{12} (\bibinfo{date}{March}
  \bibinfo{year}{2010}), \bibinfo{pages}{1414--1427}.
\newblock
\showISSN{0140-3664}


\bibitem[\protect\citeauthoryear{Machanavajjhala, Kifer, Gehrke, and
  Venkitasubramaniam}{Machanavajjhala et~al\mbox{.}}{2007}]%
        {machanavajjhala_l-diversity:_2007}
\bibfield{author}{\bibinfo{person}{Ashwin Machanavajjhala},
  \bibinfo{person}{Daniel Kifer}, \bibinfo{person}{Johannes Gehrke}, {and}
  \bibinfo{person}{Muthuramakrishnan Venkitasubramaniam}.}
  \bibinfo{year}{2007}\natexlab{}.
\newblock \showarticletitle{{L-diversity: Privacy beyond k-anonymity}}.
\newblock \bibinfo{journal}{\emph{ACM Trans. on Knowledge Discovery from Data}}
  \bibinfo{volume}{1}, \bibinfo{number}{1} (\bibinfo{date}{March}
  \bibinfo{year}{2007}), \bibinfo{pages}{3:1--3:52}.
\newblock
\showISSN{1556-4681}


\bibitem[\protect\citeauthoryear{Marcellin, Zighed, and Ritschard}{Marcellin
  et~al\mbox{.}}{2006}]%
        {marcellin_asymmetric_2006}
\bibfield{author}{\bibinfo{person}{Simon Marcellin}, \bibinfo{person}{Djamel~A.
  Zighed}, {and} \bibinfo{person}{Gilbert Ritschard}.}
  \bibinfo{year}{2006}\natexlab{}.
\newblock \showarticletitle{An asymmetric entropy measure for decision trees}.
  \bibinfo{address}{Paris, France}, \bibinfo{pages}{1292--1299}.
\newblock
\urldef\tempurl%
\url{http://archive-ouverte.unige.ch/unige:4531}
\showURL{%
\tempurl}


\bibitem[\protect\citeauthoryear{Mashey}{Mashey}{2004}]%
        {mashey_war_2004}
\bibfield{author}{\bibinfo{person}{John~R. Mashey}.}
  \bibinfo{year}{2004}\natexlab{}.
\newblock \showarticletitle{War of the {Benchmark} {Means}: {Time} for a
  {Truce}}.
\newblock \bibinfo{journal}{\emph{SIGARCH Comput. Archit. News}}
  \bibinfo{volume}{32}, \bibinfo{number}{4} (\bibinfo{date}{Sept.}
  \bibinfo{year}{2004}), \bibinfo{pages}{1--14}.
\newblock
\showISSN{0163-5964}


\bibitem[\protect\citeauthoryear{McLaughlin, McDaniel, and Aiello}{McLaughlin
  et~al\mbox{.}}{2011}]%
        {mclaughlin_protecting_2011}
\bibfield{author}{\bibinfo{person}{Stephen McLaughlin},
  \bibinfo{person}{Patrick McDaniel}, {and} \bibinfo{person}{William Aiello}.}
  \bibinfo{year}{2011}\natexlab{}.
\newblock \showarticletitle{{Protecting Consumer Privacy from Electric Load
  Monitoring}}. In \bibinfo{booktitle}{\emph{Proc. 18th {ACM} Conf. on Computer
  and Communications Security (CCS'11)}}. \bibinfo{publisher}{{ACM}},
  \bibinfo{address}{Chicago, IL}, \bibinfo{pages}{87--98}.
\newblock
\showISBNx{978-1-4503-0948-6}


\bibitem[\protect\citeauthoryear{McSherry}{McSherry}{2009}]%
        {mcsherry_privacy_2009}
\bibfield{author}{\bibinfo{person}{Frank~D. McSherry}.}
  \bibinfo{year}{2009}\natexlab{}.
\newblock \showarticletitle{{Privacy Integrated Queries: An Extensible Platform
  for Privacy-Preserving Data Analysis}}. In \bibinfo{booktitle}{\emph{Proc.
  2009 {ACM} {SIGMOD} Int. Conf. on Management of Data (SIGMOD 2004)}}.
  \bibinfo{publisher}{{ACM}}, \bibinfo{address}{Providence, RI, USA},
  \bibinfo{pages}{19--30}.
\newblock


\bibitem[\protect\citeauthoryear{Merugu and Ghosh}{Merugu and Ghosh}{2003}]%
        {merugu_privacy-preserving_2003}
\bibfield{author}{\bibinfo{person}{Srujana Merugu} {and}
  \bibinfo{person}{Joydeep Ghosh}.} \bibinfo{year}{2003}\natexlab{}.
\newblock \showarticletitle{{Privacy-preserving Distributed Clustering using
  Generative Models}}. In \bibinfo{booktitle}{\emph{Proc. 3rd Int. Conf. on
  Data Mining (ICDM'03)}}. \bibinfo{publisher}{{IEEE}},
  \bibinfo{address}{Melbourne, FL, USA}, \bibinfo{pages}{211--218}.
\newblock


\bibitem[\protect\citeauthoryear{Miklau and Suciu}{Miklau and Suciu}{2004}]%
        {miklau2004formal}
\bibfield{author}{\bibinfo{person}{Gerome Miklau} {and} \bibinfo{person}{Dan
  Suciu}.} \bibinfo{year}{2004}\natexlab{}.
\newblock \showarticletitle{A {{Formal Analysis}} of {{Information Disclosure}}
  in {{Data Exchange}}}. In \bibinfo{booktitle}{\emph{Proc. 2004 {{ACM SIGMOD
  International Conference}} on {{Management}} of {{Data}}}}
  \emph{(\bibinfo{series}{SIGMOD '04})}. \bibinfo{publisher}{{ACM}},
  \bibinfo{address}{New York, NY, USA}, \bibinfo{pages}{575--586}.
\newblock
\showISBNx{978-1-58113-859-7}


\bibitem[\protect\citeauthoryear{Mironov, Pandey, Reingold, and Vadhan}{Mironov
  et~al\mbox{.}}{2009}]%
        {mironov_computational_2009}
\bibfield{author}{\bibinfo{person}{Ilya Mironov}, \bibinfo{person}{Omkant
  Pandey}, \bibinfo{person}{Omer Reingold}, {and} \bibinfo{person}{Salil
  Vadhan}.} \bibinfo{year}{2009}\natexlab{}.
\newblock \showarticletitle{{Computational Differential Privacy}}. In
  \bibinfo{booktitle}{\emph{Proc. 29th Annu. Int. Cryptology Conf. (CRYPTO
  2009)}} \emph{(\bibinfo{series}{LNCS 5677})}. \bibinfo{publisher}{Springer},
  \bibinfo{address}{Santa Barbara, CA, USA}, \bibinfo{pages}{126--142}.
\newblock


\bibitem[\protect\citeauthoryear{Murdoch}{Murdoch}{2013}]%
        {murdoch_quantifying_2014}
\bibfield{author}{\bibinfo{person}{Steven~J. Murdoch}.}
  \bibinfo{year}{2013}\natexlab{}.
\newblock \showarticletitle{{Quantifying and Measuring Anonymity}}. In
  \bibinfo{booktitle}{\emph{7th Int. Workshop on Autonomous and Spontaneous
  Security (SETOP 2013)}} \emph{(\bibinfo{series}{LNCS})}.
  \bibinfo{publisher}{Springer}, \bibinfo{address}{Rhul, UK},
  \bibinfo{pages}{3--13}.
\newblock
\showISBNx{978-3-642-54567-2, 978-3-642-54568-9}


\bibitem[\protect\citeauthoryear{Murdoch and Watson}{Murdoch and
  Watson}{2008}]%
        {murdoch_metrics_2008}
\bibfield{author}{\bibinfo{person}{Steven~J. Murdoch} {and}
  \bibinfo{person}{Robert N.~M. Watson}.} \bibinfo{year}{2008}\natexlab{}.
\newblock \showarticletitle{{Metrics for Security and Performance in
  Low-Latency Anonymity Systems}}. In \bibinfo{booktitle}{\emph{Proc. 8th Int.
  Symp. on Privacy Enhancing Technologies (PETS)}} \emph{(\bibinfo{series}{LNCS
  5134})}. \bibinfo{address}{Leuven, Belgium}, \bibinfo{pages}{115--132}.
\newblock
\showISBNx{978-3-540-70629-8, 978-3-540-70630-4}


\bibitem[\protect\citeauthoryear{Nabar, Kenthapadi, Mishra, and Motwani}{Nabar
  et~al\mbox{.}}{2008}]%
        {nabar2008survey}
\bibfield{author}{\bibinfo{person}{Shubha~U. Nabar},
  \bibinfo{person}{Krishnaram Kenthapadi}, \bibinfo{person}{Nina Mishra}, {and}
  \bibinfo{person}{Rajeev Motwani}.} \bibinfo{year}{2008}\natexlab{}.
\newblock \showarticletitle{A {{Survey}} of {{Query Auditing Techniques}} for
  {{Data Privacy}}}.
\newblock In \bibinfo{booktitle}{\emph{Privacy-{{Preserving Data Mining}}}},
  \bibfield{editor}{\bibinfo{person}{Charu~C. Aggarwal} {and}
  \bibinfo{person}{Philip~S. Yu}} (Eds.). Number~34 in
  \bibinfo{series}{Advances in Database Systems}. \bibinfo{publisher}{{Springer
  US}}, \bibinfo{pages}{415--431}.
\newblock
\showISBNx{978-0-387-70991-8 978-0-387-70992-5}


\bibitem[\protect\citeauthoryear{Narayanan and Shmatikov}{Narayanan and
  Shmatikov}{2008}]%
        {narayanan_robust_2008}
\bibfield{author}{\bibinfo{person}{Arvind Narayanan} {and}
  \bibinfo{person}{Vitaly Shmatikov}.} \bibinfo{year}{2008}\natexlab{}.
\newblock \showarticletitle{{Robust De-anonymization of Large Sparse
  Datasets}}. In \bibinfo{booktitle}{\emph{Proc. 2008 IEEE Symp. on Security
  and Privacy (S\&P 2008)}}. \bibinfo{publisher}{{IEEE}},
  \bibinfo{address}{May}, \bibinfo{pages}{111--125}.
\newblock


\bibitem[\protect\citeauthoryear{Narayanan and Shmatikov}{Narayanan and
  Shmatikov}{2009}]%
        {narayanan_-anonymizing_2009}
\bibfield{author}{\bibinfo{person}{Arvind Narayanan} {and}
  \bibinfo{person}{Vitaly Shmatikov}.} \bibinfo{year}{2009}\natexlab{}.
\newblock \showarticletitle{{De-anonymizing Social Networks}}. In
  \bibinfo{booktitle}{\emph{Proc. 2009 30th IEEE Symp. on Security and Privacy
  (S\&P 2009)}}. \bibinfo{publisher}{IEEE}, \bibinfo{address}{Oakland, CA,
  USA}, \bibinfo{pages}{173--187}.
\newblock


\bibitem[\protect\citeauthoryear{Nergiz, Atzori, and Clifton}{Nergiz
  et~al\mbox{.}}{2007}]%
        {nergiz_hiding_2007}
\bibfield{author}{\bibinfo{person}{Mehmet~Ercan Nergiz},
  \bibinfo{person}{Maurizio Atzori}, {and} \bibinfo{person}{Chris Clifton}.}
  \bibinfo{year}{2007}\natexlab{}.
\newblock \showarticletitle{Hiding the Presence of Individuals from Shared
  Databases}. In \bibinfo{booktitle}{\emph{{Proc. 2007 {ACM} Int. Conf. on
  Management of Data (SIGMOD 2007)}}}. \bibinfo{publisher}{{ACM}},
  \bibinfo{address}{Beijing, China}, \bibinfo{pages}{665--676}.
\newblock


\bibitem[\protect\citeauthoryear{Nergiz, Clifton, and Nergiz}{Nergiz
  et~al\mbox{.}}{2009}]%
        {nergiz_multirelational_2009}
\bibfield{author}{\bibinfo{person}{Mehmet~Ercan Nergiz}, \bibinfo{person}{Chris
  Clifton}, {and} \bibinfo{person}{Ahmet~Erhan Nergiz}.}
  \bibinfo{year}{2009}\natexlab{}.
\newblock \showarticletitle{{MultiRelational k-Anonymity}}.
\newblock \bibinfo{journal}{\emph{IEEE Trans. on Knowledge Data Engineering}}
  \bibinfo{volume}{21}, \bibinfo{number}{8} (\bibinfo{date}{August}
  \bibinfo{year}{2009}), \bibinfo{pages}{1104--1117}.
\newblock


\bibitem[\protect\citeauthoryear{Nissenbaum}{Nissenbaum}{2004}]%
        {nissenbaum2004privacy}
\bibfield{author}{\bibinfo{person}{Helen Nissenbaum}.}
  \bibinfo{year}{2004}\natexlab{}.
\newblock \showarticletitle{{Privacy as Contextual Integrity}}.
\newblock \bibinfo{journal}{\emph{Washington Law Review}} \bibinfo{volume}{79},
  \bibinfo{number}{1} (\bibinfo{year}{2004}), \bibinfo{pages}{119--158}.
\newblock


\bibitem[\protect\citeauthoryear{OECD}{OECD}{2013}]%
        {oecd2013privacy}
\bibfield{author}{\bibinfo{person}{OECD}.} \bibinfo{year}{2013}\natexlab{}.
\newblock \bibinfo{booktitle}{\emph{Recommendation of the Council concerning
  Guidelines governing the Protection of Privacy and Transborder Flows of
  Personal Data}}.
\newblock \bibinfo{type}{Article} C(2013)79.
  \bibinfo{institution}{{Organisation for Economic Co-operation and
  Development}}.
\newblock


\bibitem[\protect\citeauthoryear{Oliveira and Zaïane}{Oliveira and
  Zaïane}{2003}]%
        {oliveira_privacy_2003}
\bibfield{author}{\bibinfo{person}{Stanley R.~M. Oliveira} {and}
  \bibinfo{person}{Osmar~R. Zaïane}.} \bibinfo{year}{2003}\natexlab{}.
\newblock \showarticletitle{{Privacy Preserving Clustering By Data
  Transformation}}. In \bibinfo{booktitle}{\emph{Proc. 18th Brazilian Symp. on
  Databases (SBBD'2003)}}. \bibinfo{address}{Manaus, Brazil},
  \bibinfo{pages}{304--318}.
\newblock


\bibitem[\protect\citeauthoryear{Olteanu, Huguenin, Shokri, and Hubaux}{Olteanu
  et~al\mbox{.}}{2014}]%
        {olteanu_quantifying_2014}
\bibfield{author}{\bibinfo{person}{Alexandra-Mihaela Olteanu},
  \bibinfo{person}{Kévin Huguenin}, \bibinfo{person}{Reza Shokri}, {and}
  \bibinfo{person}{Jean-Pierre Hubaux}.} \bibinfo{year}{2014}\natexlab{}.
\newblock \showarticletitle{{Quantifying the Effect of Co-location Information
  on Location Privacy}}. In \bibinfo{booktitle}{\emph{Proc. 14th Int. Symp. on
  Privacy Enhancing Technologies (PETS 2014)}} \emph{(\bibinfo{series}{LNCS
  8555})}. \bibinfo{publisher}{Springer}, \bibinfo{address}{Amsterdam,
  Netherlands}, \bibinfo{pages}{184--203}.
\newblock
\showISBNx{978-3-319-08505-0, 978-3-319-08506-7}


\bibitem[\protect\citeauthoryear{Oya, Troncoso, and Pérez-González}{Oya
  et~al\mbox{.}}{2014}]%
        {oya_dummies_2014}
\bibfield{author}{\bibinfo{person}{Simon Oya}, \bibinfo{person}{Carmela
  Troncoso}, {and} \bibinfo{person}{Fernando Pérez-González}.}
  \bibinfo{year}{2014}\natexlab{}.
\newblock \showarticletitle{{Do Dummies Pay Off? Limits of Dummy Traffic
  Protection in Anonymous Communications}}. In \bibinfo{booktitle}{\emph{Proc.
  14th Int. Symp. on Privacy Enhancing Technologies (PETS 2014)}}
  \emph{(\bibinfo{series}{LNCS 8555})}. \bibinfo{publisher}{Springer},
  \bibinfo{address}{Amsterdam, Netherlands}, \bibinfo{pages}{204--223}.
\newblock
\showISBNx{978-3-319-08505-0, 978-3-319-08506-7}


\bibitem[\protect\citeauthoryear{Preibusch}{Preibusch}{2013}]%
        {preibusch_guide_2013}
\bibfield{author}{\bibinfo{person}{Sören Preibusch}.}
  \bibinfo{year}{2013}\natexlab{}.
\newblock \showarticletitle{Guide to measuring privacy concern: {Review} of
  survey and observational instruments}.
\newblock \bibinfo{journal}{\emph{International Journal of Human-Computer
  Studies}} \bibinfo{volume}{71}, \bibinfo{number}{12} (\bibinfo{date}{Dec.}
  \bibinfo{year}{2013}), \bibinfo{pages}{1133--1143}.
\newblock
\showISSN{1071-5819}


\bibitem[\protect\citeauthoryear{Rastogi, Suciu, and Hong}{Rastogi
  et~al\mbox{.}}{2007}]%
        {rastogi_boundary_2007}
\bibfield{author}{\bibinfo{person}{Vibhor Rastogi}, \bibinfo{person}{Dan
  Suciu}, {and} \bibinfo{person}{Sungho Hong}.}
  \bibinfo{year}{2007}\natexlab{}.
\newblock \showarticletitle{{The Boundary Between Privacy and Utility in Data
  Publishing}}. In \bibinfo{booktitle}{\emph{Proc. 33rd Int. Conf. on Very
  Large Data Bases (VLDB 2007)}}. \bibinfo{publisher}{{VLDB} Endowment},
  \bibinfo{address}{September}, \bibinfo{pages}{531--542}.
\newblock


\bibitem[\protect\citeauthoryear{Reiter and Rubin}{Reiter and Rubin}{1998}]%
        {reiter_crowds:_1998}
\bibfield{author}{\bibinfo{person}{Michael~K. Reiter} {and}
  \bibinfo{person}{Aviel~D. Rubin}.} \bibinfo{year}{1998}\natexlab{}.
\newblock \showarticletitle{{Crowds: Anonymity for Web Transactions}}.
\newblock \bibinfo{journal}{\emph{{ACM} Transactions on Information and System
  Security ({TISSEC})}} \bibinfo{volume}{1}, \bibinfo{number}{1}
  (\bibinfo{date}{November} \bibinfo{year}{1998}), \bibinfo{pages}{66--92}.
\newblock


\bibitem[\protect\citeauthoryear{Rynkiewicz}{Rynkiewicz}{2015}]%
        {rynkiewicz2015private}
\bibfield{author}{\bibinfo{person}{Stephen Rynkiewicz}.}
  \bibinfo{year}{2015}\natexlab{}.
\newblock \bibinfo{title}{Private {{Data}} and {{Public Health}}: {{How Chicago
  Health Atlas Protects Identities}}}.
\newblock   (\bibinfo{date}{July} \bibinfo{year}{2015}).
\newblock
\urldef\tempurl%
\url{http://www.smartchicagocollaborative.org/health-data-privacy-security/}
\showURL{%
\tempurl}
\newblock
\shownote{Accessed Aug. 30, 2017.}


\bibitem[\protect\citeauthoryear{Samarati}{Samarati}{2001}]%
        {samarati2001protecting}
\bibfield{author}{\bibinfo{person}{Pierangela Samarati}.}
  \bibinfo{year}{2001}\natexlab{}.
\newblock \showarticletitle{Protecting respondents identities in microdata
  release}.
\newblock \bibinfo{journal}{\emph{IEEE transactions on Knowledge and Data
  Engineering}} \bibinfo{volume}{13}, \bibinfo{number}{6}
  (\bibinfo{date}{August} \bibinfo{year}{2001}), \bibinfo{pages}{1010--1027}.
\newblock


\bibitem[\protect\citeauthoryear{Samarati and Sweeney}{Samarati and
  Sweeney}{1998}]%
        {samarati_protecting_1998}
\bibfield{author}{\bibinfo{person}{Pierangela Samarati} {and}
  \bibinfo{person}{Latanya Sweeney}.} \bibinfo{year}{1998}\natexlab{}.
\newblock \showarticletitle{{Generalizing Data to Provide Anonymity when
  Disclosing Information}}. In \bibinfo{booktitle}{\emph{Proc. of the 17th ACM
  SIGACT-SIGMOD-SIGART Symposium on Principles of Database Systems (PODS
  1998)}}. \bibinfo{publisher}{ACM}, \bibinfo{address}{Seattle, WA, USA},
  \bibinfo{pages}{188}.
\newblock


\bibitem[\protect\citeauthoryear{Sampigethaya, Huang, Li, Poovendran, Matsuura,
  and Sezaki}{Sampigethaya et~al\mbox{.}}{2005}]%
        {sampigethaya_caravan:_2005}
\bibfield{author}{\bibinfo{person}{Krishna Sampigethaya},
  \bibinfo{person}{Leping Huang}, \bibinfo{person}{Mingyan Li},
  \bibinfo{person}{Radha Poovendran}, \bibinfo{person}{Kanta Matsuura}, {and}
  \bibinfo{person}{Kaoru Sezaki}.} \bibinfo{year}{2005}\natexlab{}.
\newblock \showarticletitle{{CARAVAN: Providing location privacy for VANET}}.
  In \bibinfo{booktitle}{\emph{Embedded Security in Cars (ESCAR'05)}}.
  \bibinfo{address}{Tallinn, Estonia}, \bibinfo{pages}{29--37}.
\newblock


\bibitem[\protect\citeauthoryear{Sankar, Rajagopalan, and Poor}{Sankar
  et~al\mbox{.}}{2013}]%
        {sankar2013utility-privacy}
\bibfield{author}{\bibinfo{person}{Lalitha Sankar}, \bibinfo{person}{S.~Raj
  Rajagopalan}, {and} \bibinfo{person}{H.~Vincent Poor}.}
  \bibinfo{year}{2013}\natexlab{}.
\newblock \showarticletitle{Utility-{{Privacy Tradeoffs}} in {{Databases}}:
  {{An Information}}-{{Theoretic Approach}}}.
\newblock \bibinfo{journal}{\emph{IEEE Transactions on Information Forensics
  and Security}} \bibinfo{volume}{8}, \bibinfo{number}{6} (\bibinfo{date}{June}
  \bibinfo{year}{2013}), \bibinfo{pages}{838--852}.
\newblock
\showISSN{1556-6013}


\bibitem[\protect\citeauthoryear{Serjantov and Danezis}{Serjantov and
  Danezis}{2002}]%
        {serjantov_towards_2003}
\bibfield{author}{\bibinfo{person}{Andrei Serjantov} {and}
  \bibinfo{person}{George Danezis}.} \bibinfo{year}{2002}\natexlab{}.
\newblock \showarticletitle{{Towards an Information Theoretic Metric for
  Anonymity}}. In \bibinfo{booktitle}{\emph{Proc. 2nd Int. Symp. on Privacy
  Enhancing Technologies (PETS 2002)}} \emph{(\bibinfo{series}{LNCS 2482})}.
  \bibinfo{publisher}{Springer}, \bibinfo{address}{San Francisco, CA, USA},
  \bibinfo{pages}{41--53}.
\newblock
\showISBNx{978-3-540-00565-0, 978-3-540-36467-2}


\bibitem[\protect\citeauthoryear{Seys and Preneel}{Seys and Preneel}{2009}]%
        {seys_arm:_2009}
\bibfield{author}{\bibinfo{person}{Stefaan Seys} {and} \bibinfo{person}{Bart
  Preneel}.} \bibinfo{year}{2009}\natexlab{}.
\newblock \showarticletitle{{ARM: Anonymous Routing Protocol for Mobile Ad hoc
  Networks}}.
\newblock \bibinfo{journal}{\emph{Int. Journal of Wireless and Mobile
  Computing}} \bibinfo{volume}{3}, \bibinfo{number}{3} (\bibinfo{date}{October}
  \bibinfo{year}{2009}), \bibinfo{pages}{145--155}.
\newblock


\bibitem[\protect\citeauthoryear{Shabtai, Elovici, and Rokach}{Shabtai
  et~al\mbox{.}}{2012}]%
        {shabtai2012survey}
\bibfield{author}{\bibinfo{person}{Asaf Shabtai}, \bibinfo{person}{Yuval
  Elovici}, {and} \bibinfo{person}{Lior Rokach}.}
  \bibinfo{year}{2012}\natexlab{}.
\newblock \bibinfo{booktitle}{\emph{A {{Survey}} of {{Data Leakage Detection}}
  and {{Prevention Solutions}}}}.
\newblock \bibinfo{publisher}{{Springer Science \& Business Media}}.
\newblock
\showISBNx{978-1-4614-2053-8}


\bibitem[\protect\citeauthoryear{Shannon}{Shannon}{1948}]%
        {shannon_mathematical_1948}
\bibfield{author}{\bibinfo{person}{Claude~E. Shannon}.}
  \bibinfo{year}{1948}\natexlab{}.
\newblock \showarticletitle{{A Mathematical Theory of Communication}}.
\newblock \bibinfo{journal}{\emph{Bell System Technical Journal}}
  \bibinfo{volume}{27} (\bibinfo{date}{October} \bibinfo{year}{1948}),
  \bibinfo{pages}{379--423 \& 623--656}.
\newblock


\bibitem[\protect\citeauthoryear{Shannon}{Shannon}{1949}]%
        {shannon1949communication}
\bibfield{author}{\bibinfo{person}{Claude~E. Shannon}.}
  \bibinfo{year}{1949}\natexlab{}.
\newblock \showarticletitle{Communication Theory of Secrecy Systems}.
\newblock \bibinfo{journal}{\emph{Bell system technical journal}}
  \bibinfo{volume}{28}, \bibinfo{number}{4} (\bibinfo{year}{1949}),
  \bibinfo{pages}{656--715}.
\newblock


\bibitem[\protect\citeauthoryear{Shao, Yang, Zhu, and Cao}{Shao
  et~al\mbox{.}}{2008}]%
        {shao_towards_2008}
\bibfield{author}{\bibinfo{person}{Min Shao}, \bibinfo{person}{Yi Yang},
  \bibinfo{person}{Sencun Zhu}, {and} \bibinfo{person}{Guohong Cao}.}
  \bibinfo{year}{2008}\natexlab{}.
\newblock \showarticletitle{{Towards Statistically Strong Source Anonymity for
  Sensor Networks}}. In \bibinfo{booktitle}{\emph{Proc. 27th Conf. on Computer
  Communications (INFOCOM 2008)}}. \bibinfo{publisher}{{IEEE}},
  \bibinfo{address}{Phoenix, AZ, USA}, \bibinfo{pages}{466--474}.
\newblock


\bibitem[\protect\citeauthoryear{Shi, Chan, Rieffel, Chow, and Song}{Shi
  et~al\mbox{.}}{2011}]%
        {shi_privacy-preserving_2011}
\bibfield{author}{\bibinfo{person}{Elaine Shi}, \bibinfo{person}{T-H.~Hubert
  Chan}, \bibinfo{person}{Eleanor~G. Rieffel}, \bibinfo{person}{Richard Chow},
  {and} \bibinfo{person}{Dawn Song}.} \bibinfo{year}{2011}\natexlab{}.
\newblock \showarticletitle{{Privacy-Preserving Aggregation of Time-Series
  Data}}. In \bibinfo{booktitle}{\emph{Proc. 18th Annu. Network \& Distributed
  System Security Symp. ({NDSS})}}, Vol.~\bibinfo{volume}{2}.
  \bibinfo{address}{San Diego, CA}, \bibinfo{pages}{4}.
\newblock


\bibitem[\protect\citeauthoryear{Shmatikov}{Shmatikov}{2002}]%
        {shmatikov_probabilistic_2002}
\bibfield{author}{\bibinfo{person}{Vitaly Shmatikov}.}
  \bibinfo{year}{2002}\natexlab{}.
\newblock \showarticletitle{{Probabilistic Analysis of Anonymity}}. In
  \bibinfo{booktitle}{\emph{Proc. 15th {IEEE} Computer Security Foundations
  Workshop (CSFW-15)}}. \bibinfo{publisher}{{IEEE}}, \bibinfo{address}{Cape
  Breton, Canada}, \bibinfo{pages}{119--128}.
\newblock


\bibitem[\protect\citeauthoryear{Shokri, Freudiger, and Hubaux}{Shokri
  et~al\mbox{.}}{2010a}]%
        {shokri_unified_2010}
\bibfield{author}{\bibinfo{person}{Reza Shokri}, \bibinfo{person}{Julien
  Freudiger}, {and} \bibinfo{person}{Jean-Pierre Hubaux}.}
  \bibinfo{year}{2010}\natexlab{a}.
\newblock \showarticletitle{{A Unified Framework for Location Privacy}}. In
  \bibinfo{booktitle}{\emph{Proc. 3rd Symp. on Hot Topics in Privacy Enhancing
  Technologies (HotPETs 2010)}}. \bibinfo{address}{Berlin, Germany}.
\newblock


\bibitem[\protect\citeauthoryear{Shokri, Theodorakopoulos, Le~Boudec, and
  Hubaux}{Shokri et~al\mbox{.}}{2011}]%
        {shokri_quantifying_2011}
\bibfield{author}{\bibinfo{person}{Reza Shokri}, \bibinfo{person}{George
  Theodorakopoulos}, \bibinfo{person}{Jean-Yves Le~Boudec}, {and}
  \bibinfo{person}{Jean-Pierre Hubaux}.} \bibinfo{year}{2011}\natexlab{}.
\newblock \showarticletitle{{Quantifying Location Privacy}}. In
  \bibinfo{booktitle}{\emph{Proc. 2011 32nd IEEE Symp. on Security and Privacy
  (S\&P 2011)}}. \bibinfo{publisher}{IEEE}, \bibinfo{address}{Oakland, CA,
  USA}, \bibinfo{pages}{247--262}.
\newblock


\bibitem[\protect\citeauthoryear{Shokri, Troncoso, Diaz, Freudiger, and
  Hubaux}{Shokri et~al\mbox{.}}{2010b}]%
        {shokri_unraveling_2010}
\bibfield{author}{\bibinfo{person}{Reza Shokri}, \bibinfo{person}{Carmela
  Troncoso}, \bibinfo{person}{Claudia Diaz}, \bibinfo{person}{Julien
  Freudiger}, {and} \bibinfo{person}{Jean-Pierre Hubaux}.}
  \bibinfo{year}{2010}\natexlab{b}.
\newblock \showarticletitle{{Unraveling an Old Cloak: K-anonymity for Location
  Privacy}}. In \bibinfo{booktitle}{\emph{Proc. 9th {ACM} Workshop on Privacy
  in the Electronic Society (WPES 2010)}}. \bibinfo{publisher}{{ACM}},
  \bibinfo{address}{Chicago, Illinois, USA}, \bibinfo{pages}{115--118}.
\newblock
\showISBNx{978-1-4503-0096-4}


\bibitem[\protect\citeauthoryear{Soria-Comas and Domingo-Ferrer}{Soria-Comas
  and Domingo-Ferrer}{2013}]%
        {soria-comas_differential_2013}
\bibfield{author}{\bibinfo{person}{Jordi Soria-Comas} {and}
  \bibinfo{person}{Josep Domingo-Ferrer}.} \bibinfo{year}{2013}\natexlab{}.
\newblock \showarticletitle{{Differential Privacy via t-Closeness in Data
  Publishing}}. In \bibinfo{booktitle}{\emph{Proc. 11th Annu. Conf. on Privacy,
  Security and Trust (PST2013)}}. \bibinfo{publisher}{{IEEE}},
  \bibinfo{address}{Tarragona, Spain}, \bibinfo{pages}{27--35}.
\newblock


\bibitem[\protect\citeauthoryear{Steinbrecher and Köpsell}{Steinbrecher and
  Köpsell}{2003}]%
        {steinbrecher_modelling_2003}
\bibfield{author}{\bibinfo{person}{Sandra Steinbrecher} {and}
  \bibinfo{person}{Stefan Köpsell}.} \bibinfo{year}{2003}\natexlab{}.
\newblock \showarticletitle{{Modelling Unlinkability}}. In
  \bibinfo{booktitle}{\emph{Proc. 3rd Int. Workshop on Privacy Enhancing
  Technologies (PET 2003)}} \emph{(\bibinfo{series}{LNCS 2760})}.
  \bibinfo{publisher}{Springer}, \bibinfo{address}{Dresden, Germany},
  \bibinfo{pages}{32--47}.
\newblock
\showISBNx{978-3-540-20610-1, 978-3-540-40956-4}


\bibitem[\protect\citeauthoryear{Sweeney}{Sweeney}{2002}]%
        {sweeney_k-anonymity:_2002}
\bibfield{author}{\bibinfo{person}{Latanya Sweeney}.}
  \bibinfo{year}{2002}\natexlab{}.
\newblock \showarticletitle{{k-Anonymity: A Model for Protecting Privacy}}.
\newblock \bibinfo{journal}{\emph{Int. Journal of Uncertainty, Fuzziness and
  Knowledge-Based Systems}} \bibinfo{volume}{10}, \bibinfo{number}{05}
  (\bibinfo{date}{October} \bibinfo{year}{2002}), \bibinfo{pages}{557--570}.
\newblock
\showISSN{0218-4885}


\bibitem[\protect\citeauthoryear{Syverson}{Syverson}{2013}]%
        {syverson_why_2013}
\bibfield{author}{\bibinfo{person}{Paul Syverson}.}
  \bibinfo{year}{2013}\natexlab{}.
\newblock \showarticletitle{{Why I’m Not an Entropist}}. In
  \bibinfo{booktitle}{\emph{Proc. 17th Int. Workshop on Security Protocols}}
  \emph{(\bibinfo{series}{LNCS 7028})}. \bibinfo{publisher}{Springer},
  \bibinfo{address}{Cambridge, UK}, \bibinfo{pages}{213--230}.
\newblock
\showISBNx{978-3-642-36212-5, 978-3-642-36213-2}


\bibitem[\protect\citeauthoryear{Thomas, Grier, and Nicol}{Thomas
  et~al\mbox{.}}{2010}]%
        {thomas_unfriendly_2010}
\bibfield{author}{\bibinfo{person}{Kurt Thomas}, \bibinfo{person}{Chris Grier},
  {and} \bibinfo{person}{David~M. Nicol}.} \bibinfo{year}{2010}\natexlab{}.
\newblock \showarticletitle{{unFriendly: Multi-party Privacy Risks in Social
  Networks}}. In \bibinfo{booktitle}{\emph{Proc. 10th Int. Symp. on Privacy
  Enhancing Technologies (PETS 2010)}} \emph{(\bibinfo{series}{LNCS 6205})}.
  \bibinfo{publisher}{Springer}, \bibinfo{address}{Berlin, Germany},
  \bibinfo{pages}{236--252}.
\newblock
\showISBNx{978-3-642-14526-1, 978-3-642-14527-8}


\bibitem[\protect\citeauthoryear{Tóth, Hornák, and Vajda}{Tóth
  et~al\mbox{.}}{2004}]%
        {toth_measuring_2004}
\bibfield{author}{\bibinfo{person}{Gergely Tóth}, \bibinfo{person}{Zoltán
  Hornák}, {and} \bibinfo{person}{Ferenc Vajda}.}
  \bibinfo{year}{2004}\natexlab{}.
\newblock \showarticletitle{{Measuring Anonymity Revisited}}. In
  \bibinfo{booktitle}{\emph{Proc. 9th Nordic Workshop on Secure IT Systems
  (Nordsec 2004)}}. \bibinfo{address}{Espoo, Finland}, \bibinfo{pages}{85--90}.
\newblock


\bibitem[\protect\citeauthoryear{{United Nations}}{{United Nations}}{1948}]%
        {UN1948universal}
\bibfield{author}{\bibinfo{person}{{United Nations}}.}
  \bibinfo{year}{1948}\natexlab{}.
\newblock \bibinfo{booktitle}{\emph{{The Universal Declaration of Human
  Rights}}}.
\newblock \bibinfo{type}{Resolution} 217 A.
\newblock


\bibitem[\protect\citeauthoryear{Vorob’ev}{Vorob’ev}{1977}]%
        {vorob1977infinite}
\bibfield{author}{\bibinfo{person}{Nikola\u{\i}~Nikolaevich Vorob’ev}.}
  \bibinfo{year}{1977}\natexlab{}.
\newblock \showarticletitle{Infinite antagonistic games}.
\newblock In \bibinfo{booktitle}{\emph{Game Theory}}.
  \bibinfo{series}{Applications of Mathematics}, Vol.~\bibinfo{volume}{7}.
  \bibinfo{publisher}{Springer}, \bibinfo{pages}{56--89}.
\newblock


\bibitem[\protect\citeauthoryear{Vratonjic, Huguenin, Bindschaedler, and
  Hubaux}{Vratonjic et~al\mbox{.}}{2013}]%
        {vratonjic_how_2013}
\bibfield{author}{\bibinfo{person}{Nevena Vratonjic}, \bibinfo{person}{Kévin
  Huguenin}, \bibinfo{person}{Vincent Bindschaedler}, {and}
  \bibinfo{person}{Jean-Pierre Hubaux}.} \bibinfo{year}{2013}\natexlab{}.
\newblock \showarticletitle{{How Others Compromise Your Location Privacy: The
  Case of Shared Public {IPs} at Hotspots}}. In \bibinfo{booktitle}{\emph{Proc.
  13th Int. Symp. on Privacy Enhancing Technologies (PETS 2013)}}
  \emph{(\bibinfo{series}{LNCS 7981})}. \bibinfo{publisher}{Springer},
  \bibinfo{address}{Bloomington, IN, USA}, \bibinfo{pages}{123--142}.
\newblock
\showISBNx{978-3-642-39076-0, 978-3-642-39077-7}


\bibitem[\protect\citeauthoryear{Wagner}{Wagner}{2015}]%
        {wagner2015genomic}
\bibfield{author}{\bibinfo{person}{Isabel Wagner}.}
  \bibinfo{year}{2015}\natexlab{}.
\newblock \showarticletitle{{Genomic Privacy Metrics: A Systematic
  Comparison}}. In \bibinfo{booktitle}{\emph{36th IEEE Symposium on Security
  and Privacy (S{\&}P): 2nd International Workshop on Genome Privacy and
  Security (GenoPri'15)}}. \bibinfo{address}{San Jose, CA}.
\newblock


\bibitem[\protect\citeauthoryear{Wagner}{Wagner}{2017}]%
        {wagner2017evaluating}
\bibfield{author}{\bibinfo{person}{Isabel Wagner}.}
  \bibinfo{year}{2017}\natexlab{}.
\newblock \showarticletitle{{Evaluating the Strength of Genomic Privacy
  Metrics}}.
\newblock \bibinfo{journal}{\emph{ACM Transactions on Privacy and Security}}
  \bibinfo{volume}{20}, \bibinfo{number}{1} (\bibinfo{date}{January}
  \bibinfo{year}{2017}), \bibinfo{pages}{Article 2}.
\newblock
\showISSN{2471-2566}


\bibitem[\protect\citeauthoryear{Wagner and Eckhoff}{Wagner and
  Eckhoff}{2014}]%
        {wagner2014privacy}
\bibfield{author}{\bibinfo{person}{Isabel Wagner} {and} \bibinfo{person}{David
  Eckhoff}.} \bibinfo{year}{2014}\natexlab{}.
\newblock \showarticletitle{{Privacy Assessment in Vehicular Networks Using
  Simulation}}. In \bibinfo{booktitle}{\emph{Proc. Winter Simulation Conf. (WSC
  '14)}}. \bibinfo{address}{Savannah, GA, USA}.
\newblock


\bibitem[\protect\citeauthoryear{Wang and Fung}{Wang and Fung}{2006}]%
        {wang_anonymizing_2006}
\bibfield{author}{\bibinfo{person}{Ke Wang} {and} \bibinfo{person}{Benjamin
  Fung}.} \bibinfo{year}{2006}\natexlab{}.
\newblock \showarticletitle{{Anonymizing Sequential Releases}}. In
  \bibinfo{booktitle}{\emph{Proc. 12th {ACM} {SIGKDD} Int. Conf. on Knowledge
  Discovery and data Mining (KDD'06)}}. \bibinfo{publisher}{{ACM}},
  \bibinfo{address}{Philadelphia, PA, USA}, \bibinfo{pages}{414--423}.
\newblock


\bibitem[\protect\citeauthoryear{Wang, Fung, and Philip}{Wang
  et~al\mbox{.}}{2007}]%
        {wang_handicapping_2007}
\bibfield{author}{\bibinfo{person}{Ke Wang}, \bibinfo{person}{Benjamin~CM
  Fung}, {and} \bibinfo{person}{S.~Yu Philip}.}
  \bibinfo{year}{2007}\natexlab{}.
\newblock \showarticletitle{{Handicapping Attacker's Confidence: An Alternative
  to k-Anonymization}}.
\newblock \bibinfo{journal}{\emph{Knowledge and Information Systems}}
  \bibinfo{volume}{11}, \bibinfo{number}{3} (\bibinfo{date}{April}
  \bibinfo{year}{2007}), \bibinfo{pages}{345--368}.
\newblock


\bibitem[\protect\citeauthoryear{Wang, Wang, Li, Tang, Reiter, and Dong}{Wang
  et~al\mbox{.}}{2009}]%
        {wang_privacy-preserving_2009}
\bibfield{author}{\bibinfo{person}{Rui Wang}, \bibinfo{person}{XiaoFeng Wang},
  \bibinfo{person}{Zhou Li}, \bibinfo{person}{Haixu Tang},
  \bibinfo{person}{Michael~K. Reiter}, {and} \bibinfo{person}{Zheng Dong}.}
  \bibinfo{year}{2009}\natexlab{}.
\newblock \showarticletitle{{Privacy-Preserving Genomic Computation Through
  Program Specialization}}. In \bibinfo{booktitle}{\emph{Proc. 16th {ACM} Conf.
  on Computer and Communications Security (CCS'09)}}.
  \bibinfo{publisher}{{ACM}}, \bibinfo{address}{Chicago, IL, USA},
  \bibinfo{pages}{338--347}.
\newblock
\showISBNx{978-1-60558-894-0}


\bibitem[\protect\citeauthoryear{Westin}{Westin}{1967}]%
        {westin1967privacy}
\bibfield{author}{\bibinfo{person}{Alan Westin}.}
  \bibinfo{year}{1967}\natexlab{}.
\newblock \bibinfo{booktitle}{\emph{{Privacy and Freedom}}}.
\newblock \bibinfo{publisher}{Atheneum}.
\newblock


\bibitem[\protect\citeauthoryear{Wong, Li, Fu, and Wang}{Wong
  et~al\mbox{.}}{2006}]%
        {wong__2006}
\bibfield{author}{\bibinfo{person}{Raymond Chi-Wing Wong},
  \bibinfo{person}{Jiuyong Li}, \bibinfo{person}{Ada Wai-Chee Fu}, {and}
  \bibinfo{person}{Ke Wang}.} \bibinfo{year}{2006}\natexlab{}.
\newblock \showarticletitle{({$\alpha$}, k)-Anonymity: An Enhanced k-Anonymity
  Model for Privacy Preserving Data Publishing}. In
  \bibinfo{booktitle}{\emph{Proc. 12th {ACM} {SIGKDD} Int. Conf. on Knowledge
  Discovery and data Mining (KDD'06)}}. \bibinfo{publisher}{{ACM}},
  \bibinfo{address}{Philadelphia, PA, USA}, \bibinfo{pages}{754--759}.
\newblock


\bibitem[\protect\citeauthoryear{Wright, Adler, Levine, and Shields}{Wright
  et~al\mbox{.}}{2002}]%
        {wright_analysis_2002}
\bibfield{author}{\bibinfo{person}{Matthew Wright}, \bibinfo{person}{Micah
  Adler}, \bibinfo{person}{Brian~Neil Levine}, {and} \bibinfo{person}{Clay
  Shields}.} \bibinfo{year}{2002}\natexlab{}.
\newblock \showarticletitle{{An Analysis of the Degradation of Anonymous
  Protocols}}. In \bibinfo{booktitle}{\emph{Proc. Network and Distributed
  System Security Symp. (NDSS)}}, Vol.~\bibinfo{volume}{2}.
  \bibinfo{address}{San Diego, CA}, \bibinfo{pages}{39--50}.
\newblock


\bibitem[\protect\citeauthoryear{Wright, Adler, Levine, and Shields}{Wright
  et~al\mbox{.}}{2003}]%
        {wright_defending_2003}
\bibfield{author}{\bibinfo{person}{Matthew Wright}, \bibinfo{person}{Micah
  Adler}, \bibinfo{person}{Brian~Neil Levine}, {and} \bibinfo{person}{Clay
  Shields}.} \bibinfo{year}{2003}\natexlab{}.
\newblock \showarticletitle{{Defending Anonymous Communications Against Passive
  Logging Attacks}}. In \bibinfo{booktitle}{\emph{Proc. IEEE Symp. on Research
  in Security and Privacy (S\&P)}}. \bibinfo{address}{Oakland, CA},
  \bibinfo{pages}{28--41}.
\newblock


\bibitem[\protect\citeauthoryear{Xiao and Tao}{Xiao and Tao}{2006}]%
        {xiao_personalized_2006}
\bibfield{author}{\bibinfo{person}{Xiaokui Xiao} {and} \bibinfo{person}{Yufei
  Tao}.} \bibinfo{year}{2006}\natexlab{}.
\newblock \showarticletitle{{Personalized Privacy Preservation}}. In
  \bibinfo{booktitle}{\emph{Proc. 2006 {ACM} {SIGMOD} Int. Conf. Management of
  Data (SIGMOD 2004)}}. \bibinfo{publisher}{ACM}, \bibinfo{address}{Chicago,
  IL, USA}, \bibinfo{pages}{229--240}.
\newblock


\bibitem[\protect\citeauthoryear{Xiao and Tao}{Xiao and Tao}{2007}]%
        {xiao_m-invariance:_2007}
\bibfield{author}{\bibinfo{person}{Xiaokui Xiao} {and} \bibinfo{person}{Yufei
  Tao}.} \bibinfo{year}{2007}\natexlab{}.
\newblock \showarticletitle{{M-invariance: Towards Privacy Preserving
  Re-Publication of Dynamic Datasets}}. In \bibinfo{booktitle}{\emph{Proc. ACM
  SIGMOD Int. Conf. on Management of data (SIGMOD '07)}}.
  \bibinfo{publisher}{ACM}, \bibinfo{address}{Beijing, China},
  \bibinfo{pages}{689--700}.
\newblock
\showISBNx{978-1-59593-686-8}


\bibitem[\protect\citeauthoryear{Xu and Cai}{Xu and Cai}{2009}]%
        {xu_feeling-based_2009}
\bibfield{author}{\bibinfo{person}{Toby Xu} {and} \bibinfo{person}{Ying Cai}.}
  \bibinfo{year}{2009}\natexlab{}.
\newblock \showarticletitle{{Feeling-based Location Privacy Protection for
  Location-based Services}}. In \bibinfo{booktitle}{\emph{Proc. 16th {ACM}
  Conf. on Computer and Communications Security (CCS'09)}}.
  \bibinfo{publisher}{{ACM}}, \bibinfo{address}{Chicago, IL, USA},
  \bibinfo{pages}{338--347}.
\newblock
\showISBNx{978-1-60558-894-0}


\bibitem[\protect\citeauthoryear{Xu, Ma, Tang, and Tian}{Xu
  et~al\mbox{.}}{2014}]%
        {xu2014survey}
\bibfield{author}{\bibinfo{person}{Yang Xu}, \bibinfo{person}{Tinghuai Ma},
  \bibinfo{person}{Meili Tang}, {and} \bibinfo{person}{Wei Tian}.}
  \bibinfo{year}{2014}\natexlab{}.
\newblock \showarticletitle{A {{Survey}} of {{Privacy Preserving Data
  Publishing}} Using {{Generalization}} and {{Suppression}}}.
\newblock \bibinfo{journal}{\emph{Applied Mathematics \& Information Sciences}}
  \bibinfo{volume}{8}, \bibinfo{number}{3} (\bibinfo{date}{May}
  \bibinfo{year}{2014}), \bibinfo{pages}{1103--1116}.
\newblock


\bibitem[\protect\citeauthoryear{Yang, Lutes, Li, Luo, and Liu}{Yang
  et~al\mbox{.}}{2012}]%
        {yang_stalking_2012}
\bibfield{author}{\bibinfo{person}{Yuhao Yang}, \bibinfo{person}{Jonathan
  Lutes}, \bibinfo{person}{Fengjun Li}, \bibinfo{person}{Bo Luo}, {and}
  \bibinfo{person}{Peng Liu}.} \bibinfo{year}{2012}\natexlab{}.
\newblock \showarticletitle{{Stalking Online: On User Privacy in Social
  Networks}}. In \bibinfo{booktitle}{\emph{Proc. 2nd ACM Conf. on Data and
  Application Security and Privacy (CODASPY)}}. \bibinfo{address}{San Antonio,
  TX}, \bibinfo{pages}{37--48}.
\newblock
\showISBNx{978-1-4503-1091-8}


\bibitem[\protect\citeauthoryear{Yang, Shao, Zhu, Urgaonkar, and Cao}{Yang
  et~al\mbox{.}}{2008}]%
        {yang_towards_2008}
\bibfield{author}{\bibinfo{person}{Yi Yang}, \bibinfo{person}{Min Shao},
  \bibinfo{person}{Sencun Zhu}, \bibinfo{person}{Bhuvan Urgaonkar}, {and}
  \bibinfo{person}{Guohong Cao}.} \bibinfo{year}{2008}\natexlab{}.
\newblock \showarticletitle{{Towards Event Source Unobservability with Minimum
  Network Traffic in Sensor Networks}}. In \bibinfo{booktitle}{\emph{Proc. 1st
  ACM Conf. on Wireless Network Security (WiSec'08)}}.
  \bibinfo{publisher}{{ACM}}, \bibinfo{address}{Alexandria, VA, USA},
  \bibinfo{pages}{77--88}.
\newblock
\showISBNx{978-1-59593-814-5}


\bibitem[\protect\citeauthoryear{Yao, Frikken, Atallah, and Tamassia}{Yao
  et~al\mbox{.}}{2008}]%
        {yao_private_2008}
\bibfield{author}{\bibinfo{person}{Danfeng Yao}, \bibinfo{person}{Keith~B.
  Frikken}, \bibinfo{person}{Mikhail~J. Atallah}, {and}
  \bibinfo{person}{Roberto Tamassia}.} \bibinfo{year}{2008}\natexlab{}.
\newblock \showarticletitle{{Private Information: To Reveal or Not to Reveal}}.
\newblock \bibinfo{journal}{\emph{ACM Transactions on Information and Systems
  Security}} \bibinfo{volume}{12}, \bibinfo{number}{1} (\bibinfo{date}{October}
  \bibinfo{year}{2008}), \bibinfo{pages}{6:1--6:27}.
\newblock
\showISSN{1094-9224}


\bibitem[\protect\citeauthoryear{Yu, Fienberg, Slavković, and Uhler}{Yu
  et~al\mbox{.}}{2014}]%
        {yu_scalable_2014}
\bibfield{author}{\bibinfo{person}{Fei Yu}, \bibinfo{person}{Stephen~E.
  Fienberg}, \bibinfo{person}{Aleksandra~B. Slavković}, {and}
  \bibinfo{person}{Caroline Uhler}.} \bibinfo{year}{2014}\natexlab{}.
\newblock \showarticletitle{Scalable privacy-preserving data sharing
  methodology for genome-wide association studies}.
\newblock \bibinfo{journal}{\emph{Journal of Biomedical Informatics}}
  \bibinfo{volume}{50} (\bibinfo{date}{Aug.} \bibinfo{year}{2014}),
  \bibinfo{pages}{133--141}.
\newblock
\showISSN{1532-0464}


\bibitem[\protect\citeauthoryear{Zeadally, Pathan, Alcaraz, and Badra}{Zeadally
  et~al\mbox{.}}{2013}]%
        {zeadally_towards_2013}
\bibfield{author}{\bibinfo{person}{Sherali Zeadally},
  \bibinfo{person}{Al-Sakib~Khan Pathan}, \bibinfo{person}{Cristina Alcaraz},
  {and} \bibinfo{person}{Mohamad Badra}.} \bibinfo{year}{2013}\natexlab{}.
\newblock \showarticletitle{{Towards Privacy Protection in Smart Grid}}.
\newblock \bibinfo{journal}{\emph{Wireless Personal Communications}}
  \bibinfo{volume}{73}, \bibinfo{number}{1} (\bibinfo{date}{November}
  \bibinfo{year}{2013}), \bibinfo{pages}{23--50}.
\newblock
\showISSN{0929-6212, 1572-834X}


\bibitem[\protect\citeauthoryear{Zhang, Zhou, Chen, Wang, and Ruan}{Zhang
  et~al\mbox{.}}{2011}]%
        {zhang_sedic:_2011}
\bibfield{author}{\bibinfo{person}{Kehuan Zhang}, \bibinfo{person}{Xiaoyong
  Zhou}, \bibinfo{person}{Yangyi Chen}, \bibinfo{person}{XiaoFeng Wang}, {and}
  \bibinfo{person}{Yaoping Ruan}.} \bibinfo{year}{2011}\natexlab{}.
\newblock \showarticletitle{{Sedic: Privacy-aware Data Intensive Computing on
  Hybrid Clouds}}. In \bibinfo{booktitle}{\emph{Proc. 18th {ACM} Conf. on
  Computer and Communications Security (CCS)}}. \bibinfo{address}{Chicago, IL},
  \bibinfo{pages}{515--526}.
\newblock
\showISBNx{978-1-4503-0948-6}


\bibitem[\protect\citeauthoryear{Zhang, Jajodia, and Brodsky}{Zhang
  et~al\mbox{.}}{2007a}]%
        {zhang_information_2007}
\bibfield{author}{\bibinfo{person}{Lei Zhang}, \bibinfo{person}{Sushil
  Jajodia}, {and} \bibinfo{person}{Alexander Brodsky}.}
  \bibinfo{year}{2007}\natexlab{a}.
\newblock \showarticletitle{{Information Disclosure Under Realistic
  Assumptions: Privacy Versus Optimality}}. In \bibinfo{booktitle}{\emph{14th
  ACM Conf. on Computer and Communications Security (CCS'07)}}.
  \bibinfo{address}{Alexandria, VA}, \bibinfo{pages}{573--583}.
\newblock
\showISBNx{978-1-59593-703-2}


\bibitem[\protect\citeauthoryear{Zhang, Koudas, Srivastava, and Yu}{Zhang
  et~al\mbox{.}}{2007b}]%
        {zhang_aggregate_2007}
\bibfield{author}{\bibinfo{person}{Qing Zhang}, \bibinfo{person}{Nick Koudas},
  \bibinfo{person}{Divesh Srivastava}, {and} \bibinfo{person}{Ting Yu}.}
  \bibinfo{year}{2007}\natexlab{b}.
\newblock \showarticletitle{{Aggregate Query Answering on Anonymized Tables}}.
  In \bibinfo{booktitle}{\emph{Proc. IEEE 23rd Int. Conf. on Data Engineering
  (ICDE 2007)}}. \bibinfo{publisher}{IEEE}, \bibinfo{address}{Istanbul,
  Turkey}, \bibinfo{pages}{116--125}.
\newblock


\end{thebibliography}

\end{document}